\documentclass[twocolumn]{aastex62}

\usepackage{amssymb, amsfonts, amsmath,framed}

\usepackage{bm}
\expandafter\ifx\csname package@font\endcsname\relax\else
 \expandafter\expandafter
 \expandafter\usepackage
 \expandafter\expandafter
 \expandafter{\csname package@font\endcsname}
\fi
\expandafter\ifx\csname package@font\endcsname\relax\else
 \expandafter\expandafter
 \expandafter\usepackage
 \expandafter\expandafter
 \expandafter{\csname package@font\endcsname}
\fi
\def\bq{\begin{equation}}
\def\eq{\end{equation}}
\def\bqy{\begin{eqnarray}}
\def\eqy{\end{eqnarray}}
\begin{document}
\title{The interplay of magnetically-dominated turbulence and magnetic reconnection in producing nonthermal particles}

\correspondingauthor{Luca Comisso, Lorenzo Sironi}
\email{luca.comisso@columbia.edu, lsironi@astro.columbia.edu}

\author{Luca Comisso}
\affiliation{Department of Astronomy and Columbia Astrophysics Laboratory, Columbia University, New York, NY 10027, USA}

\author{Lorenzo Sironi}
\affiliation{Department of Astronomy and Columbia Astrophysics Laboratory, Columbia University, New York, NY 10027, USA}

\begin{abstract}
Magnetized turbulence and magnetic reconnection are often invoked to explain the nonthermal emission observed from a wide variety of astrophysical sources. By means of fully-kinetic 2D and 3D particle-in-cell simulations, we investigate the interplay between turbulence and reconnection in generating nonthermal particles in magnetically-dominated (or, equivalently, ``relativistic'') pair plasmas. A generic by-product of the turbulence evolution is the generation of a nonthermal particle spectrum with a power-law energy range.  The power-law slope $p$ is harder for larger magnetizations and stronger turbulence fluctuations, and it can be as hard as $p \lesssim 2$. The Larmor radius of particles at the high-energy cutoff is comparable to the size $l$ of the largest turbulent eddies. Plasmoid-mediated  reconnection, which self-consistently occurs in the turbulent plasma, controls the physics of particle injection. Then, particles are further accelerated by stochastic scattering off turbulent fluctuations. The work done by parallel electric fields --- naturally expected in reconnection layers --- is responsible for most of the initial energy increase, and is proportional to the magnetization $\sigma$ of the system, while the subsequent energy gain, which  dominates the overall energization of high-energy particles, is powered by the perpendicular electric fields of turbulent fluctuations. The two-stage acceleration process leaves an imprint in the particle pitch-angle distribution: low-energy particles are aligned with the field, while the highest energy particles move preferentially orthogonal to it. The energy diffusion coefficient of stochastic acceleration scales as $D_\gamma\sim 0.1\sigma(c/l)\gamma^2$, where $\gamma$ is the particle Lorentz factor. This results in fast acceleration timescales $t_{acc}\sim (3/\sigma)\,l/c$. Our findings have important implications for understanding the generation of nonthermal particles in high-energy astrophysical sources.

$\,$

$\,$

\end{abstract}

\section{Introduction}
Generation of energetic particles far exceeding thermal energies is ubiquitous in the collisionless plasmas found in space and astrophysical environments. 
Thus, it is not surprising that over the last several decades, significant efforts have been made to understand the mechanisms of particle acceleration. Among such mechanisms, plasma turbulence has been often invoked to explain nonthermal particles in a variety of astrophysical systems \citep[e.g.][]{Melrose80,Petrosian2012SSRv,Lazarian2012SSRv}. Indeed, turbulence is ubiquitous in astrophysics, in systems as diverse as stellar coronae and winds \citep{Matthaeus99,Cranmer07}, the interstellar medium \citep{Armstrong95,Lithwick2001}, supernova remnants \citep{Weiler88,Roy09}, pulsar wind nebulae \citep{Porth2014MNRAS,LyutikovMNRAS2019}, black hole accretion disks \citep{Balbus98RvMP,Brandenburg05}, jets from active galactic nuclei \citep{Marscher08,MacDonald2018}, radio lobes \citep{Vogt2005,OSullivan09}, gamma-ray bursts \citep{Piran04,Kumar2009MNRAS}, and galaxy clusters \citep{Zweibel97Natur,Subramanian2006MNRAS}.

A characteristic feature of magnetized turbulence is the tendency to develop sheets of strong electric current density that are prone to magnetic reconnection \citep{Matthaeus86,Biskamp89,Carbone_1990,Politano95,Dmitruk2006PhPl,Retino07,Sundkvist07,Servidio2009}. These reconnecting current sheets are natural sites of magnetic energy dissipation and particle acceleration \citep{Arzner2004,Dmitruk2004,Matsumoto15}. At the same time, it has long been known that particles can gain energy through random scattering by turbulence fluctuations \citep[e.g.][]{KulsrudFerrari71}. Therefore, turbulence fluctuations and magnetic reconnection operate in synergy, and a comprehensive understanding of the particle acceleration physics in a turbulent environment will require a detailed investigation of their interplay.

Here, we want to study the physics of the generation of energetic particles in magnetically-dominated turbulence \citep{ThompsonBlaes98,Cho2005,Inoue11,Zrake12,Cho2014,Zrake14}. In this case, the magnetic energy density exceeds not only the pressure, but also the rest mass energy of the plasma, and the Alfv{\'e}n speed approaches the speed of light. Understanding the process of particle acceleration in this turbulence regime is important to shed light on the bright nonthermal synchrotron and inverse Compton signatures that are routinely observed from  high-energy astrophysical sources such as pulsar magnetospheres and winds \citep{Buhler2014}, jets from active galactic nuclei (AGNs) \citep{Begelman84}, or coronae of accretion disks \citep{Yuan14}. In particular, there are several crucial questions that need to be answered: (i) how efficient is the turbulence acceleration process in these systems? (ii) what is the slope of a (potential) power-law high-energy tail generated by  turbulence? (iii) what is the maximum attainable particle energy? (iv) which physical mechanism governs the injection of particles from the thermal pool to higher energies? and (v) on what timescales particle acceleration proceeds?

Given the complexity of the problem, an analytic treatment is often insufficient, and one must rely on numerical simulations. In this case, most of the previous works have used test particle simulations, where turbulence was represented by prescribed fields \citep[e.g.][]{Michalek1996,Arzner2006,Fraschetti2008,OSullivan09,TerakiAsano2019} or it was provided by turbulent fields obtained from MHD simulations \citep[e.g.][]{AmbrosianoJGR88,Dmitruk2004,Kowal12,DalenaApJ2014,Lynn2014,KimuraApJ16,Beresnyak2016,IslikerPRL17,Gonzalez17,KimuraMNRAS19}. These approaches offer a useful strategy to study the problem of particle acceleration with relatively inexpensive computational simulations. On the other hand, they have also some limitations, e.g., the absence of back reaction to the imposed electromagnetic fields and ad-hoc particle injection prescriptions. These limitations are overcome by recent hybrid (kinetic ions and fluid electrons) \citep{Servidio2012,Kunz2016,PecoraJPP18} and fully-kinetic PIC simulations \citep{Zhdankin17,ComissoSironi18,Zhdankin18,Zhdankin19,Wong2019arXiv,Nattila2019arXiv,Zhdankin2019arXiv}, where the particle acceleration process can be followed self-consistently durying the turbulence evolution. These simulations have confirmed in a self-consistent way that in a collisionless plasma, turbulence can drive particles out of thermal equilibrium.

In our earlier work \citep{ComissoSironi18}, we performed large-scale fully-kinetic simulations to show that decaying turbulence in magnetically-dominated plasmas can generate a large fraction of nonthermal particles with a power-law distribution that extends to very high energies. The simulation domains were large enough to capture both the MHD cascade at large scales and the kinetic cascade at small scales, and in this astrophysically-relevant setting we found that the power-law slope attains an asymptotic, system-size-independent value, while the high-energy cutoff increases linearly with the system size. \citet{Zhdankin17, Zhdankin18} found that driven plasma turbulence is also a viable astrophysical particle accelerator. Indeed, they showed that nonthermal energy distributions produced by driven turbulence converge to a system-size-independent power-law slope for sufficiently large domains. In order to explain the formation of nonthermal particle populations in magnetically-dominated turbulence, in \citet{ComissoSironi18} we analyzed  self-consistent particle trajectories from one of the PIC simulations, finding that most of the particles enter into the acceleration process through an injection phase that occurs at reconnecting current sheets which form self-consistently in the turbulent system. 
However, we also found that this initial energy gain, mediated by reconnection, is relatively small. At higher energies,  particles were stochastically accelerated by scattering off the turbulent fluctuations, thereby experiencing a biased random walk in momentum space.

In this paper, we extend our previous analysis of the particle acceleration process to a suite of large-scale PIC simulations. In particular, we analyze in a more extended way the impact of magnetic reconnection on the initial stage of particle acceleration, the properties of the particle diffusion process in energy space due to stochastic scattering off turbulence fluctuations, and the signatures of these acceleration processes on the particle distribution. We show that elongated current sheets are prone to the rapid development of the plasmoid instability and break up into plasmoids/flux ropes separated by secondary current sheets, which gives rise to fast reconnection and efficient particle injection. Plasmoids/flux ropes are ubiquitous in both  2D and 3D simulations, as a consequence of the large scale separation betwen the energy-containing eddies and the plasma skin depth. 
The initial energization of  particles (i.e., at injection) is controlled by the work done by the electric field parallel to the local magnetic field, which is nonzero at reconnecting current sheets. On the other hand, after the first energization phase, the work done by the perpendicular electric field takes over and eventually dominates the overall energization for high-energy particles. Indeed, also the slope of the power-law high-energy tail is controlled by energization via perpendicular electric fields.
We show that the particle pitch-angle distribution bears memory of the different energization processes, showing that particle velocities are preferentially aligned with the magnetic field at low  energies, while they are preferentially oriented in the direction perpendicular to the magnetic field at high particle energies. We also determine the diffusion coefficient in energy space that characterizes the physics of stochastic acceleration by turbulent fluctuations. In both 2D and 3D simulations, in the energy interval pertaining to the nonthermal power-law tail, the energy diffusion coefficient increases linearly with the plasma magnetization and quadratically with the particle energy. For high plasma magnetizations, this yields a fast rate of particle energy gain, which can be comparable or even higher then the  particle energy gain rate from fast magnetic reconnection.

This paper is organized as follows. In Section \ref{SecMethodSetup}, we describe our computational method and simulations setup. This is followed, in Section \ref{SecTurb}, by a description of the fully developed turbulence state and the resulting particle energy spectra for  different plasma conditions. The following sections are mostly devoted to the analysis of the acceleration mechanisms and their signature on the particle distribution function. In particular, in Section \ref{SecInjection} we investigate the role of magnetic reconnection in providing an efficient particle injection mechanism. In Section \ref{SecEnergiz} we study the different contributions of the parallel \emph{vs} perpendicular electric field in driving the energization of particles. The properties of pitch angle particle distributions, four-velocity distribution functions, and mixing of the energized particles, are presented in Section \ref{SecAnisotropy}. Then, in Section \ref{SecEnergyDiff}, we study the properties of diffusion in energy space of the particles that are accelerated by stochastic scattering off the turbulent fluctuations. Finally, in Section \ref{SecConclusions} we summarize our findings.

\section{Numerical Method and Setup}  \label{SecMethodSetup}

In order to study the particle acceleration process from first principles, we solve the full Vlasov-Maxwell system of equations through the particle-in-cell (PIC) method \citep{birdsall_langdon_85}, which evolves electromagnetic fields via Maxwell's equations and particle trajectories via the Lorentz force. 
To this purpose, we employ the electromagnetic fully-relativistic PIC code TRISTAN-MP \citep{buneman_93,spitkovsky_05}, which allows us to perform large-scale two-dimensional (2D) and three-dimensional (3D) simulations of plasma turbulence. 
In 2D our computational domain is a  square of size $L^2$ in the $xy$-plane, while in 3D it is a  cube of size $L^3$. We use periodic boundary conditions in all directions. For both 2D and 3D domains, all three components of particle momenta and electromagnetic fields are evolved in time.

We initialize a uniform electron-positron plasma with total particle density $n_0$ according to a Maxwell-J\"{u}ttner distribution  
\begin{equation} \label{}
{f_0}({\bm{p}}) = \frac{1}{{4\pi {m^3}{c^3}{\theta_0}{K_2}(1/{\theta_0})}}\exp \left( { - \frac{\gamma ({\bm{p}})}{\theta_0}} \right) \, ,
\end{equation}
where $\gamma({\bm{p}})=\sqrt{1+ ({\bm{p}}/mc)^2}$ is the particle Lorentz factor, $\theta_0 = {k_B T_0}/{m c^2}$ is the dimensionless temperature, and $K_2(z)$ is the modified Bessel function of the second kind. Here, as usual, $k_B$ indicates the Boltzmann constant, $T_0$ is the initial plasma temperature, $m$ denotes the particle mass, ${\bm{p}}$ is the particle momentum and $c$ is the speed of light in vacuum. In all the simulations, we set up a uniform mean magnetic field along the $z$ direction, ${\bm{B}}_0  = B_0 {\bm{\hat z}}$. The initial equilibrium is  perturbed by magnetic fluctuations of the form 
\begin{equation} \label{}
\delta {\bm{B}}({\bm{x}}) = \sum\limits_{\bm{k}} {\delta B({\bm{k}}) {\bm{\hat \xi}}({\bm{k}}) \exp \left[ {i \left( {{\bm{k}} \cdot {\bm{x}} + {\phi_{\bm{k}}}} \right)} \right]}  \, ,
\end{equation}
where $\delta B({\bm{k}})$ is the Fourier amplitude of the mode with wave vector ${\bm{k}}$, ${\bm{\hat \xi}}({\bm{k}}) = i \, {\bm{k}} \times {\bm{B}}_0/|{\bm{k}} \times {\bm{B}}_0|$ are Alfv{\'e}nic polarization unit vectors, and $\phi_{\bm{k}}$ are random phases.  By setting $\delta B(-{\bm{k}}) = \delta B({\bm{k}}) $ and $\phi_{-{\bm{k}}} = -\phi_{{\bm{k}}}$ we ensure that $ \delta {\bm{B}}({\bm{x}})$ is a real function. 
We adopt equal amplitude per mode and wave vector components $k_j = 2\pi n_j/L$ with mode numbers in the interval $n_j \in \{ {1, \ldots ,N_j} \}$. We set $N_x=N_y=8$ in 2D simulations, while $N_x=N_y=4$ and $N_z=2$ in 3D simulations. The choice of perturbing lower mode numbers in 3D simulations is due to the smaller domain size affordable in 3D and the desire to maximize the inertial range of the turbulent cascade. With these choices, the initial magnetic energy spectrum peaks near $k_N=2\pi N_{\rm{max}}/L$ (where $N_{\rm{max}}=8$ in 2D and $N_{\rm{max}}=4$ in 3D). In the following, we will use $l=2 \pi/k_N$ as our unit length, which we also refer to as the energy-carrying scale.

The strength of the initial magnetic field fluctuations is parameterized by the magnetization 
\begin{equation} \label{initial_magnetization_fluctuations}
\sigma_0  = \frac{\delta B_{{\rm{rms}}0}^2}{{4\pi n_0 w_0 m c^2}}  \, ,
\end{equation}
where  $\delta B_{{\rm{rms}}0} = \langle {\delta {B^2} (t=0)} \rangle^{1/2}$ is the space-averaged root-mean-square value of the initial magnetic field fluctuations and $w_0 m c^2 = [K_3(1/\theta_0)/K_2(1/\theta_0)] m c^2$ is the initial enthalpy per particle, with $K_n(z)$ indicating the modified Bessel function of the second kind of order $n$. 
Since we are interested in magnetically-dominated environments, we present results from  simulations with different values of $\sigma_0$ (from $2.5$ to $80$) in the regime $\sigma_0 \gg 1 $. In this case, the Alfv{\'e}n speed defined with the fluctuating fields is ${v_{A0}} = c\sqrt {{\sigma _0}/(1 + {\sigma _0})}  \sim c$. We find that with our definition of $\sigma_0$, our results do not depend on the choice of the initial dimensionless temperature $\theta_0$, apart from an overall energy rescaling (see Section \ref{SecTurb}).

%%%%%%%%%%%%%%%%%%%%%%%%%%%%%%  

%\startlongtable
\begin{deluxetable}{lccccccc}
\tablewidth{0.8\columnwidth} 
\tablecaption{Simulation parameters.\label{tab:param}}
\tablehead{
\colhead{Sim} & \colhead{$L/d_{e0}$} & \colhead{$\sigma_0$} & \colhead{$\delta B_{{\rm{rms}}0} /B_0$} & \colhead{$\theta_0$} & \colhead{$N_{\rm{max}}$}
} 
\startdata
 3D[a]&	$820$	     & $5$	     & $1$     & $0.3$      & $4$	\\
 3D[b]*&  $820$	     & $10$    & $1$	    & $0.3$      & $4$	\\
 3D[c]&	$820$	     & $20$	 & $1$     & $0.3$      & $4$	\\
 3D[d]&	$820$	     & $40$	 & $1$	     & $0.3$      & $4$	\\
 \hline
 2D[a]&	$1640$	& $2.5$   &   $1$	   & $0.3$       & $8$   \\
 2D[b]&	$1640$	& $5$  	&   $1$	   & $0.3$       & $8$	  \\
 2D[c]*&	$1640$	& $10$ 	&   $1$	   & $0.3$       & $8$	  \\
 2D[d]&	$1640$	& $20$ 	&   $1$       & $0.3$       & $8$	  \\
 2D[e]&	$1640$	& $40$ 	&   $1$	   & $0.3$       & $8$	  \\
 2D[f]&	$1640$	& $80$ 	&   $1$	   & $0.3$		 & $8$   \\
 2D[g]&	$1640$	& $2.5$	&   $2$	   & $0.3$		 & $8$   \\
 2D[h]&	$1640$	& $5$     &   $2$	       & $0.3$		 & $8$   \\
 2D[i]&  	$1640$	& $10$   &   $2$	       & $0.3$		 & $8$   \\
 2D[j]&  	$1640$	& $20$   &   $2$	       & $0.3$		 & $8$   \\
 2D[k]&	$1640$	& $40$   &   $2$	       & $0.3$		 & $8$   \\
 2D[l]&	    $1640$	& $80$ 	&  $2$	       & $0.3$		 & $8$	   \\
 2D[m]&	$1640$	& $10$   &   $1$	       & $0.1$ 	   & $8$	\\ 
 2D[n]&	$1640$	& $10$   &   $1$	       & $1$ 		   & $8$	\\ 
 2D[o]&	$1640$	& $10$   &   $1$	       & $3$ 		   & $8$	\\ 
 2D[p]&	$1640$	& $10$   &   $1$	       & $10$       & $8$	\\ 
 2D[q]&	$3280$	& $40$   &   $1$	       & $0.3$ 		& $8$	\\ 
 2D[r]&	$3280$	& $40$   &   $2$	       & $0.3$ 		& $8$	\\ 
 2D[s]&	$3280$	& $40$   &   $4$	       & $0.3$ 		& $8$	\\ 
 2D[t]&	$6560$	& $10$   &   $1$	       & $0.3$ 	    & $8$	\\ 
\hline
\enddata
\tablecomments{We mark the reference simulations with an asterisk (*). The magnetization parameter $\sigma_0$ is defined with the initial magnetic field fluctuations, $\sigma_0  = {\delta B_{{\rm{rms}}0}^2}/{{4\pi n_0 w_0 m c^2}}$, where $\delta B_{{\rm{rms}}0} = \langle {\delta {B^2} (t=0)} \rangle^{1/2}$. In this paper we use also the instantaneous magnetization parameter $\sigma  = {\delta B_{{\rm{rms}}}^2}/{{4\pi n_0 w m c^2}}$, where $\delta B_{{\rm{rms}}} = \langle {\delta {B^2}} \rangle^{1/2}$ (and $w m c^2$ is the instantaneous enthalpy per particle), and the magnetization associated with the mean magnetic field, $\sigma_z = {B_0^2}/{4\pi n_0 w_0 m c^2} = \sigma_0  \left( {B_0}/{\delta B_{{\rm{rms}}0}} \right)^2$. }
\end{deluxetable}

%%%%%%%%%%%%%%%%%%%%%%%%%%%%%%  

We resolve the initial plasma skin depth $d_{e0}=c/\omega_{p0}$ with 10 cells in 2D and 3 cells in 3D (in 2D we have checked that runs with $d_{e0} = 3$ or 10 cells give identical results, including the development of turbulent structures, as can be seen in the Appendix). Note that the initial plasma skin depth is defined with the relativistic plasma frequency $\omega_{p0} =\sqrt {4\pi n_0 {e^2}/\gamma_{th0} {m}}$, where $\gamma_{th0} = w_0 - \theta_0$ is the initial mean particle Lorentz factor.

In order to capture the full plasma turbulence cascade from macroscopic MHD scales to kinetic scales, we solve the kinetic system of equations on large computational domains. This is achieved by adopting a box of $2460^3$ cells in 3D simulations and $16400^2$ cells in 2D simulations. For the 2D analysis, we also present results from three simulations with $32800^2$ cells and one simulation with $65600^2$ cells. In our reference 2D simulation we employ 64 particles per cell, while 16 particles per cell are adopted for our reference 3D simulation. For the other runs, we employ 16 particles per cell in 2D and 4 particles per cell in 3D. We have tested that in the magnetically-dominated regime of interest here, the discussed results are the same when using up to 256 particles per cell (see a particle spectrum comparison in the Appendix). 

The simulation timestep is controlled by the numerical speed of light of 0.45 cells/timestep. The simulations are run for $ct/l=12-15$, at which point most of the turbulent magnetic energy has been transferred to the particles.  Our study is focused on magnetically-dominated turbulence, and for this purpose we have performed several simulations at different magnetizations $\sigma_0$. In 2D we have investigated $\sigma_0 \in \left\{ {2.5,5,10,20,40,80} \right\}$. In 3D we have explored $\sigma_0 \in \left\{ {5,10,20,40} \right\}$. If not otherwise specified, the simulations start with $\delta B_{{\rm{rms}}0} /B_0 = 1$ and $\theta_0=0.3$. 
Cases with different $\delta B_{{\rm{rms}}0} /B_0$ and $\theta_0$ have also been performed in 2D.  For convenience, we have summarized the physical parameters of the presented simulations in Table \ref{tab:param}. Our reference 2D and 3D simulations are indicated with an asterisk.

%%%%%%%%%%%%%%%%%%%%%%%%%%%%%%  
\begin{figure*}
 \centering 
  \includegraphics[width=0.47\textwidth]{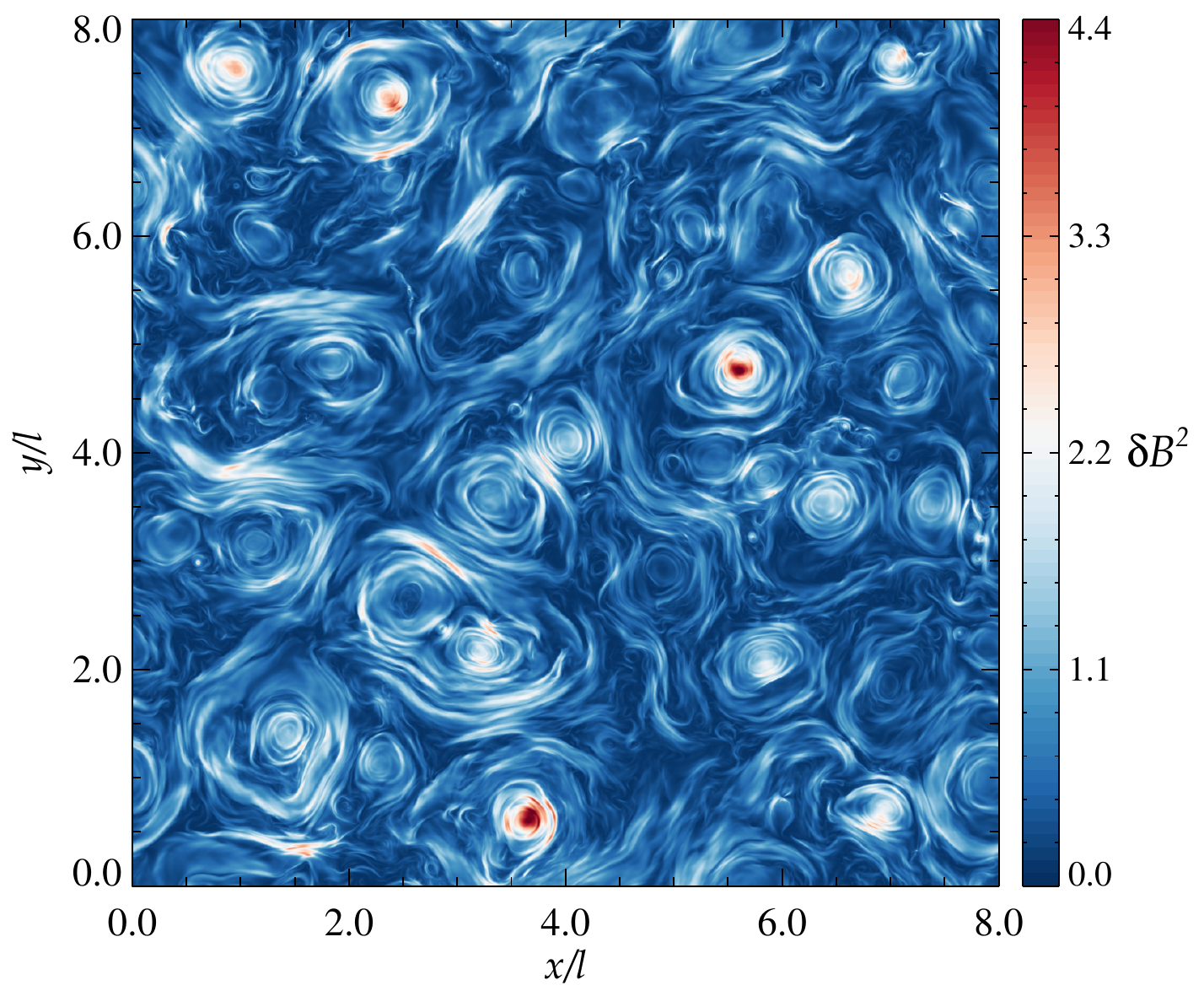}
  \includegraphics[width=0.47\textwidth]{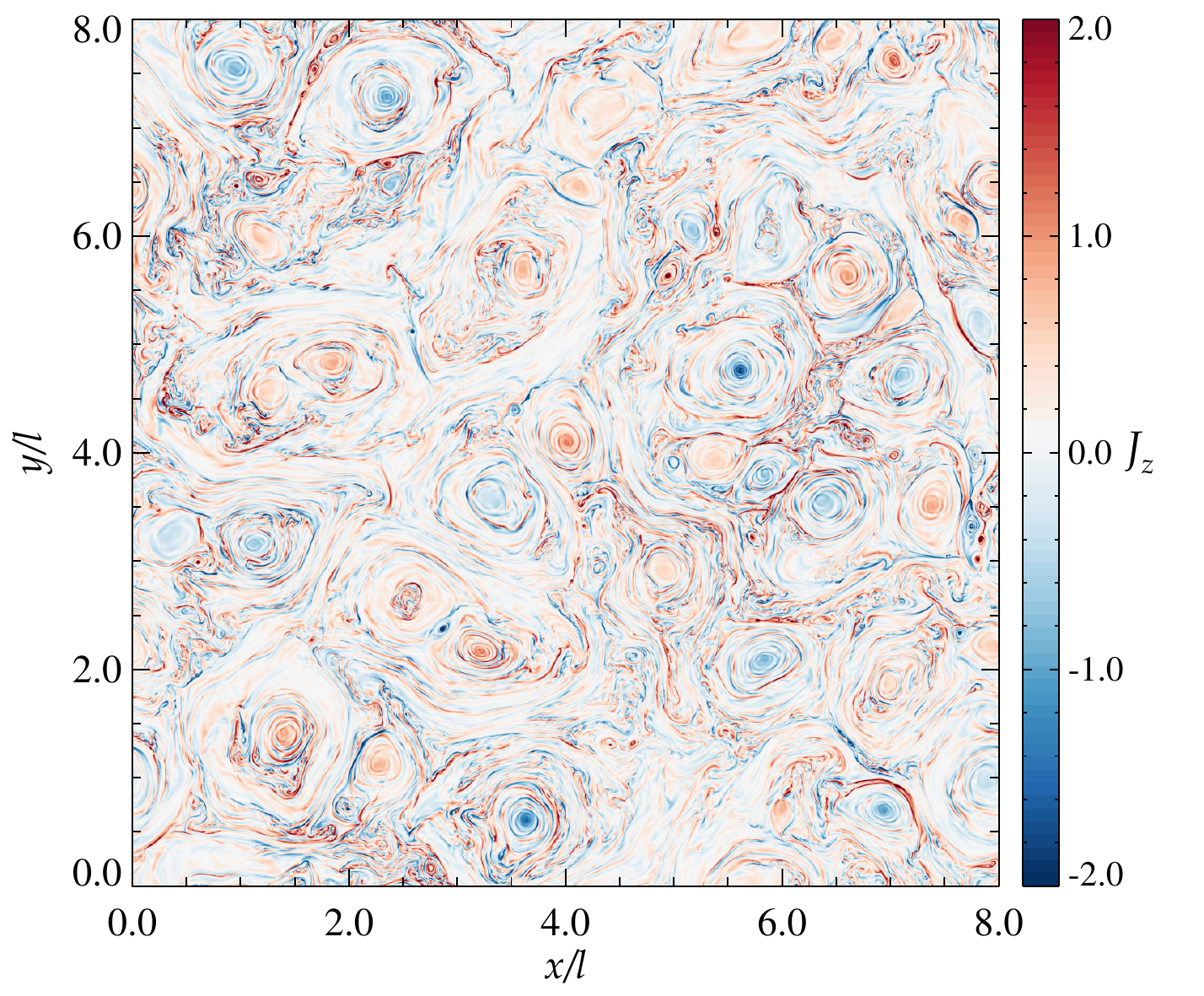}
  \hfill
  \includegraphics[width=0.47\textwidth]{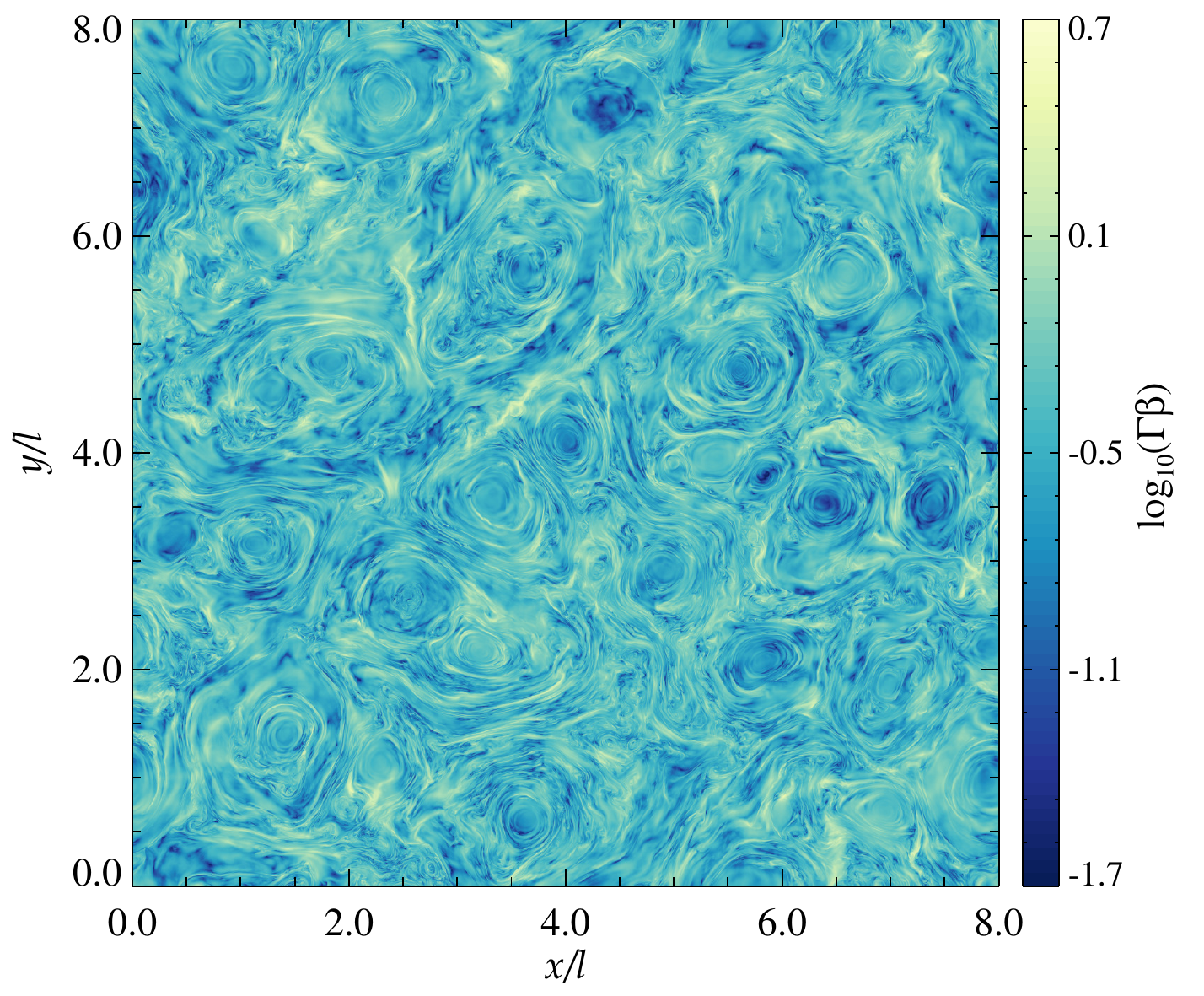}
  \includegraphics[width=0.47\textwidth]{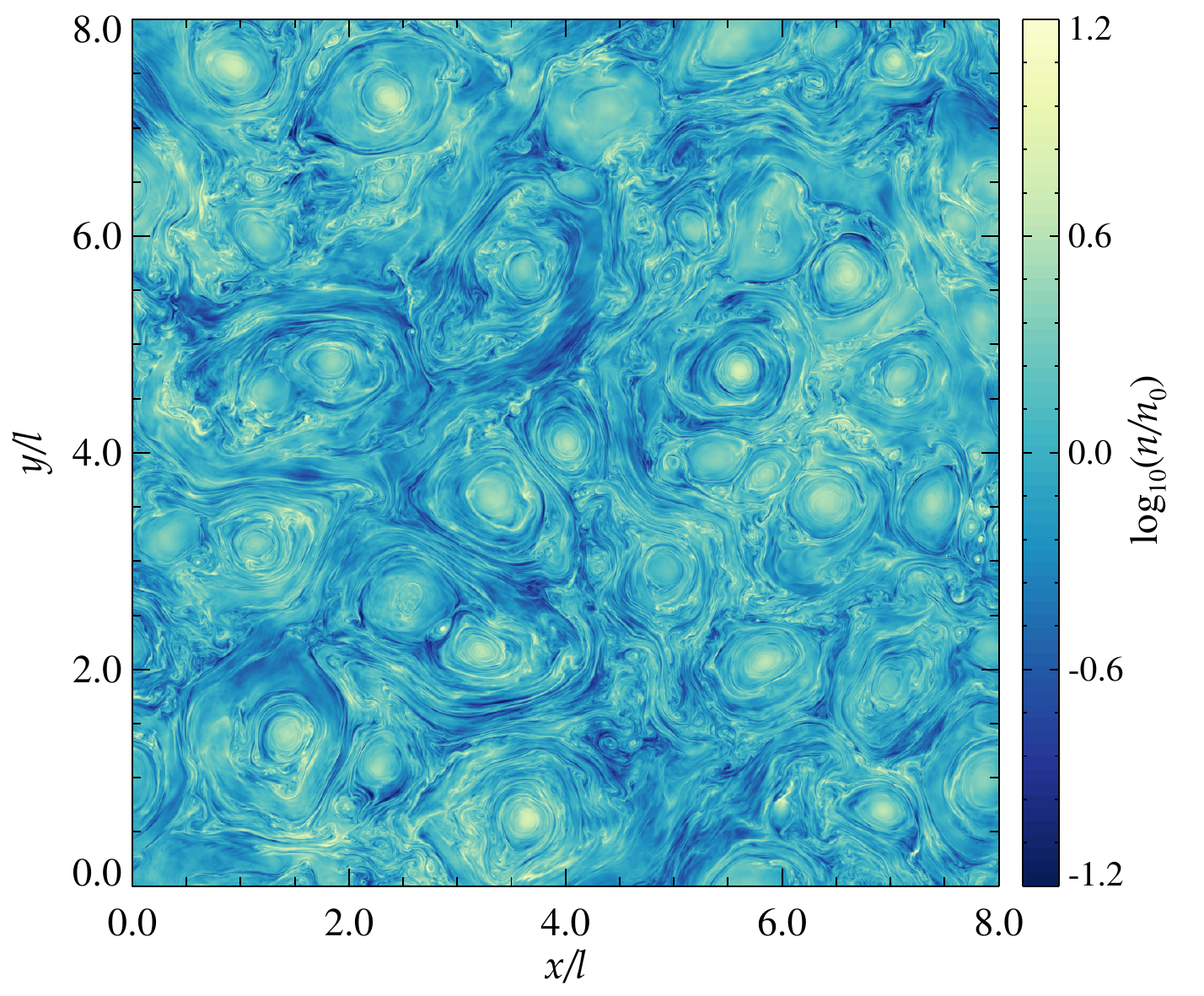}
 \caption{2D plots of different fluid structures in fully developed 2D turbulence (at $ct/l=4.6$) with $\sigma_0=10$, $\delta B_{{\rm{rms}}0}/ B_0=1$, and $L/d_{e0}=1640$ (with $l=L/8$). The displayed quantities are (from left to right, top to bottom) the fluctuation magnetic energy density in units of $B_0^2/8\pi$, the current density $J_z$ along the mean magnetic field in units of $e n_0 c$, the bulk dimensionless four-velocity $\Gamma \beta$, and the particle density ratio $n/n_0$. Note that the color bars for $\Gamma \beta$ and $n/n_0$ are in logarithmic scale.} 
 \label{fig1}
\end{figure*} 
%%%%%%%%%%%%%%%%%%%%%%%%%%%%%%

\section{Plasma Turbulence and Particle Spectrum} \label{SecTurb}

In this section, we give an overview of the plasma turbulence state in 2D and 3D PIC simulations, with a particular focus on the particle energy spectrum that develops self-consistently. 
We first present the characteristic fluid structures of the magnetized turbulence state, and the time evolution of the magnetic power spectrum. Then we show the time evolution of the particle energy spectrum and we discuss its dependence on the main physical parameters.

%%%%%%%%%%%%%%%%%%%%%%%%%%%%%  
\begin{figure*}
 \centering 
  \includegraphics[width=0.497\textwidth]{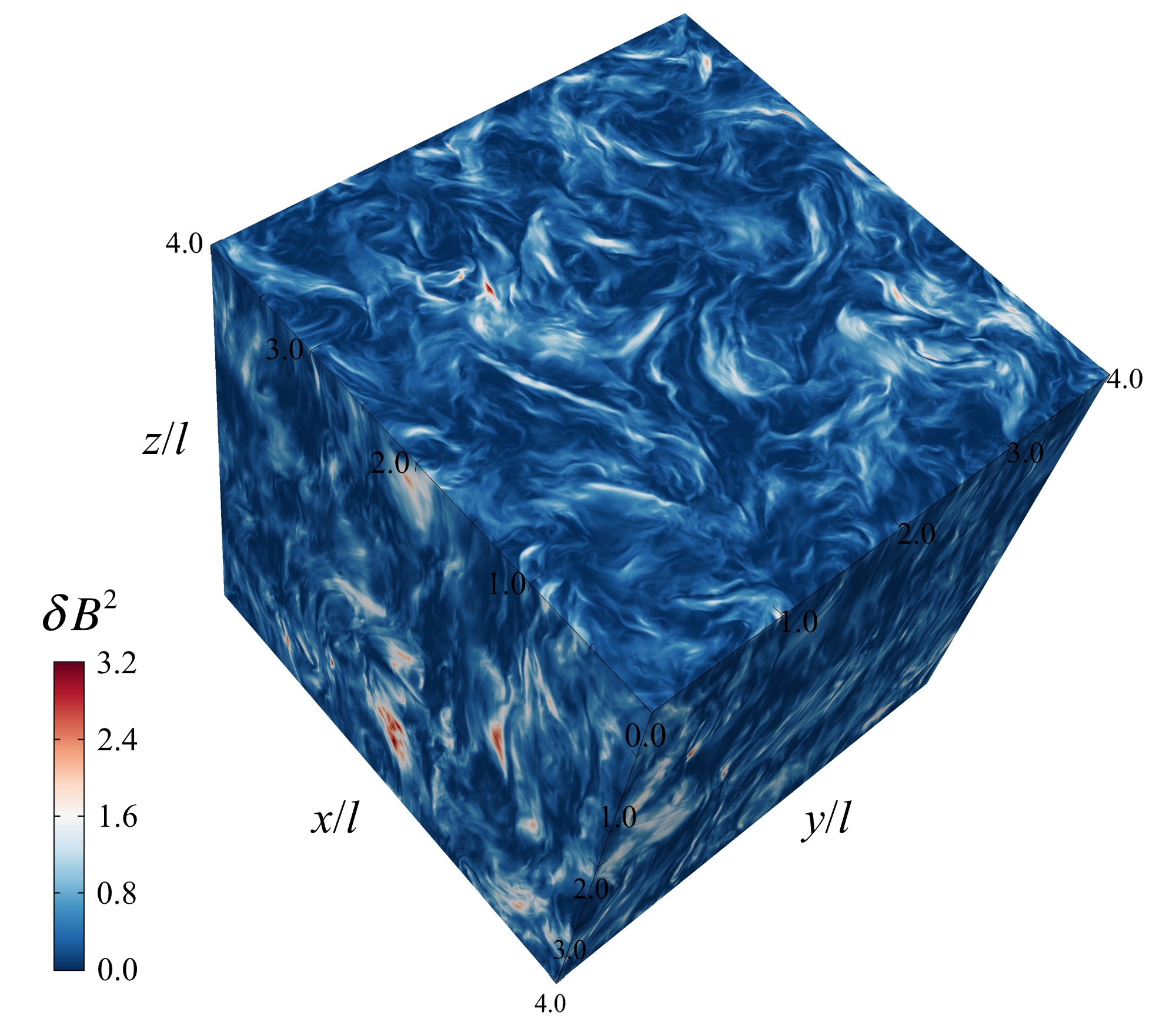}
  \includegraphics[width=0.497\textwidth]{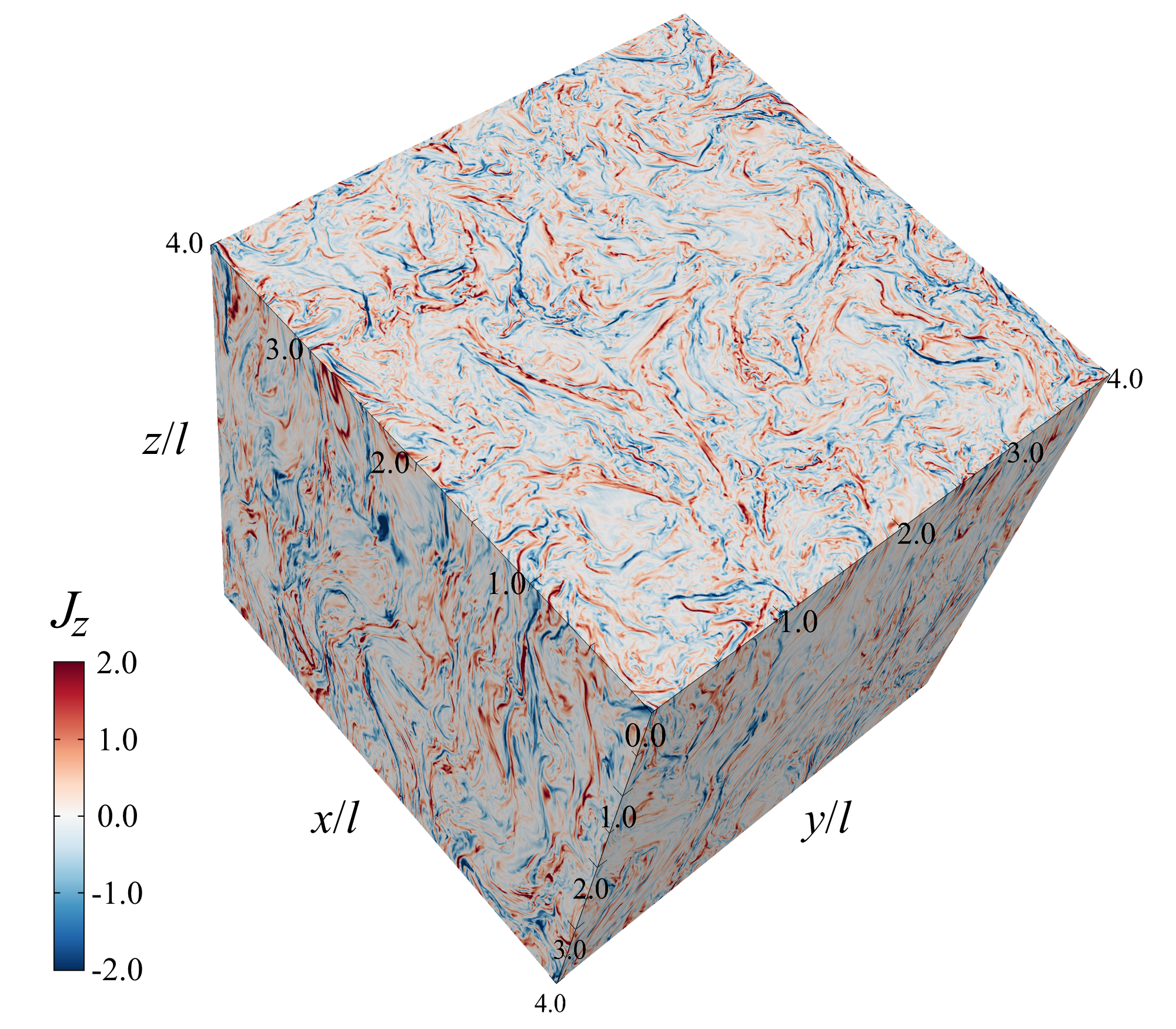}
  \hfill
  \includegraphics[width=0.497\textwidth]{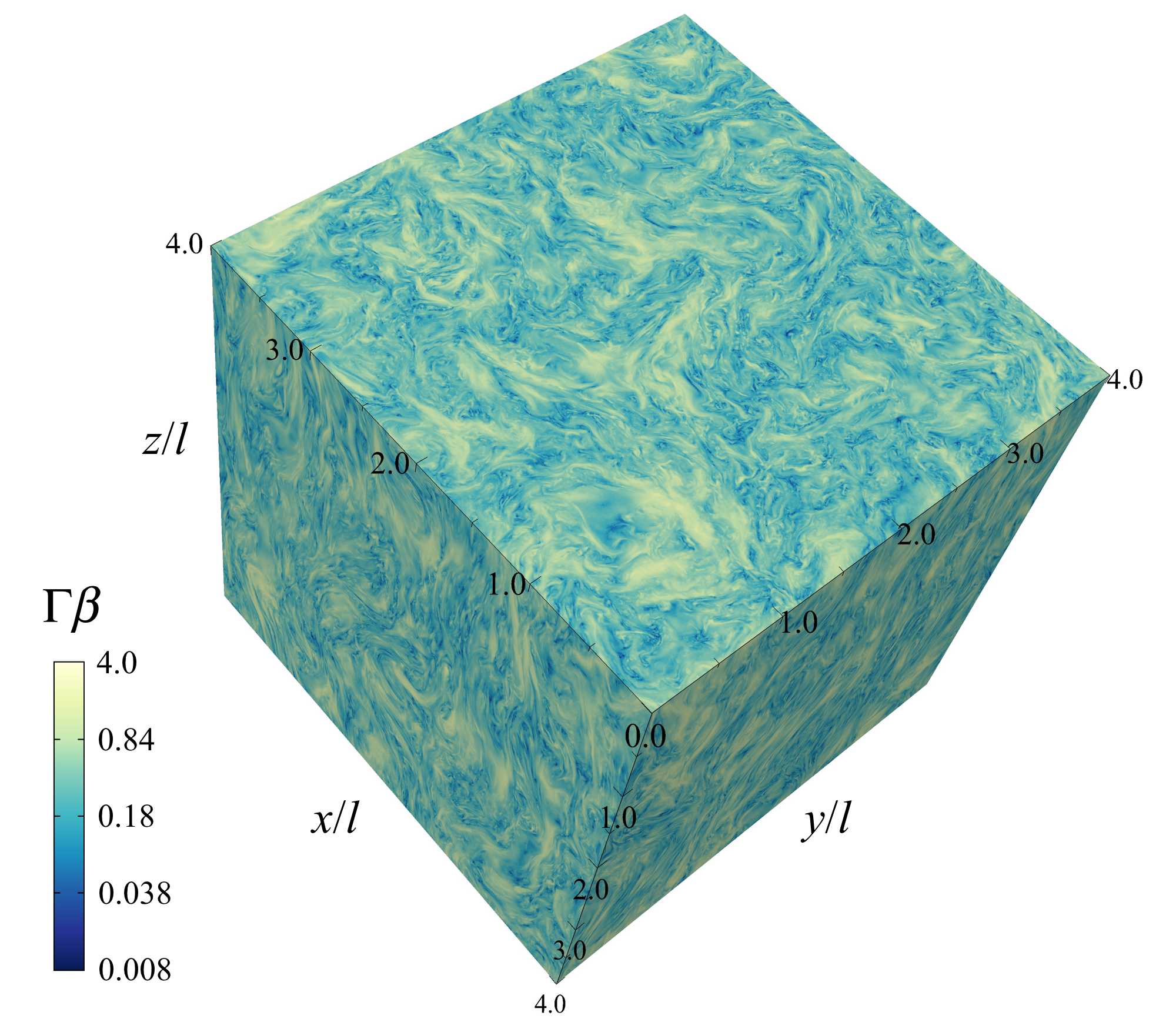}
  \includegraphics[width=0.497\textwidth]{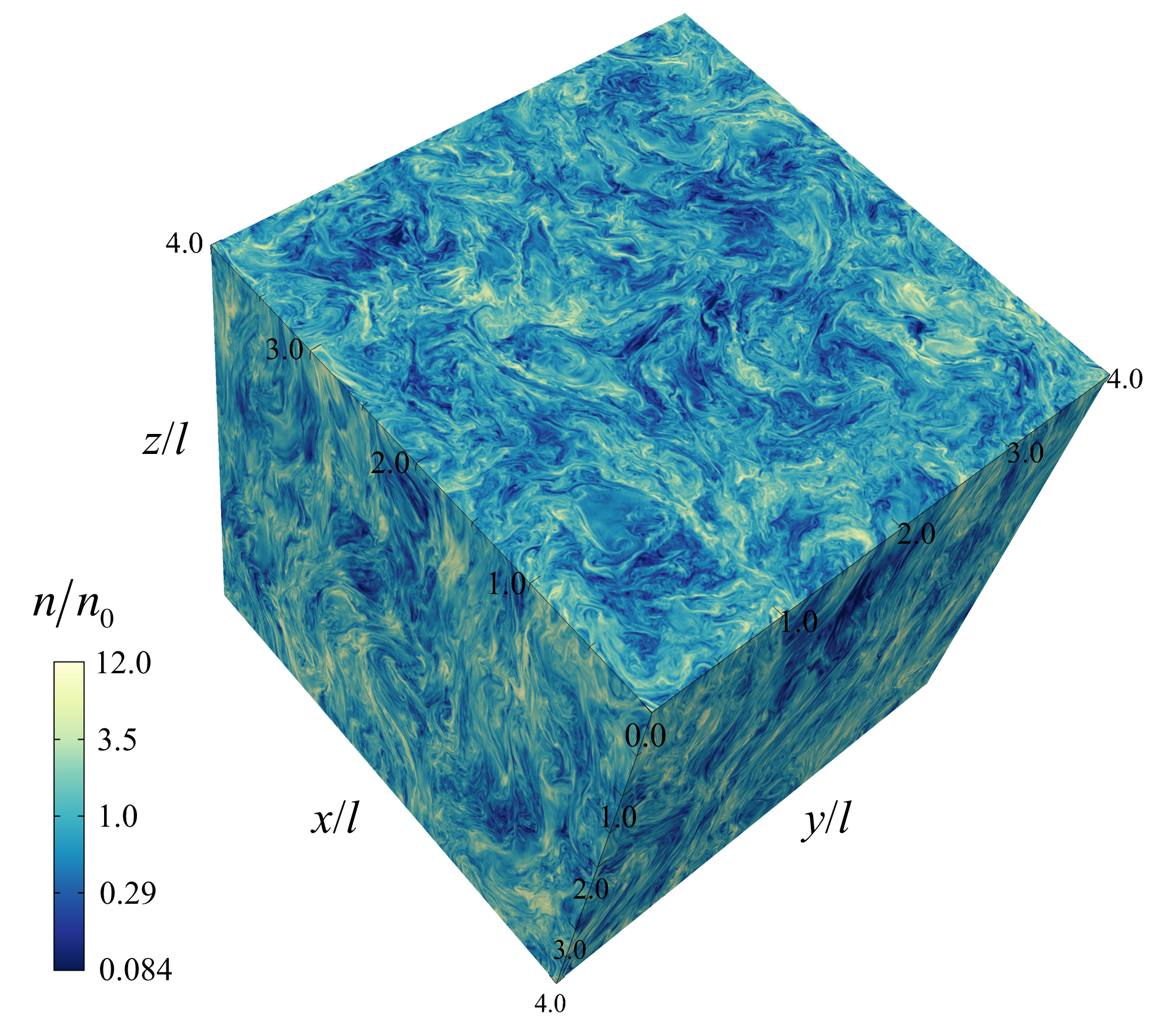}
 \caption{3D plots of different fluid structures in fully developed 3D turbulence (at $ct/l=2.7$) with $\sigma_0=10$, $\delta B_{{\rm{rms}}0}/ B_0=1$, and $L/d_{e0}=820$ (with $l=L/4$). The displayed quantities are (from left to right, top to bottom) the fluctuation magnetic energy density in units of $B_0^2/8\pi$, the current density $J_z$ along the mean magnetic field in units of $e n_0 c$, the bulk dimensionless four-velocity $\Gamma \beta$, and the particle density ratio $n/n_0$. Note that the color bars for $\Gamma \beta$ and $n/n_0$ are in logarithmic scale. An animation showing the current density $J_z$ in different $xy$ slices can be found at \url{https://doi.org/10.7916/d8-prt9-kn88}.} 
\label{fig6}
\end{figure*} 
%%%%%%%%%%%%%%%%%%%%%%%%%%%%%

\subsection{Plasma turbulence} 

Turbulence structures from our reference 2D simulation are illustrated in Fig.~\ref{fig1}. We plot the magnetic field squared fluctuations $\delta B^2$, the out-of-plane electric current density $J_z$, the bulk dimensionless four-velocity $\Gamma \beta$, and the particle density ratio $n/n_0$. Here, $\Gamma = 1/\sqrt {1 - {({\bm{V}}/c)}^2}$ indicates the plasma bulk Lorentz factor and $\beta  = |{\bm{V}}|/c$ is the dimensionless plasma bulk speed obtained by averaging the velocities of individual particles. 
We can see that the fluctuations $\delta B^2$ are generally stronger in large-scale flux tubes (see the circular structures of size comparable to the energy-carrying scale $l$), but high values of $\delta B^2$  are also obtained in small-scale structures identified with reconnection plasmoids (see the circular structures with size $\ll l$). These are  ``secondary'' magnetic islands (flux ropes in 3D) that are produced by magnetic reconnection \citep{Biskamp2000}. In such plasmoids, the particle number density $n$ exhibits strong enhancements in excess of $n \sim 15\, n_0$. High values of particle number density occur in large-scale flux tubes as well. In general, the particle density displays strong compressions in coherent quasi-circular structures spanning a range of scales.

In between flux tubes, reconnection layers reveal the formation of plasmoids within narrow current sheets. Indeed, current sheets with high aspect ratio tend to fragment into plasmoids and secondary current sheets as a result of magnetic reconnection. Smaller-size current sheets are also ubiquitous, spanning a wide range of scales. We will see in the following sections that reconnecting current sheets, which are a natural by-product of turbulent cascades in magnetized plasmas  \citep[e.g.][]{Servidio2009,Wan2013PhPl,Cerri2017NJPh,Franci2017ApJ,Haggerty17,Dong_2018,ComissoSironi18,Papini2019}, play an important role for particle injection into the acceleration process \citep{ComissoSironi18}. 
Finally, we also point out that in the strongly magnetized regime of plasma turbulence investigated here, the plasma bulk speed can reach very high values. In particular, we observe ultra-relativistic flows with bulk Lorentz factor as high as $\Gamma \sim 5$. Such high speeds develop predominantly in between the large-scale flux tubes, although high-velocity fluctuations occur all over the spatial domain.

\begin{figure}
\begin{center}
\hspace*{-0.185cm}\includegraphics[width=8.75cm]{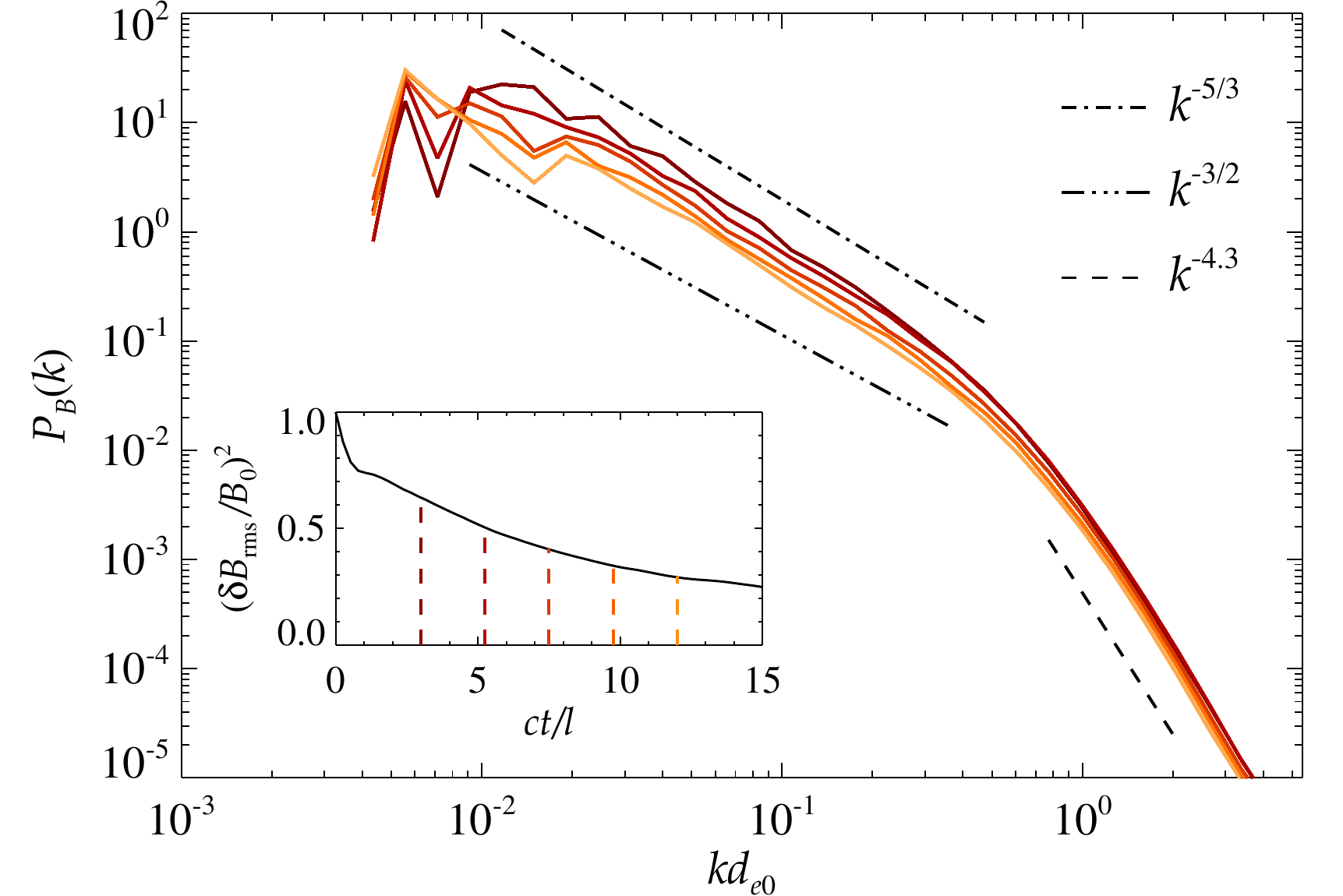}
\end{center}
\caption{Power spectrum of the magnetic field for the 2D simulation in Fig.~\ref{fig1}, showing a well-developed inertial range and a kinetic range scaling roughly as $P_B(k) \propto k^{-4.3}$. The inset shows the time evolution of $\delta B_{\rm rms}^2=\langle {\delta {B^2}}\rangle$ normalized to $B_0^2$, with vertical dashed lines indicating the times when the magnetic power spectra presented in the main panel are computed (same color coding).}
\label{fig2}
\end{figure}

We now consider the fluid structures that develop in 3D plasma turbulence. Our reference 3D simulation has $L/d_{e0}=820$, which is half the size of the reference 2D simulation. However, since in 3D we adopt perturbation numbers up to $N_{\rm max}=4$ (as compared to $N_{\rm max}=8$ in 2D),  we still have a well-extended turbulence inertial range. In fact, the ratio of the initial energy-carrying scale $l = 2 \pi/k_N$ to the plasma skin depth $d_{e0}$ remains the same between our reference 2D and 3D simulations, leading to the same high-energy cutoff of the particle energy spectrum (see \citet{ComissoSironi18} as well as Eq. (\ref{gamma_c}) in the following subsection).

The turbulence structures from our reference 3D simulation are displayed in Fig.~\ref{fig6}. The magnetic field squared fluctuations $\delta B^2$ present both large-scale and small-scale structures. However, in this case, the large-scale fluctuations are not organized in coherent flux tubes (as it was in 2D, where they were  a result of the constrained 2D dynamics). Despite differences in the large-scale structure of the magnetic field, there is still a copious presence of current sheets (current ribbons when considering the third direction). Due to the presence of the mean magnetic field ${\bm{B}}_0  = B_0 {\bm{\hat z}}$, current ribbons are mostly elongated along ${\bm{\hat z}}$. We can see that the $z$ component of the electric current, $J_z$, displays a variety of current sheets of different sizes. Some of these current layers break into plasmoids (see also Sec. \ref{SecInjection}), as highly elongated layers cannot be stable against the plasmoid instability, also in  3D geometry \citep[e.g.][]{Daughton11,ss_14,guo_15a,Huang16,Ebrahimi17,werner_17,Baalrud18,Stanier19}. Here, we show that plasmoids/flux ropes are self-consistently created in  fully 3D plasma turbulence (see Sec. \ref{SecInjection}), where current sheets are self-consistently generated by the turbulence itself. As for 2D plasma turbulence, we will see that these current sheets play an important role in the initial stages of  particle acceleration  (Sections \ref{SecInjection}-\ref{SecAnisotropy}).

Locations characterized by strong electric current densities are typically accompanied by strong gradients in particle density. In localized regions, the particle density can exceed $n \sim 12\, n_0$, similar to the 2D case. On the other hand, large-scale structures like the overdense regions at the core of  2D large-scale flux tubes are missing in 3D. Finally, we observe that also in 3D, due to the high magnetization of the system, the plasma flow speed is generally very high. We can see regions with ultra-relativistic flow speeds having bulk Lorentz factor as high as $\Gamma \sim 4$.

We now present the time evolution of the magnetic power spectrum from the reference 2D and 3D simulations. In our simulations, turbulence develops from the initialized magnetic fluctuations. The magnetic energy decays in time, as no continuous driving is imposed, and a well-developed inertial range and kinetic range of the turbulence cascade develop within the outer-scale nonlinear timescale. In Fig.~\ref{fig2}, we show the time evolution of the magnetic power spectrum $P_B(k)$ for the reference 2D simulation, where 
\begin{equation} \label{}
P_B(k) dk = \sum\limits_{{\bm{k}} \in dk} {\frac{{\delta {\bm{B}}_{\bm{k}} \cdot \delta {\bm{B}}_{\bm{k}}^*}}{{B_0^2}}}
\end{equation}
is computed from the discrete Fourier transform $\delta {\bm{B}}_{\bm{k}}$ of the fluctuating magnetic field. Each curve refers to a different time (from brown to orange), as indicated by the corresponding vertical dashed lines in the inset, where we present the temporal decay of the energy in turbulent fluctuations $\delta B^2_{\rm rms}/B_0^2$. 
We can see that at MHD scales ($kd_{e0}\lesssim 0.5$) the magnetic power spectrum is consistent with a Kolmogorov scaling $P_B(k) \propto k^{-5/3}$ \citep{Biskamp2003} (compare with the dot-dashed line), while the Iroshnikov-Kraichnan scaling $P_B(k) \propto k^{-3/2}$ \citep{Iroshnikov63,Kraichnan65} (triple-dot-dashed line) is possibly approached at late times. At kinetic scales ($kd_{e0}\gtrsim 0.5$), the spectrum steepens and approaches a power-law slope $P_B(k) \propto k^{-4.3}$ (compare with the dashed line). A similar slope was proposed for magnetized turbulence at sub-inertial scales in a  cold plasma \citep{Abdelhamid2016,Passot2017}. 
We finally observe that the turbulence integral scale 
\begin{equation} \label{int_length}
{\ell}(t) = \frac{{2\pi }}{{{k_I}(t)}} = 2\pi \frac{{\int_0^\infty  {{k^{ - 1}}{P_B}(k,t)dk} }}{{\int_0^\infty  {{P_B}(k,t)dk} }} \, ,
\end{equation}
which is close to the energy-carrying scale associated to the wavenumber where ${P_B}(k,t)$ peaks, increases as the magnetic energy decays in time. This is due to the merging of the large-scale magnetic flux tubes, which drives an inverse energy transfer to scales larger than the initial integral scale \citep[e.g.][]{BiskampSchwarz01}.

\begin{figure}
\begin{center}
\hspace*{-0.185cm}\includegraphics[width=8.75cm]{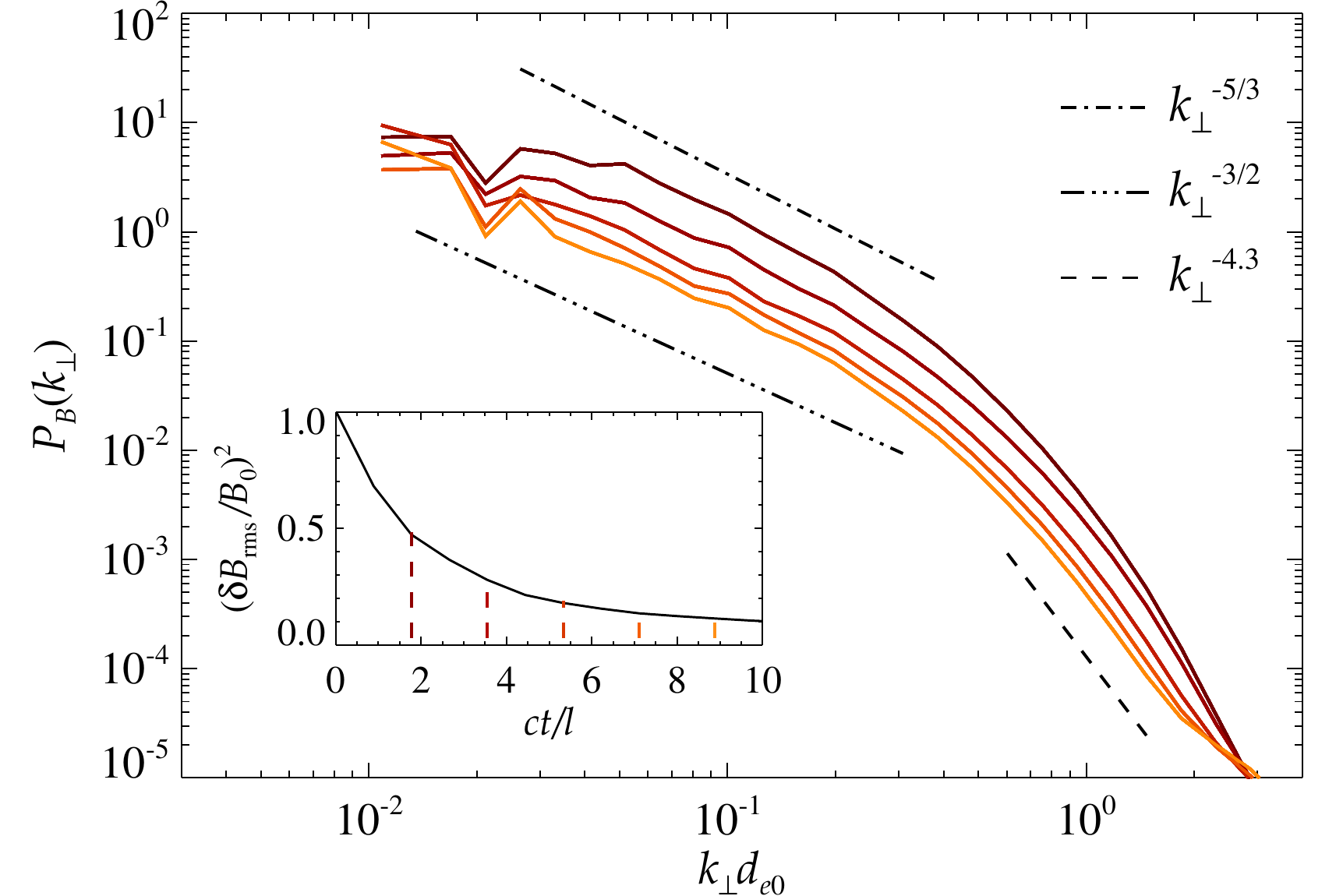}
\end{center}
\caption{Power spectrum of the magnetic field for the 3D simulation in Fig.~\ref{fig6}, showing a well-developed inertial range and a kinetic range scaling roughly as $P_B(k) \propto k^{-4.3}$. The inset shows the time evolution of $\delta B_{\rm rms}^2=\langle {\delta {B^2}}\rangle$ normalized to $B_0^2$, with vertical dashed lines indicating the times when the magnetic power spectra presented in the main panel are computed (same color coding).}
\label{fig7}
\end{figure} 

We now consider the time evolution of the magnetic power spectrum in 3D. Due to the presence of the large-scale mean magnetic field ${\bm{B}}_0$, turbulence becomes increasingly anisotropic toward small scales, within the inertial range. To account for this global anisotropy with respect to ${\bm{B}}_0$, we consider the magnetic power spectrum with respect to the wavenumber $k_\bot = {(k_x^2 + k_y^2)^{1/2}}$ perpendicular to the mean field, obtained from the discrete Fourier transform of the fluctuating magnetic field as 
\begin{equation} \label{}
P_B(k_\bot) d k_\bot = \sum\limits_{{\bm{k}} \in d k_\bot} {\frac{{\delta {\bm{B}}_{\bm{k}} \cdot \delta {\bm{B}}_{\bm{k}}^*}}{{B_0^2}}} \, .
\end{equation}
Figure~\ref{fig7} shows the time evolution of the magnetic power spectrum $P_B(k_\bot)$, which does exhibit inertial and kinetic ranges of the turbulence cascade at times $ct/l \gtrsim 1$. As the magnetic energy decays (see inset), the inertial range ($k_\bot d_{e0} \lesssim 0.5$) of the magnetic power spectrum tends to flatten from $P_B(k_\bot) \propto k_\bot^{-5/3}$ \citep{gs95,ThompsonBlaes98} (dot-dashed line)  to $P_B(k_\bot) \propto k_\bot^{-3/2}$ \citep{Boldyrev2006} (triple-dot-dashed line). At kinetic scales ($k_\bot d_{e0} \gtrsim 0.5$), the spectrum steepens to a power law $P_B(k_\bot) \propto k_\bot^{-4.3}$ (dashed line), similar to the 2D result and in agreement with theoretical predictions for magnetized turbulence at sub-inertial scales in  cold plasmas \citep{Abdelhamid2016,Passot2017}. Note also that in the 3D case, the magnetic energy decays faster than in the 2D case (compare insets of Figs.~\ref{fig2} and \ref{fig7}). We will show that this leads to a reduced particle acceleration rate at late times.

\subsection{Particle spectrum}

\begin{figure}
\begin{center}
\hspace*{-0.185cm}\includegraphics[width=8.75cm]{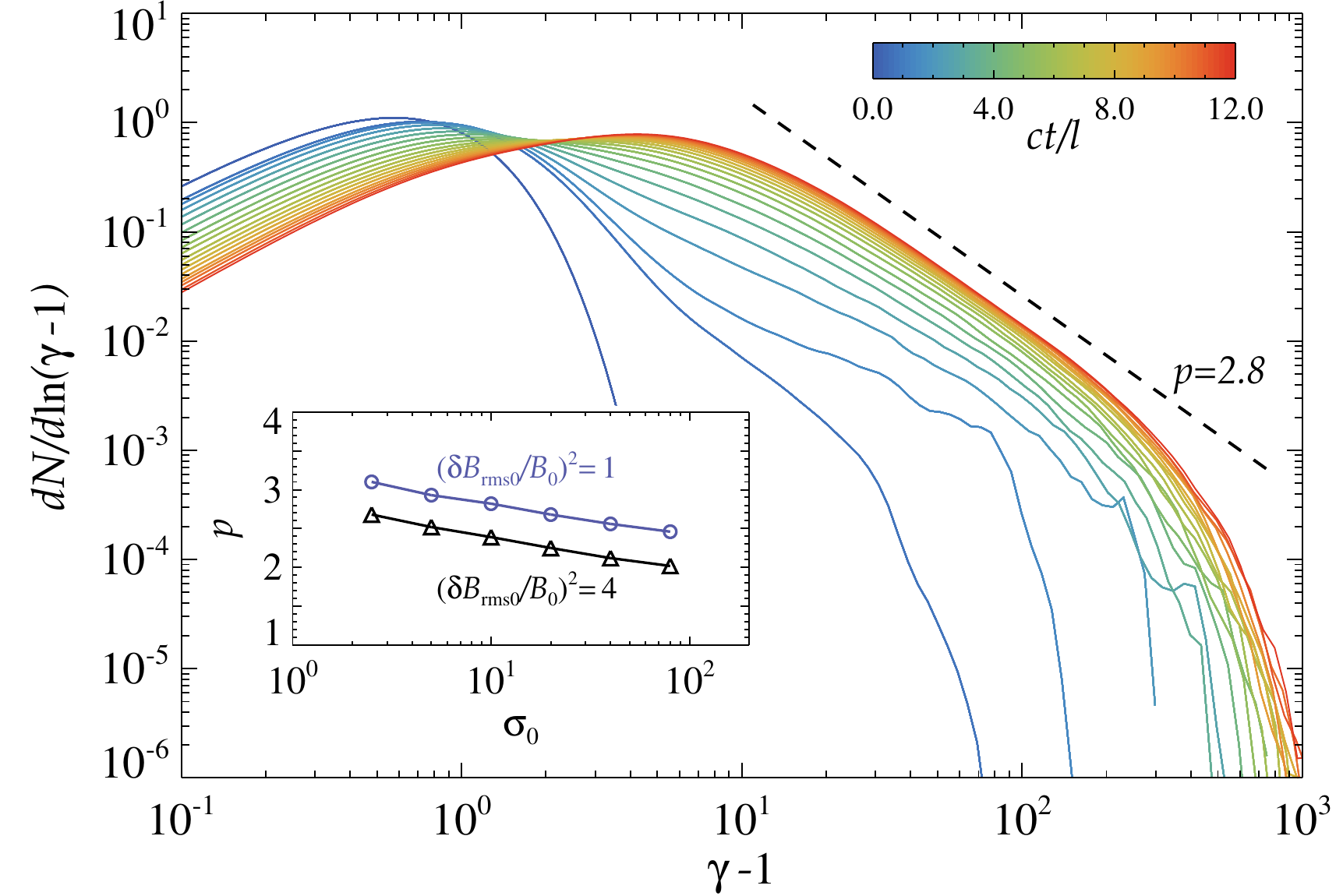}
\end{center}
\caption{Time evolution of the particle spectrum $dN/d\ln(\gamma-1)$ for the simulation in Fig.~\ref{fig1}. At late times, the particle spectrum displays a power-law tail with index $p = - d\log N/d\log (\gamma -1)\sim 2.8$. 
About $17 \%$ of the particles have $\gamma \geq 12$ at $ct/l=12$ (twice the peak of the particle energy spectrum at that time), which gives an indication of the percentage of nonthermal particles. The inset shows the power-law index $p$ as a function of the magnetization $\sigma_0$ for two values of $\delta B_{{\rm{rms}}0}/ B_0$.}
\label{fig3}
\end{figure}     

The most interesting outcome of the turbulent cascade is the generation of a large population of nonthermal particles. This is shown in Fig.~\ref{fig3} (for the 2D setup), where the time evolution of the particle energy spectrum $dN/d\ln(\gamma-1)$ is presented ($\gamma - 1 = E_k/mc^2$ is the normalized particle kinetic energy). As a result of turbulent field dissipation, the spectrum shifts to energies much larger than the initial Maxwellian, which is shown by the blue line peaking at $\gamma-1\sim \gamma_{th0}-1\simeq 0.6$. At late times, when most of the turbulent energy has decayed, the spectrum stops evolving (orange and red lines): it peaks at  $\gamma-1\sim 5$, and extends well beyond the peak into a nonthermal tail of ultra-relativistic particles that can be described by a power-law
\begin{equation} \label{plaw}
\frac{dN}{d \gamma} = N_0 \, {\left( {\frac{\gamma -1}{\gamma_{st} -1}} \right)^{-p}} \, , \quad {\rm{for}} \; \gamma_{st}  <  \gamma  <  \gamma_c \, ,
\end{equation}
and a sharp cutoff for $\gamma  \geq \gamma_c$. 
Here, $N_0$ is the normalization of the power-law and $p$ is the power-law index, which is about $2.8$ for the simulation results presented in the main frame of Fig.~\ref{fig3} (note that in our figures we plot $dN/d\ln(\gamma-1)$ to emphasize the particle content, proportional to $(\gamma  - 1)^{-p+1} d \gamma$ for the distribution in Eq.~(\ref{plaw})). The percentage of the particles in the nonthermal tail (measured as the number of particles with Lorentz factor exceeding twice the thermal peak) is high, $\zeta_{\rm{nt}} \sim 17 \%$, and it corresponds to a high value of the normalization $N_0$, which is close to the thermal peak. The starting point of the power-law, $\gamma_{st}$, is roughly only a factor of two larger than the peak of the particle energy spectrum at late times.  Therefore, dropping  $O(1)$ factors, the starting point of the power-law can be estimated as 
\begin{equation} \label{gamma_st}
\gamma_{st} \sim {\gamma_\sigma} = \left( {1 + \frac{{{\sigma _0}}}{2}} \right) \gamma_{th0} \, ,
\end{equation}
since most of the magnetic energy is converted to particle energy by the time the particle energy spectrum has saturated (see inset of Fig.~\ref{fig2}). On the other hand, the high-energy cutoff $\gamma_c$ depends on the system size. As discussed in the following Sections, stochastic acceleration by turbulent fluctuations dominates the energy gain of the most energetic particles. High-energy particles cease to be efficiently scattered by turbulent fluctuations when their Larmor radius $\rho_L = \left( {\gamma mc/eB} \right) v_\bot$ exceeds the integral length scale $\ell =2\pi/k_I$, implying an upper limit to their Lorentz factor of 
\begin{equation} \label{gamma_c}
\gamma_c \sim e \sqrt{\langle B^2 \rangle} \frac{\ell}{m c^2}  \sim   \frac{2 \pi}{k_I d_{e0}} \sqrt{\sigma_z} \gamma_{th0} \, ,
\end{equation}
where $\langle {B^2} \rangle$ is the space-averaged mean-square value of the magnetic field, and 
\begin{equation} \label{sigma_z}
\sigma_z = \frac{B_0^2}{4\pi n_0 w_0 m c^2} = \sigma_0  \left( \frac{B_0}{\delta B_{{\rm{rms}}0}} \right)^2  \, ,
\end{equation}
This argument assumes that the turbulence survives long enough to allow the particles to reach this upper limit (we also assumed $B_0/\delta B_{\rm rms} \gtrsim 1$). A numerical confirmation of Eq.~(\ref{gamma_c}), with $k_I \sim k_N$, was presented in \citet{ComissoSironi18} by performing simulations with different domain sizes. We point out that inverse magnetic energy transfer \citep[e.g.][]{BiskampSchwarz01,Zrake14,Brandenburg15} can possibly drive a substantial decrease in time of $k_I$, which, in turn, can allow the most energetic particles to reach even higher energies.

We observe that the slope of the power-law is not universal, but it depends on the magnetization $\sigma_0$ and the ratio $\delta B_{{\rm{rms0}}}/B_0$ \citep{ComissoSironi18}. The inset of Fig.~\ref{fig3} shows how the power-law index changes with the magnetization $\sigma_0$ from two series of simulations having $\delta B_{{\rm{rms0}}}/B_0=1$ and $\delta B_{{\rm{rms0}}}/B_0=2$. We can see that the slope of the power-law becomes harder for larger magnetization, and that for fixed $\sigma_0$ it is harder when increasing $\delta B_{{\rm{rms0}}}/B_0$ (see also Fig.~\ref{fighardspec}). The decrease of the power-law index $p$ for increasing magnetization $\sigma_0$ (see also \citet{Zhdankin17,ComissoSironi18}) is in analogy with the results of PIC simulations of relativistic magnetic reconnection \citep{ss_14,guo_14,werner_16,lyutikov_17,PetropoulouSironi2018}. We will see that magnetic reconnection plays an important role also in the turbulence scenario considered here. However,  as we show below, its role is confined to the initial stages of  particle acceleration, while the dominant acceleration process is given by stochastic scattering off turbulent fluctuations, which determines the slope and the cutoff of the high-energy power-law tail.

%%%%%%%%%%%%%%%%%%%%%%%%%%%%%%%%

\begin{figure}
\begin{center}
\hspace*{-0.185cm}\includegraphics[width=8.75cm]{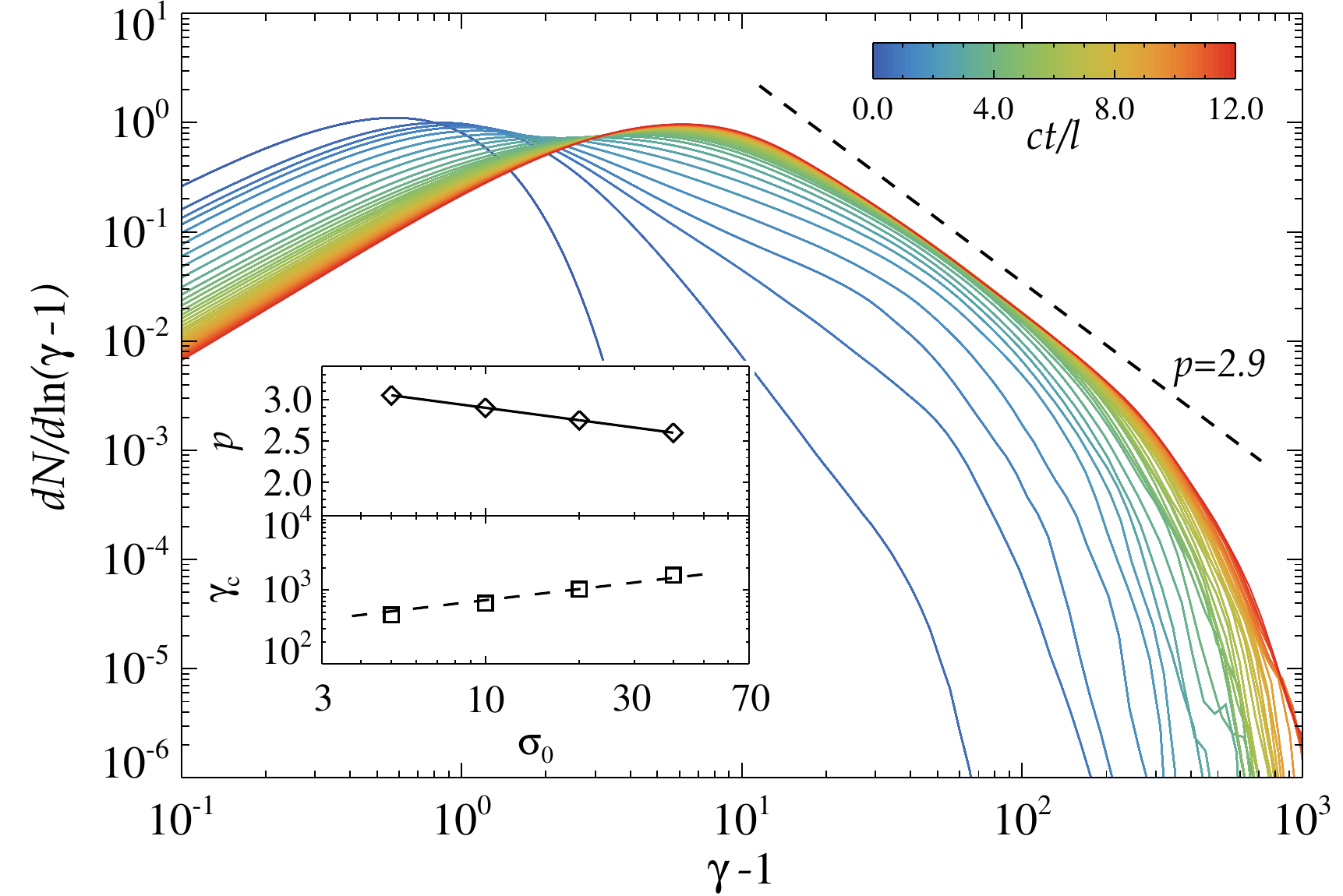}
\end{center}
\caption{Time evolution of the particle spectrum $dN/d\ln(\gamma-1)$ for the simulation in Fig.~\ref{fig6}. At late times, the spectrum displays a power-law tail with index $p = - d\log N/d\log (\gamma -1)\sim 2.9$. About $16 \%$ of the particles have $\gamma \geq 15$ at $ct/l=12$ (twice the peak of the particle energy spectrum), which gives an indication of the percentage of nonthermal particles. The inset shows the power-law index $p$ and the cutoff Lorentz factor $\gamma_c$ as a function of the magnetization $\sigma_0$. The dashed line indicates the scaling $\gamma_c \propto \sigma_0^{1/2}$ expected for a $\sigma_0$-independent domain size $L/d_{e0}=820$.}
\label{fig8}
\end{figure}

A similar picture holds in 3D, i.e., a generic by-product of the magnetized turbulence cascade is the production of a large number of nonthermal particles. 
Figure~\ref{fig8} shows the evolution of the particle energy spectrum $dN/d\ln(\gamma-1)$ starting from the initial Maxwellian peaked at $\gamma-1\sim \gamma_{th0}-1\simeq 0.6$. As time progresses, the particle energy spectrum shifts to higher energies and develops a high-energy tail containing a large fraction of particles. At late times, when most of the turbulent energy has decayed, the particle energy spectrum stops evolving (orange and red lines) and it peaks at  $\gamma-1\sim 7$. It extends well beyond the peak into a nonthermal tail of ultra-relativistic particles that can be described by a power-law with an index $p \sim 2.9$ (main frame of Fig.~\ref{fig8}). As in the 2D case, the normalization of the power-law is close to the peak of the spectrum, giving a large fraction of nonthermal particles. 
At $ct/l=12$ we find that about $16 \%$ of  particles have or exceed twice the energy of the spectral peak, which provides an indication of the percentage of particles in the nonthermal tail $\zeta_{\rm{nt}}$.

In order to understand the dependence of the high-energy power-law slope on the initial magnetization in 3D, we performed four large-scale 3D simulations with $\sigma_0 \in \left\{ {5,10,20,40} \right\}$ and same $\delta B_{{\rm{rms}}0}/ B_0 = 1$, $L/d_{e0}=820$. The power-law index $p$ decreases for increasing $\sigma_0$ (see top inset in Fig.~\ref{fig8}), with values that are close to the ones from the corresponding 2D simulations with $\delta B_{{\rm{rms}}0}/ B_0 = 1$ (blue curve from the inset in Fig.~\ref{fig3}).
Here we show also the scaling of the high-energy cutoff $\gamma_c$ (bottom inset in Fig.~\ref{fig8}), defined as the Lorentz factor where the spectrum drops one order of magnitude below the power-law best fit. The high-energy cutoff $\gamma_c$ increases as $\gamma_c \propto \sigma_0^{1/2}$ (compare with dashed line in the inset), which is consistent with the expectation from Eqs. (\ref{gamma_c}) and (\ref{sigma_z}) for a $\sigma_0$-independent domain size $L/d_{e0}$ and fixed $\delta B_{{\rm{rms}}0}/ B_0$.

%%%%%%%%%%%%%%%%%%%%%%%%%%%%%%%%

\begin{figure}
\begin{center}
\hspace*{-0.185cm}\includegraphics[width=8.75cm]{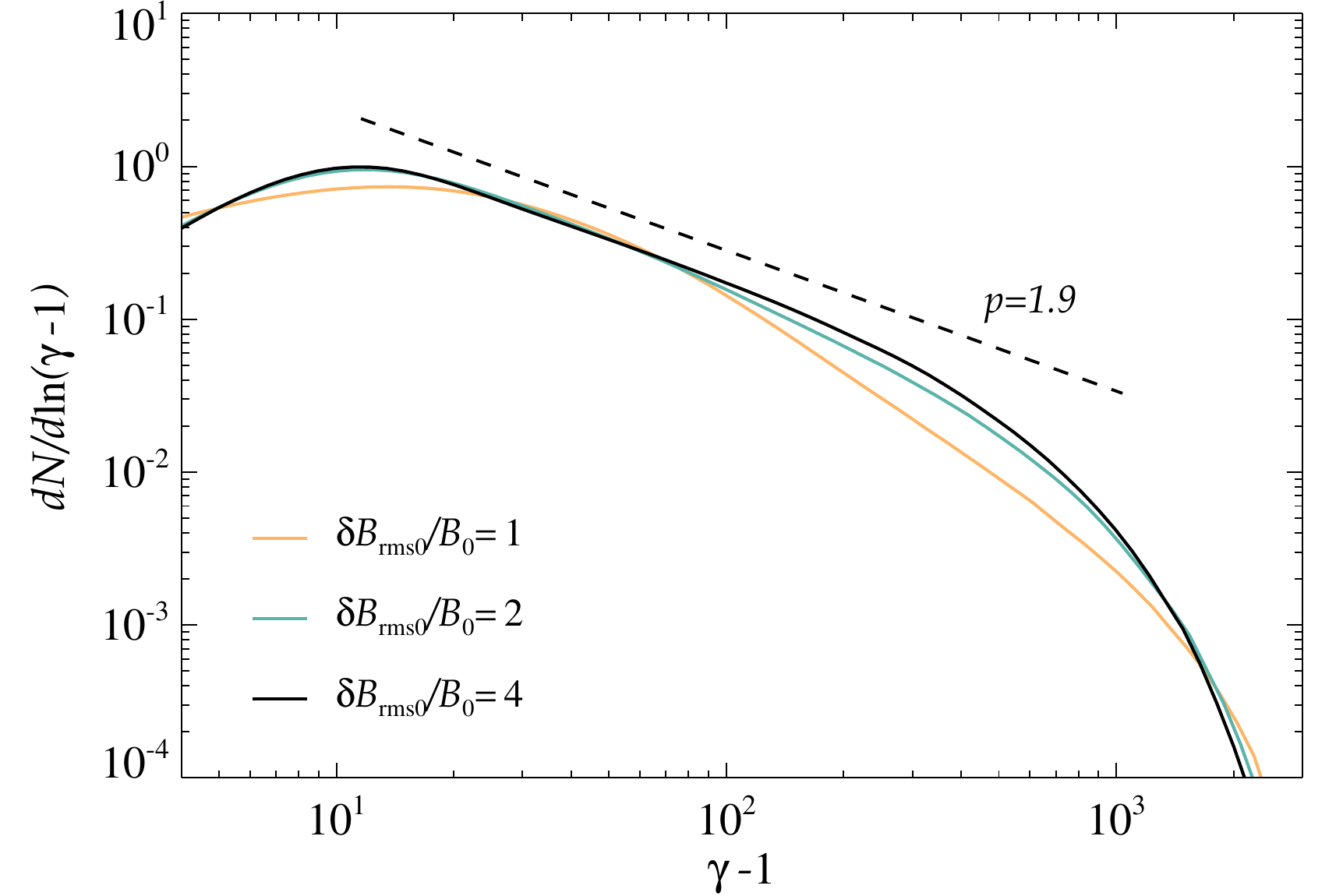}
\end{center}
\caption{Particle spectra $dN/d\ln(\gamma-1)$ at late times for simulations with magnetization $\sigma_0=40$, system size $L/d_{e0}=3280$ (with $l=L/8$), and different values of initial fluctuations $\delta B_{{\rm{rms}}0}/ B_0 \in \left\{ {1,2,4} \right\}$. For the case with larger initial fluctuations, the late time particle spectrum displays a power-law tail with index $p = - d\log N/d\log (\gamma -1)\sim 1.9$ and about $31 \%$ of the particles have $\gamma \geq 25$ at $ct/l=12$ (twice the peak of the particle energy spectrum at that time), which which gives an indication of the percentage of particles in the nonthermal tail.}
\label{fighardspec}
\end{figure}   
\begin{figure}
\begin{center}
\hspace*{-0.185cm}\includegraphics[width=8.75cm]{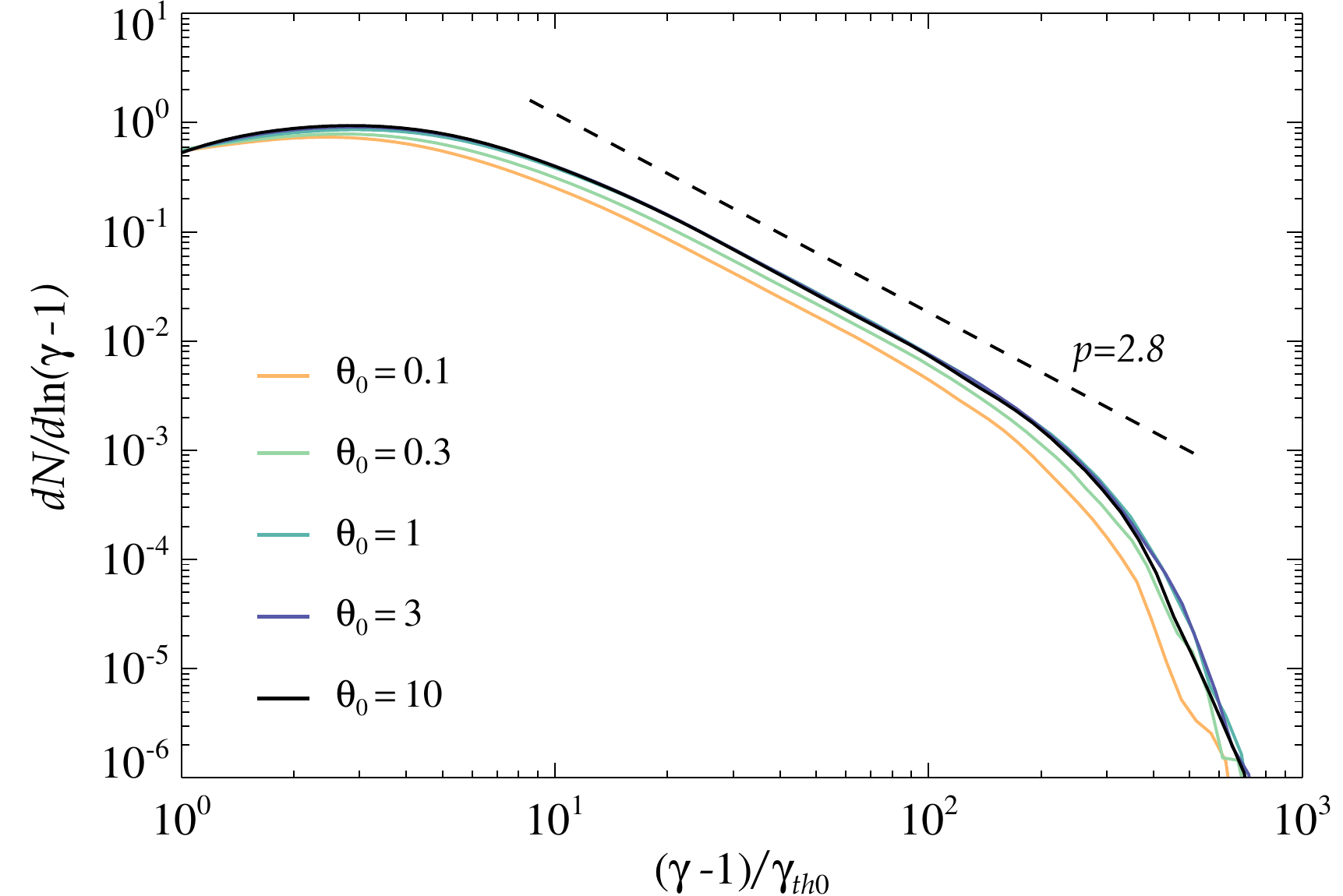}
\end{center}
\caption{Particle spectra $dN/d\ln(\gamma-1)$ at $ct/l=12$ for simulations with fixed $\sigma_0=10$, $\delta B_{{\rm{rms}}0}/ B_0=1$, and $L/d_{e0}=1640$ (with $l=L/8$), but different normalized initial temperature $\theta_0 = {k_B T_0}/{m c^2} \in \left\{ {0.1,0.3,1,3,10} \right\}$. The $x$-axis has been normalized to the initial thermal Lorentz factor $\gamma_{th0}$ to facilitate comparison among the different cases.}
\label{fig4}
\end{figure}    

Several astrophysical systems are thought to have $\delta B_{{\rm{rms}}}/ B_0$ larger than unity (e.g.,  $\delta B_{{\rm{rms}}}^2/ B_0^2 \sim 6$ in some regions of the Crab Nebula, \citealt{LyutikovMNRAS2019}). Therefore, we have performed  three additional 2D simulations with initial ratios $\delta B_{{\rm{rms}}0}/ B_0 = 1, \, 2, \, 4$, with fixed initial magnetization $\sigma_0=40$, and a larger domain size $L/d_{e0}=3280$. Fig.~\ref{fighardspec} shows that the power-law becomes harder with increasing $\delta B_{{\rm{rms}}0}/ B_0$, with $p < 2$ for large initial fluctuations. In this case, both Eq. (\ref{gamma_st}) and Eq. (\ref{gamma_c}) should be understood as upper limits which are subject to energy constraints, as we now discuss. The starting point of the power-law tail, $\gamma_{st}$, could be lower than indicated in Eq. (\ref{gamma_st}), if only a minor fraction of the available energy goes into thermal particles, while most of the energy goes into the nonthermal tail. Also, while in the case $p > 2$ one can have from Eq. (\ref{gamma_c}) that $\gamma_c \to \infty$ as $k_I d_{e0} \to 0$, the case $1<p<2$ has a lower attainable high-energy cutoff $\gamma_c$, since the mean energy per particle in the power-law tail has to be \citep{ss_14}
\begin{equation} \label{power_law_constraint}
\frac{{1 - p}}{{2 - p}} \, \frac{{{{({\gamma _c} - 1)}^{2 - p}} - {{({\gamma _{st}} - 1)}^{2 - p}}}}{{{{({\gamma _c} - 1)}^{1 - p}} - {{({\gamma _{st}} - 1)}^{1 - p}}}} = \left( {1 + \chi \frac{{{\sigma _0}}}{2}} \right){\gamma_{th0}} \, ,
\end{equation}
where $\chi$ is the fraction of turbulent magnetic energy converted into particles belonging to the power-law tail.

We conclude this section with the results of 2D simulations having different initial plasma temperature $\theta_0$. From Fig.~\ref{fig4}, we can see that the slope $p$, the fraction of particles in the nonthermal tail, and the extent of the nonthermal tail $\gamma_c/\gamma_{st}$ do not depend on $\theta_0$. Indeed, this plot shows that  spectra obtained from simulations with different $\theta_0$ nearly overlap, when shifted by an amount equal to the initial thermal Lorentz factor $\gamma_{th0}$. The role of the initial choice of temperature is only to produce an energy rescaling, since both $\gamma_{st}$ and $\gamma_c$ are proportional to $\gamma_{th0}$, as can be seen from the relations (\ref{gamma_st}) and (\ref{gamma_c}), and the definitions of $\sigma_0$ and $\sqrt{\sigma_z}/d_{e0}$ already take into account relativistic thermal effects.

Up to this point, we have discussed general features of the particle spectrum generated as a by-product of the plasma turbulence. We have found that despite some differences between  2D and 3D settings, the produced particle spectrum does not depend on the dimensionality of the simulation domain (see also \citet{ComissoSironi18}). In both cases, the high-energy  power-law  range extends from (about) the thermal peak to a maximum energy set by the energy-containing scale of  turbulence. These common features, combined with the fact that  the slope of the power-law is also similar, yield a similar percentage of particles in the power-law tail. In the next Sections, we will shed  light on the particle acceleration mechanisms that  produce the nonthermal particle spectrum.

\section{Particle Injection and Fast Reconnection} \label{SecInjection}

In this section, we investigate the physics behind the initial rapid acceleration of particles from low energies ($\gamma m c^2 \sim \gamma_{th} m c^2$), to energies well above the thermal peak ($\gamma m c^2 \gg \gamma_{th} m c^2$), which is usually referred to as the injection mechanism. The investigation of the injection mechanism will not be limited to this section, but it will be pursued also in parts of Sections \ref{SecEnergiz} and \ref{SecAnisotropy}. Here, specifically, as a continuation of our earlier analysis \citep{ComissoSironi18}, we want to examine the spatial locations where the injection process occurs, and understand what is special about these locations. To this aim, we have tracked the time evolution of a large sub-sample of particles that were randomly selected from our reference PIC simulations. 
Following in time their trajectory and energy evolution, we can analyze, for the fraction of particles that experience an injection process, the physical conditions at the moment of their rapid initial acceleration phase. Then we calculate the conditions for having efficient particle injection by reconnection, which are linked to the onset of fast magnetic reconnection mediated by the plasmoid instability. Indeed, despite their small filling fraction, we show that reconnecting current sheets can inject a large fraction of particles in a few outer-scale eddy turnover times.

\subsection{Particle injection at reconnecting current sheets} 

\begin{figure}
\begin{center}
\hspace*{-0.085cm}\includegraphics[width=8.75cm]{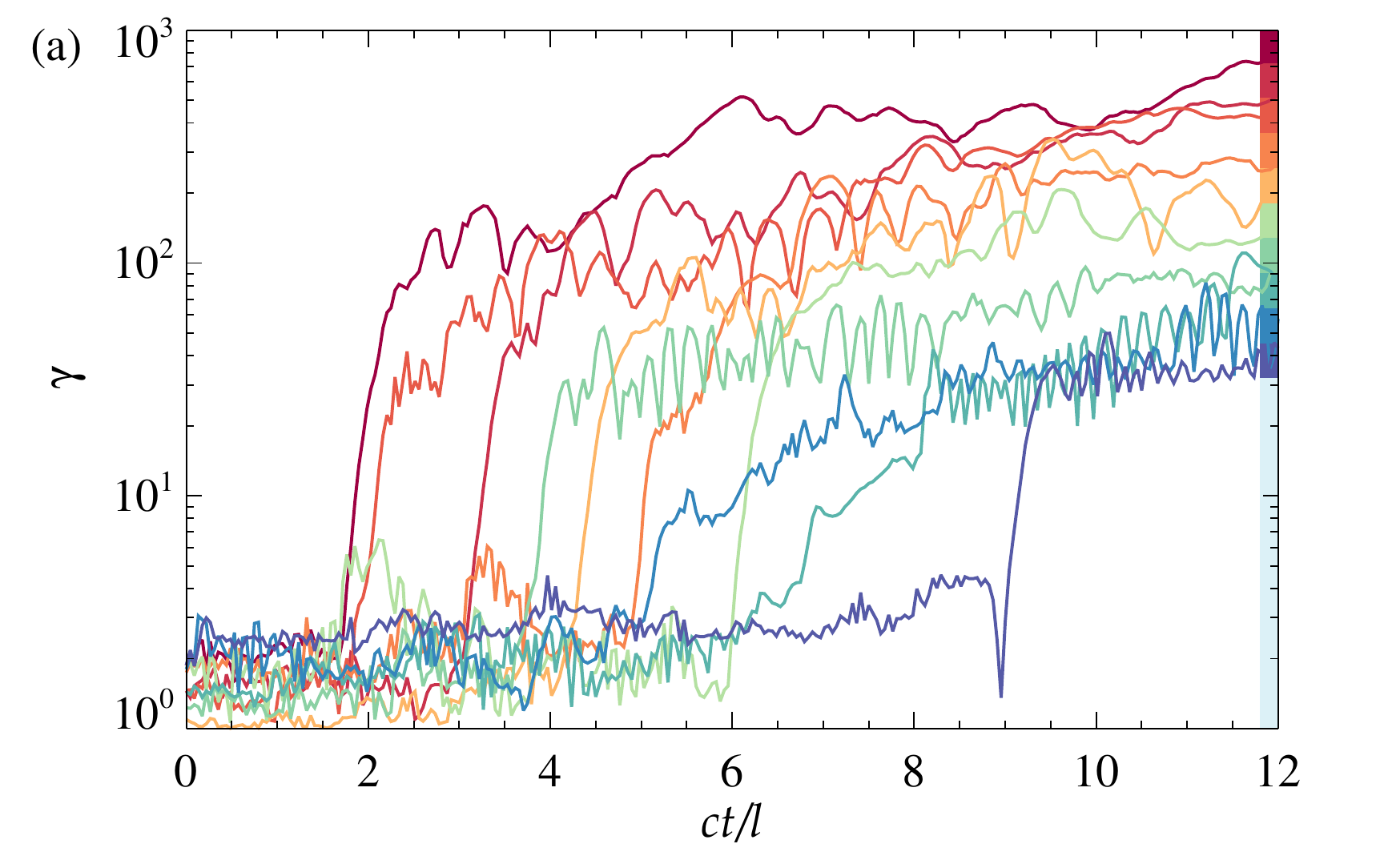}
\hspace*{-0.085cm}\includegraphics[width=8.75cm]{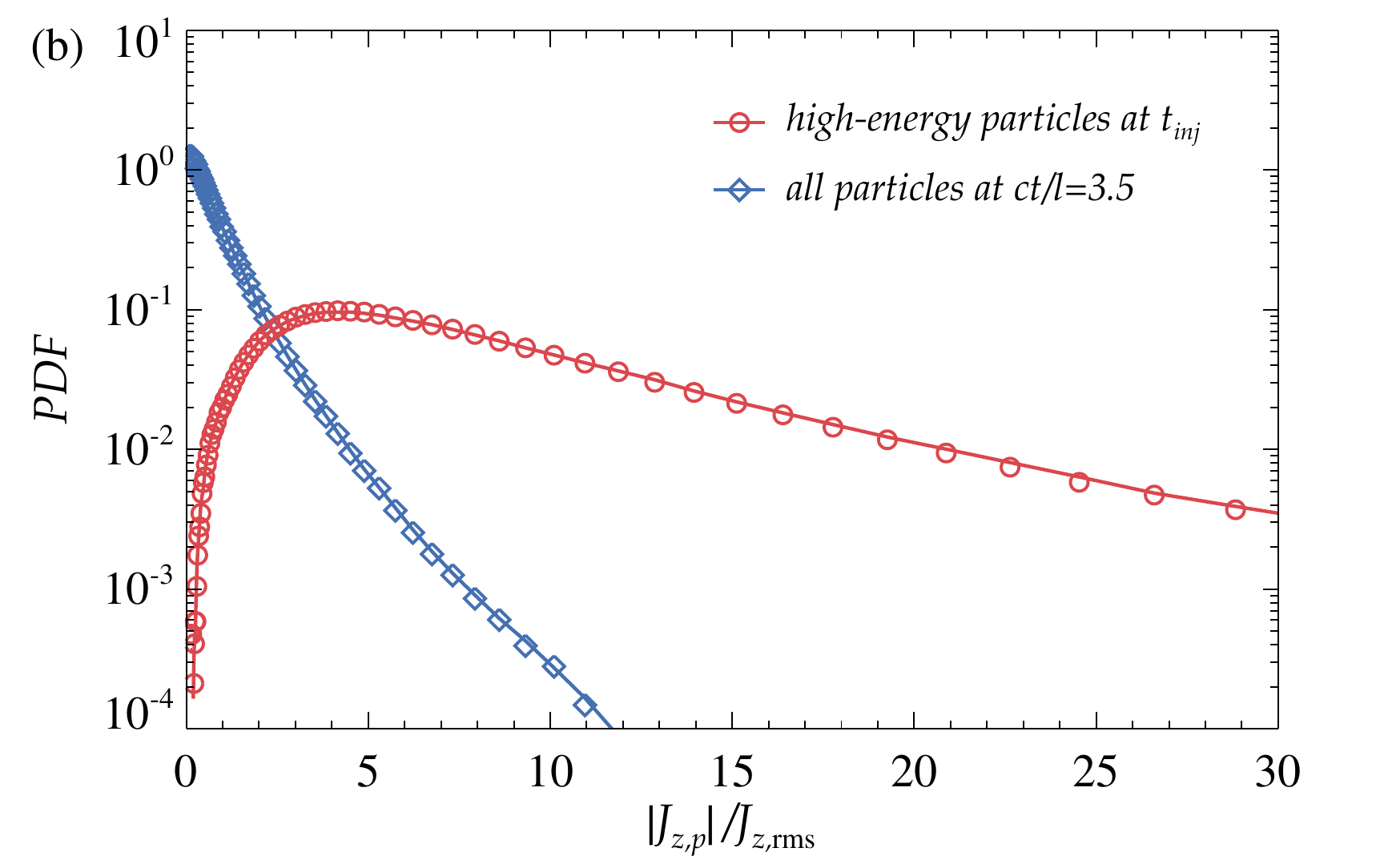}
\end{center}
\caption{Relation between particle injection and electric current density from the 2D simulation with $\sigma_0=10$, $\delta B_{{\rm{rms}}0}/ B_0=1$, and $L/d_{e0}=1640$. Top frame: Time evolution of the Lorentz factor for $10$ representative particles selected to end up in different energy bins at $ct/l=12$ (matching the different colors in the color bar on the right). Bottom frame: Probability density functions of $|{J_{z,p}}|/{J_{z,{\rm{rms}}}}$ experienced by  high-energy particles at their injection time $t_{in\!j}$ (red circles) and by all our tracked particles at $ct/l=3.5$ (blue diamonds). About $95\%$ of the high-energy particles are injected at locations with $|{J_{z,p}}| \ge 2{J_{z,{\rm{rms}}}}$.}
\label{fig_PartInj1}
\end{figure}  

\begin{figure}
\begin{center}
\includegraphics[width=8.00cm]{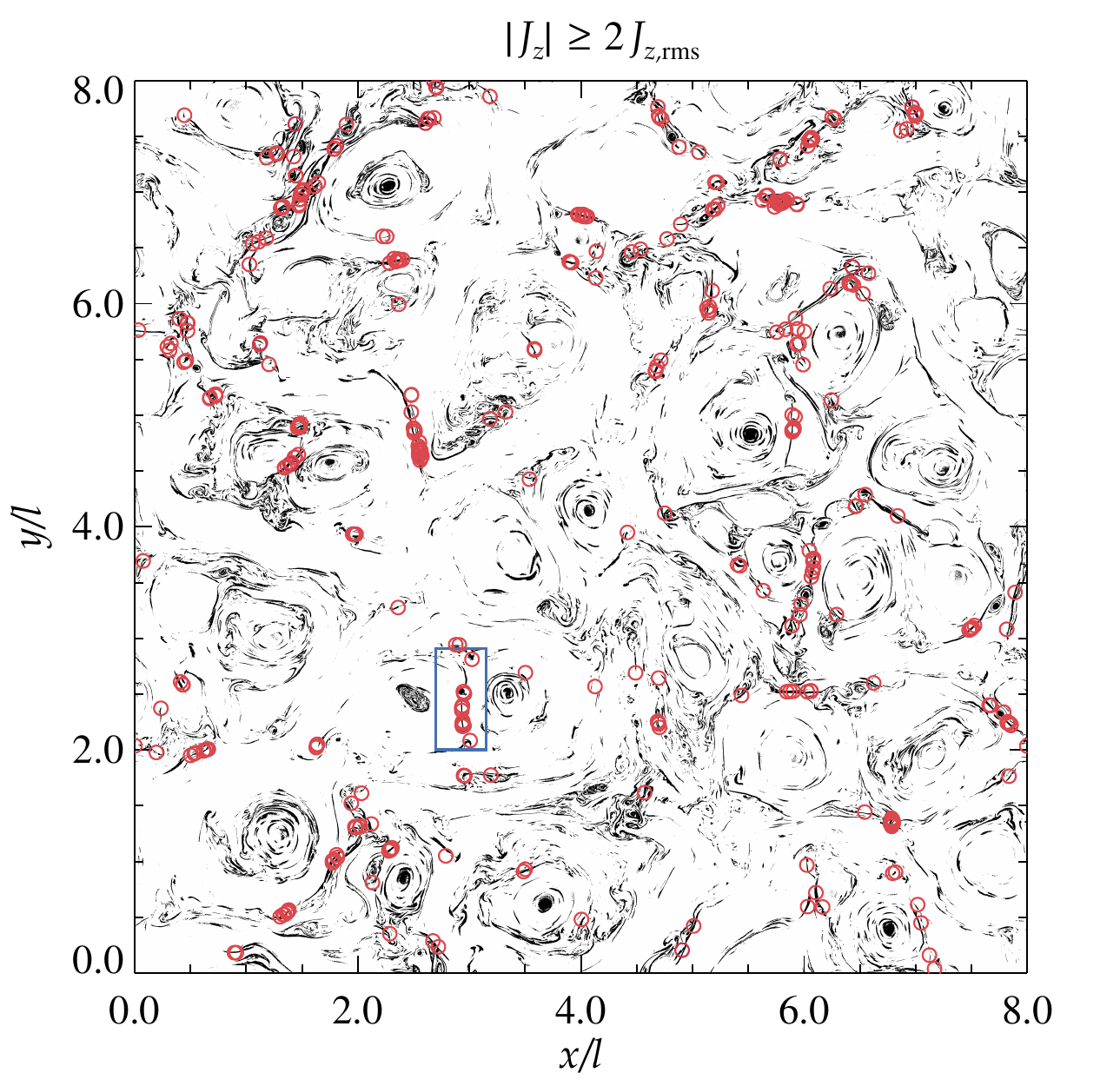}
\includegraphics[width=8.00cm]{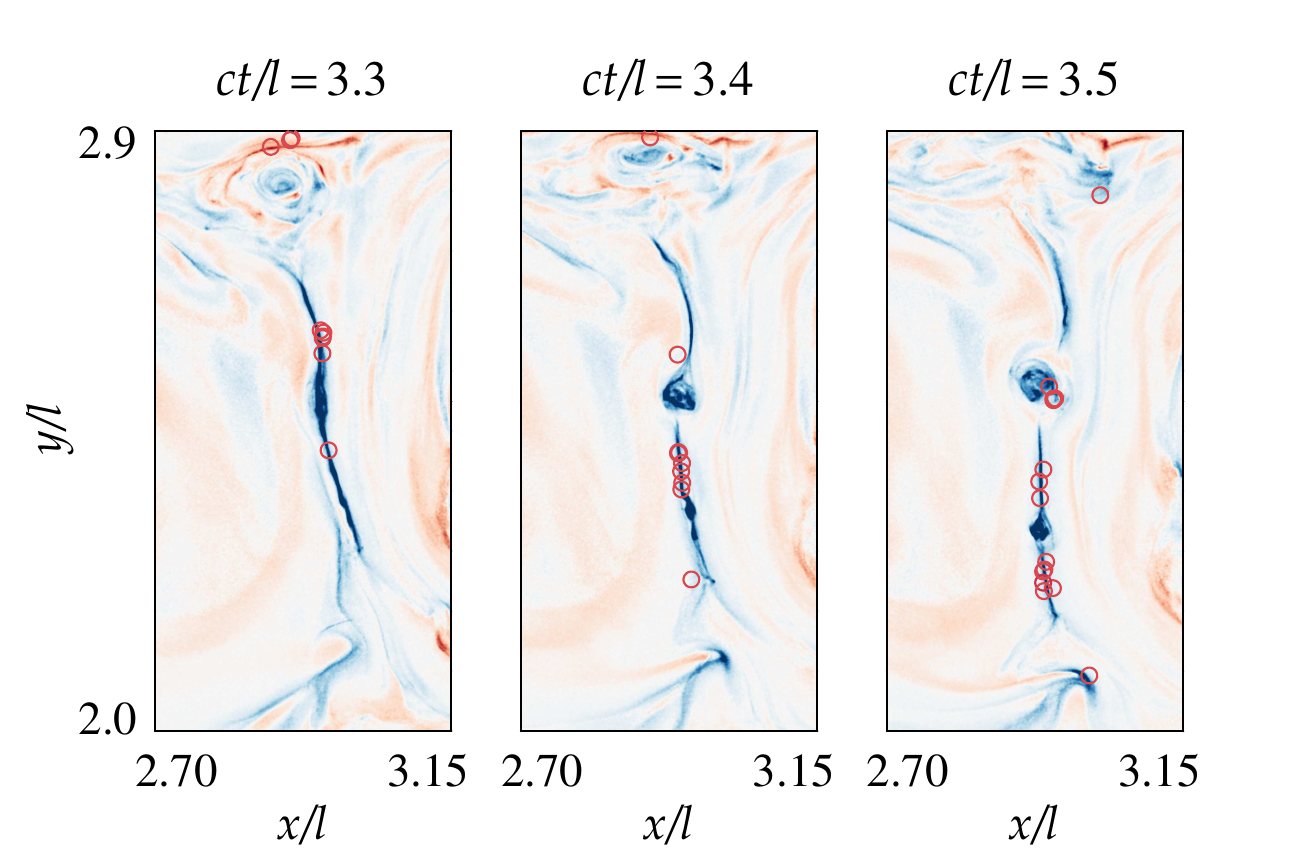}
\end{center}
\caption{Spatial correlation between particle injection and reconnecting current sheets for the same simulation as in Fig.~\ref{fig_PartInj1}. Top frame: Regions of space with $|{J_z}| \geq 2\, {\big\langle J_z^2 \big\rangle}^{1/2}$ (shown in black) at $ct/l=3.5$, with red circles indicating the positions of the particles undergoing injection around this time. 
Bottom frames: Shaded isocontours of $J_z$ in the spatial domain $(x/l,y/l) \in [2.70,3.15] \times [2.0,2.9]$ (corresponding to the area within the rectangular blue contour in the top frame) at times $ct/l=3.3$ (left), $ct/l=3.4$ (center), and $ct/l=3.5$ (right). The red circles indicate the positions of particles undergoing injection around this time. The color scheme for the shaded isocontours is such that blue indicates  regions with $J_z < 0$, while red indicates  regions with $J_z > 0$. }
\label{fig_PartInj2}
\end{figure}

We begin our analysis from the reference 2D case, and then we extend the analysis to the reference 3D case. For the injection analysis presented in this section, we employed a sub-sample of $\sim10^6$ tracked particles for the 2D case, and a sub-sample of $\sim10^7$ tracked particles for the 3D case.

We show in Fig. \ref{fig_PartInj1}(a) the time evolution of the Lorentz factor for $10$ representative particles that eventually populate the nonthermal tail at $ct/l=12$ (see particle spectrum in Fig. \ref{fig3}). These particles have a distinct moment in which they are ``extracted'' from the thermal pool at $\gamma \sim \gamma_{th}$ and injected to higher Lorentz factors $\gamma \gg \gamma_{th}$. To identify this moment, that we call injection time $t_{in\!j}$, we evaluate when the rate of increase of the particle Lorentz factor (averaged over $c \Delta{t} / d_{e0} =45$)  satisfies $\Delta\gamma/\Delta{t} \ge {\dot\gamma}_{thr}$, and prior to this time the particle Lorentz factor was $\gamma \le 4\gamma_{th0}\sim6$. We take the threshold ${\dot\gamma}_{thr}\simeq0.01\sqrt{\sigma_0}\gamma_{th0}\omega_{p0}$, but we have verified that our identification of $t_{in\!j}$ is nearly the same when varying ${\dot\gamma}_{thr}$ around this value by up to a factor of three (the factor $0.01$ is much lower than the typical collisionless reconnection rate [$\sim 0.1$, in units of the Alfv\'en speed], which is the appropriate reference scaling here,  as showed in \citet{ComissoSironi18} and below).

Once $t_{in\!j}$ is determined for the population of particles at hand, it is possible to explore the properties of the electromagnetic fields at the injection location. In this case, by analyzing the fields at injection, we find that the out-of-plane current density $J_z$ is particularly revealing. In particular, $J_z$ has, in general, high values at  injection locations. To provide a statistical measure of the likelihood of this occurrence, we can construct the probability density function (PDF) of the magnitude of the out-of-plane electric current density experienced by the particles at their injection time, $|{J_{z,p}}|$, normalized by $J_{z,{\rm{rms}}}$, i.e., the standard deviation of the current density ${J_{z,{\rm{rms}}}} = {\big\langle J_z^2 \big\rangle}^{1/2}$ in the whole domain at that time. The outcome of this analysis is shown in Fig. \ref{fig_PartInj1}(b) by the red circles. The PDF of the high-energy particles at injection should be contrasted with the PDF of the entire population of particles at a representative time (here, $ct/l=3.5$), shown by the blue diamonds in Fig. \ref{fig_PartInj1}(b). The difference between the two PDFs is striking. The main difference is that the PDF of the overall particle population is peaked around zero, while the PDF of the high-energy particles at injection is peaked at much higher values corresponding to $|{J_{z,p}}| \sim 4\, {J_{z,{\rm{rms}}}}$. 
In particular, approximately $\sim95\%$ of the high-energy particles are injected at locations with $|{J_{z,p}}| \ge 2\, {J_{z,{\rm{rms}}}}$. On the other hand, by taking all the particles at the representative time $ct/l=3.5$, only $\sim9\%$ of them happen to be in regions where $|{J_{z,p}}| \ge 2\,{J_{z,{\rm{rms}}}}$. Note also that the PDF of the overall particle population does not follow Gaussian statistics due to the intermittent nature of current sheets in turbulence \citep[e.g.][]{Servidio2009,CerriApJL17,Haggerty17,Dong_2018}, which is therefore reflected in the PDF of the particles that sample the entire domain.

To obtain further insight, we look now at the morphology of regions with out-of-plane current density $|{J_z}| \ge 2\,{J_{z,{\rm{rms}}}}$, and we correlate it with the spatial locations of the particles undergoing injection at $t_{in\!j}$. This is shown in Fig. \ref{fig_PartInj2}(a), where we can see that the vast majority of the structures with $|{J_z}| \ge 2\,{J_{z,{\rm{rms}}}}$ are sheet-like structures, namely current sheets, and the overwhelming majority of particles at injection resides in these regions. A large fraction of these current sheets are active reconnection layers, fragmenting into plasmoids. A typical case of such reconnecting current sheets is illustrated in Fig. \ref{fig_PartInj2}(b), where we show a small portion of the domain, corresponding to the area within the rectangular blue contour in Fig. \ref{fig_PartInj2}(a), at different times $ct/l=3.3,3.4,3.5$. The reconnecting current sheet evolves in time and breaks up in shorter sheets due to the formation of plasmoids. During this period of time, particles are constantly injected up to nonthermal energies, as shown by the red circles in Fig. \ref{fig_PartInj2}(b).

%%%%%%%%%%%%%%%%%%%%%%%%%%%%%%
\begin{figure}
\begin{center}
\hspace*{-0.085cm}\includegraphics[width=8.75cm]{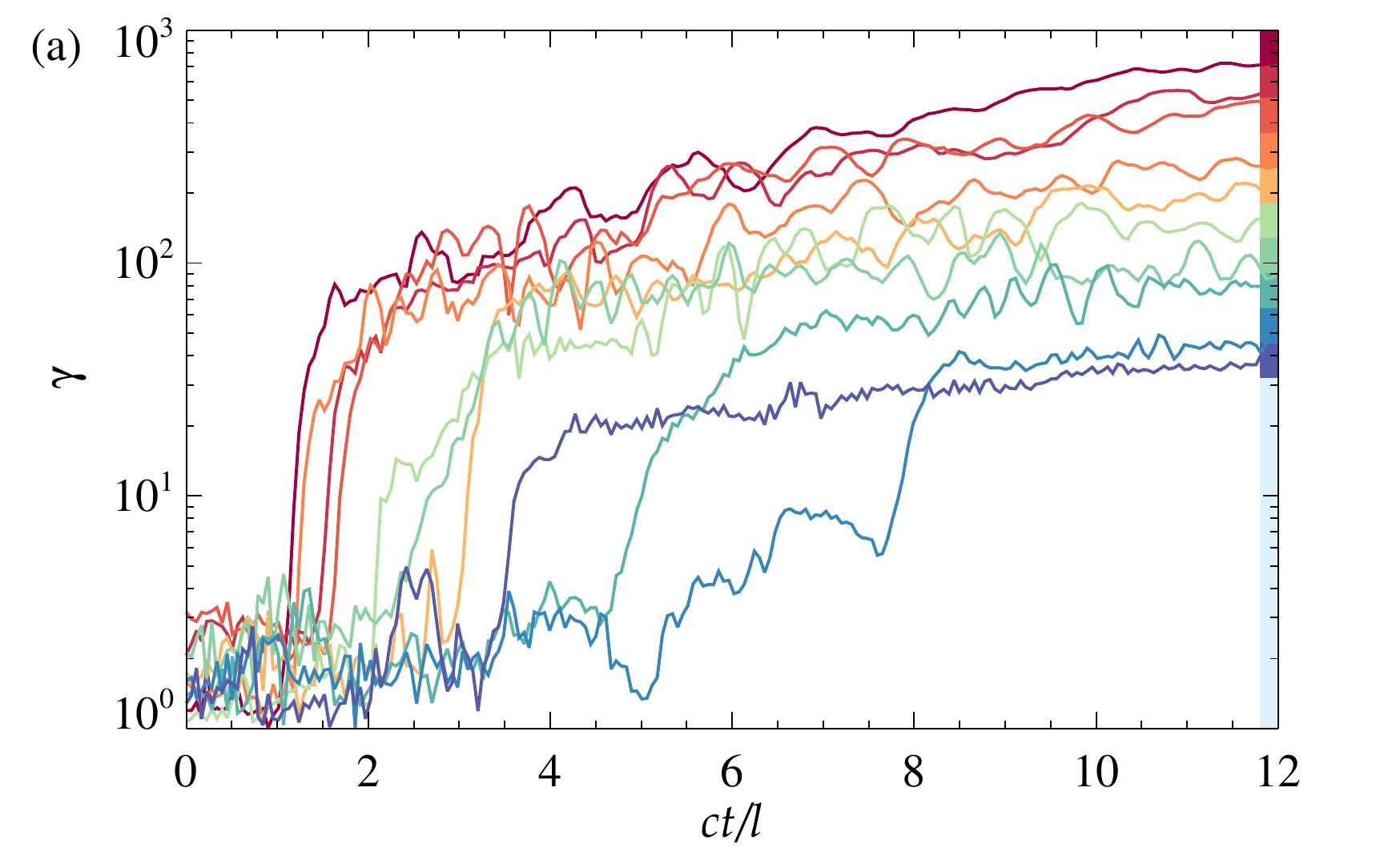}
\hspace*{-0.085cm}\includegraphics[width=8.75cm]{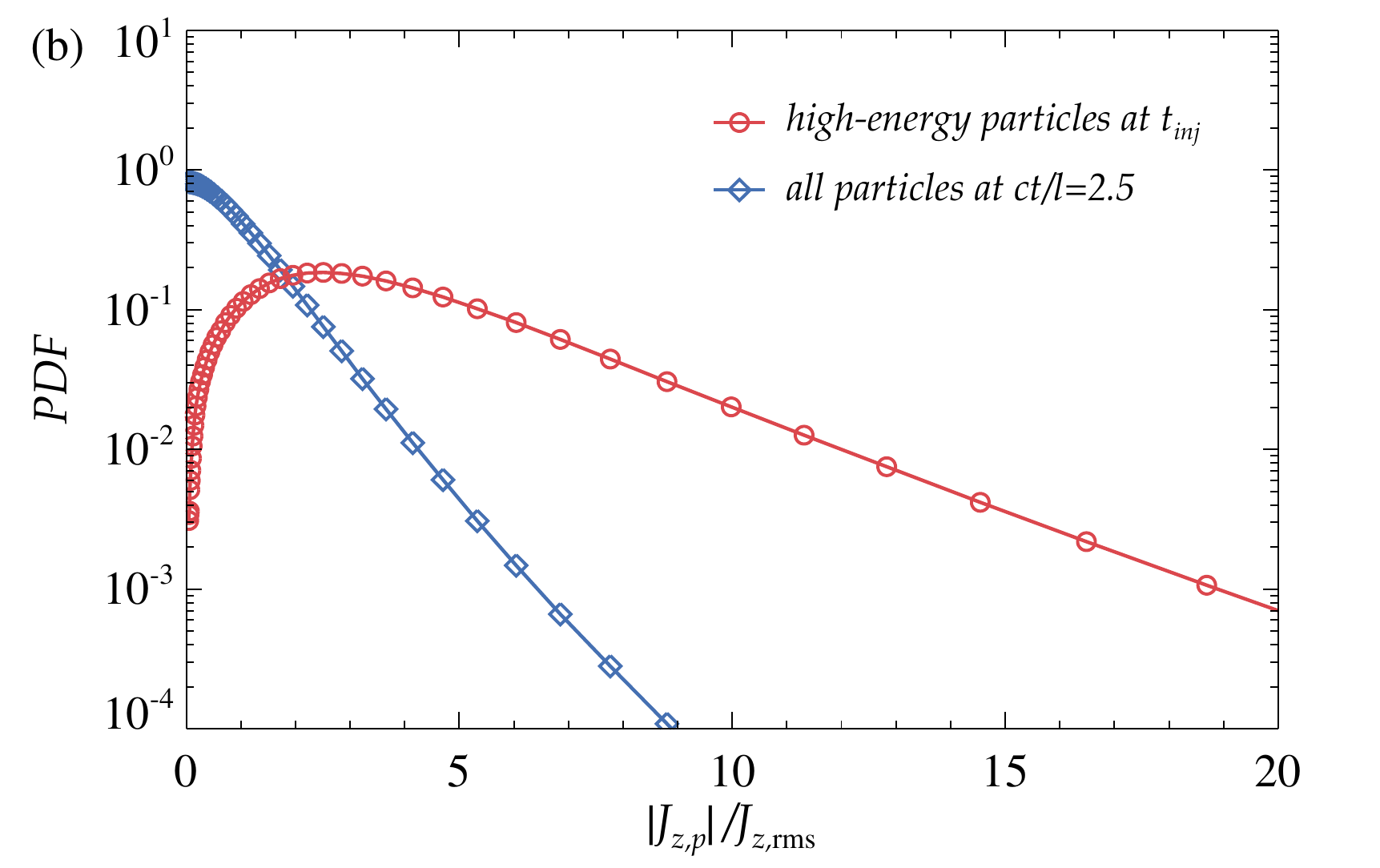}
\end{center}
\caption{Relation between particle injection and electric current density from the 3D simulation with $\sigma_0=10$, $\delta B_{{\rm{rms}}0}/ B_0=1$, and $L/d_{e0}=820$. Top frame: Time evolution of the Lorentz factor for $10$ representative particles selected to end up in different energy bins at $ct/l=12$ (matching the different colors in the color bar on the right). Bottom frame: Probability density functions of $|{J_{z,p}}|/{J_{z,{\rm{rms}}}}$ experienced by the high-energy particles at their $t_{in\!j}$ (red circles) and by all our tracked particles at $ct/l=2.5$ (blue diamonds). About $80 \%$ of the high-energy particles are injected at regions with $|{J_{z,p}}| \ge 2{J_{z,{\rm{rms}}}}$.}
\label{fig_PartInj5}
\end{figure}  
%%%%%%%%%%%%%%%%%%%%%%%%%%%%%%

%%%%%%%%%%%%%%%%%%%%%%%%%%%%%%
\begin{figure}
\begin{center}
\includegraphics[width=8.00cm]{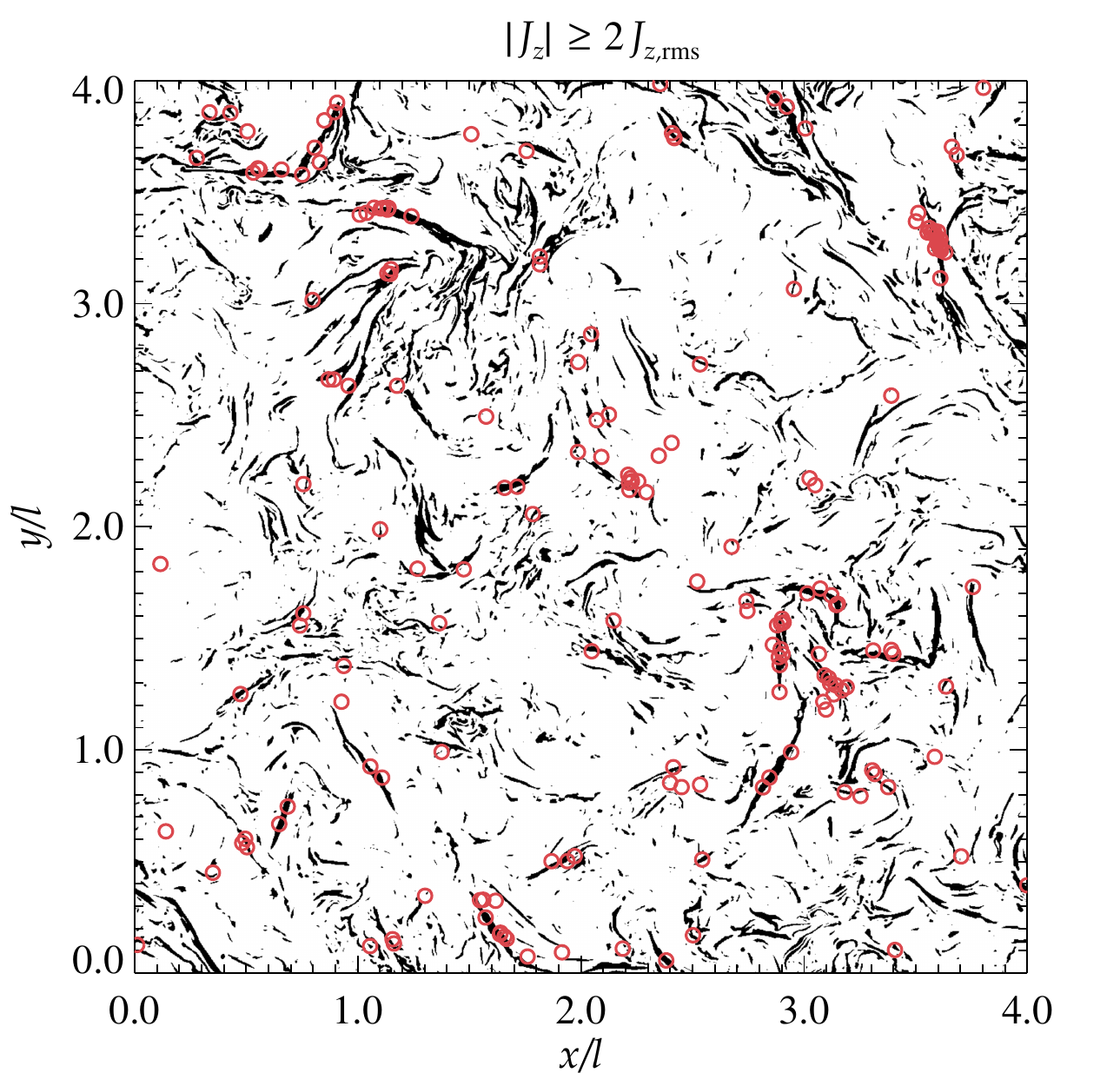}
\includegraphics[width=8.00cm]{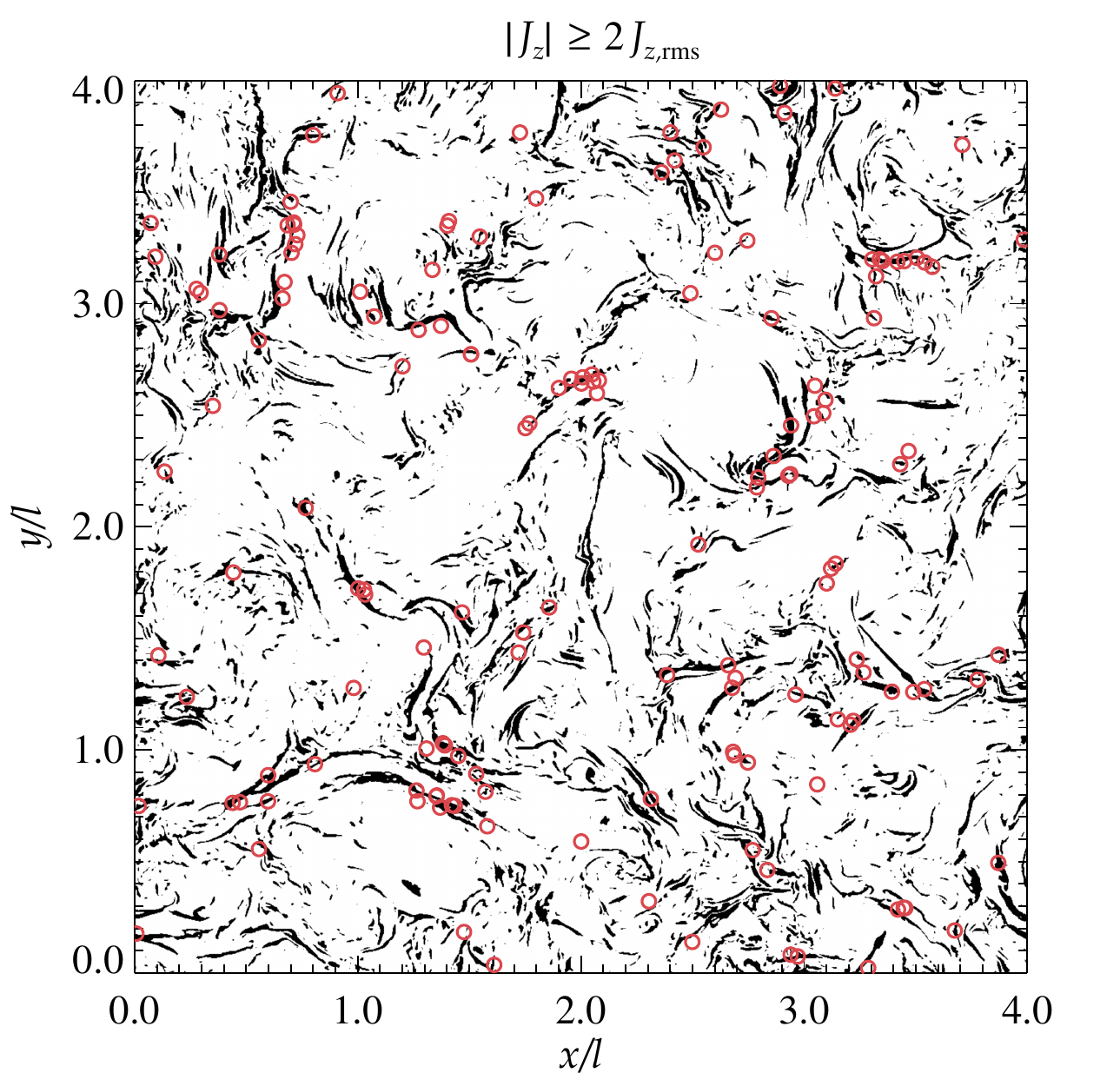}
\end{center}
\caption{Spatial correlation between particle injection and reconnecting current sheets for the same 3D simulation as in Fig.~\ref{fig_PartInj5}. In black, we show regions of space with strong current density $|{J_z}| \geq 2\, {\big\langle J_z^2 \big\rangle}^{1/2}$ at $ct/l=2.5$, for two representative planes of the 3D domain, taken at $z/l=0.6$ (top frame) and $z/l=3.4$ (bottom frame). The large-scale mean magnetic field ${\bm{B}}_0$ is in the out-of-plane direction. The red circles indicate the positions of  particles undergoing injection around this time.}
\label{fig_PartInj6}
\end{figure}  
%%%%%%%%%%%%%%%%%%%%%%%%%%%%%%

%%%%%%%%%%%%%%%%%%%%%%%%%%%%%%
\begin{figure}
\begin{center}
\includegraphics[width=8.55cm]{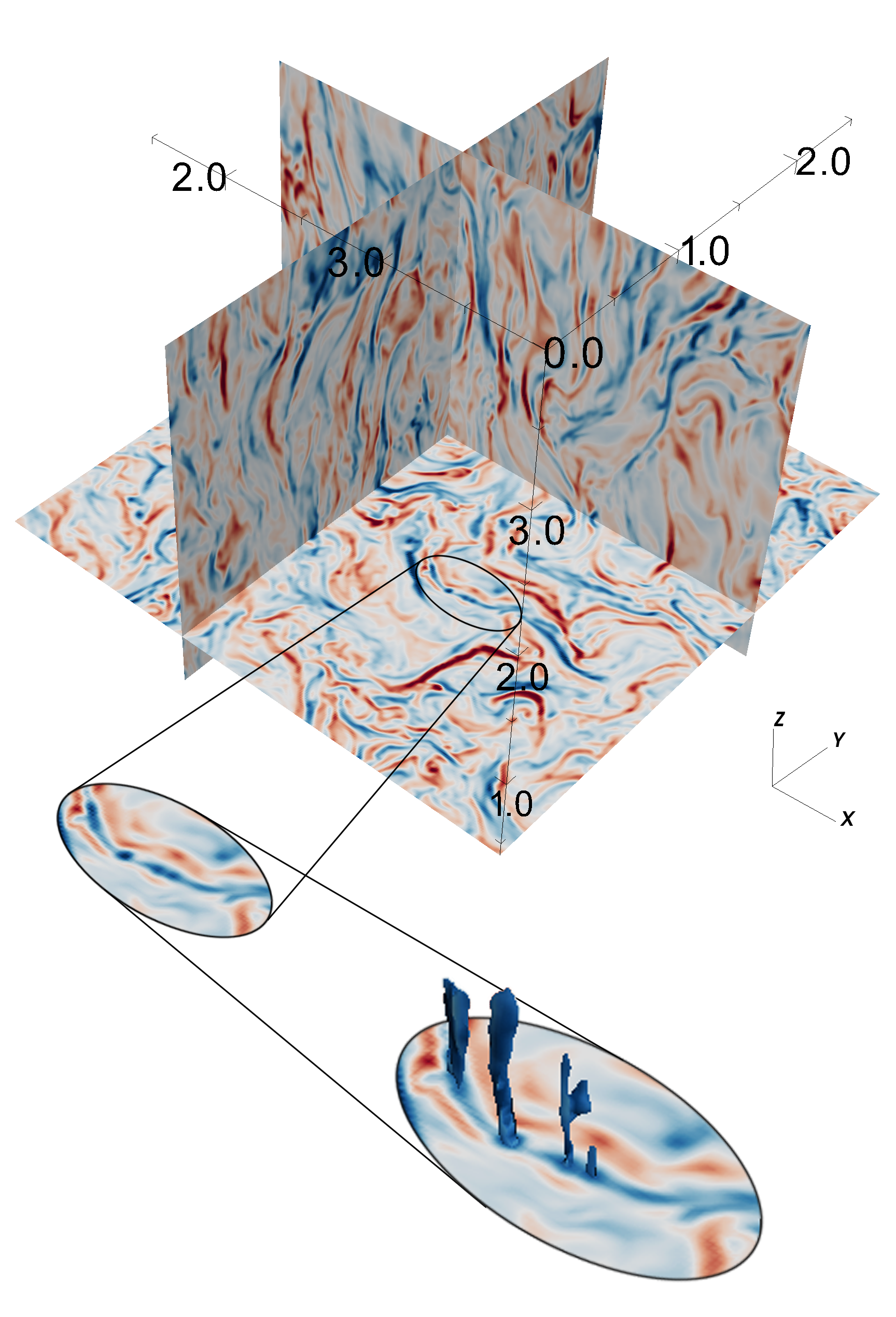}
\end{center}
\caption{Chain of flux ropes formed in a reconnecting current sheet that self-consistently develops in 3D turbulence (with $\sigma_0=10$, $\delta B_{{\rm{rms}}0}/ B_0=1$, and $L/d_{e0}=820$). Isosurfaces of the current density $J_z$ are shown in blue color in the zoomed region, highlighting four flux ropes (3D plasmoids) elongated along ${\bm{\hat z}}$, i.e., the direction of the mean magnetic field. The color scheme for the shaded isocontours is such that blue indicates  regions with $J_z < 0$, while red indicates  regions with $J_z > 0$.}
\label{fig_PartInj7}
\end{figure}  
%%%%%%%%%%%%%%%%%%%%%%%%%%%%%%

These results are also robust in 3D, for which we have performed the same type of analysis. Fig. \ref{fig_PartInj5}(a) shows the time evolution of the Lorentz factor for $10$ representative particles selected to end up in different energy bins of the nonthermal tail at $ct/l=12$ (see particle spectrum in Fig. \ref{fig8}). As in 2D, we can see a sudden acceleration episode with particles that are extracted from the thermal pool and injected into the acceleration process. We identify the  injection time $t_{in\!j}$ as for the 2D case, by evaluating when the Lorentz factor increases at a rate exceeding the same threshold ${\dot\gamma}_{thr}$ adopted for 2D, starting from a value that is $\gamma \le 5\gamma_{th0} \sim 8$ (this value is slightly higher than the 2D case, since in 3D a larger fraction of magnetic energy is dissipated by the end of the simulation). Note that, as in 2D, after the injection phase the particles continue to gain energy due to stochastic scattering off turbulent fluctuations. We will discuss in detail this second acceleration stage in Sec. \ref{SecEnergyDiff}.

By constructing the PDF of $|{J_{z,p}}|/J_{z,{\rm{rms}}}$ for the high-energy particles at injection and  for all particles at a representative time (taken at $ct/l=2.5$), we find results that are similar to the ones we have obtained for the 2D case. Fig. \ref{fig_PartInj5}(b) indeed shows that the PDF of the particles at injection (red circles) peaks at $|{J_{z,p}}|/{J_{z,{\rm{rms}}}} \sim 2.5$, as opposed to the PDF of the entire population of particles at the representative time $ct/l=2.5$  (blue diamonds), which peaks at $|{J_{z,p}}|/{J_{z,{\rm{rms}}}} \sim 0$. Again, particles at injection feel a substantial electric current density in the direction of the mean magnetic field. The peak of the PDF for the particles at injection is at a lower value of $|{J_{z,p}}|/{J_{z,{\rm{rms}}}}$  than in 2D, and in general there are weaker $|{J_{z,p}}|/{J_{z,{\rm{rms}}}}$ wings for both the PDF of all particles and the PDF of particles experiencing injection. This can be attributed to the lower levels of intermittency that characterize 3D magnetized turbulence with respect to its 2D counterpart \citep[e.g.][]{Biskamp2003}. Nevertheless, about $80 \%$ of the particles are injected in regions with $|{J_{z,p}}| \ge 2 \, {J_{z,{\rm{rms}}}}$. On the other hand, only approximately $11 \%$ of the entire population of particles (at the representative time $ct/l=2.5$) reside at $|{J_{z,p}}| \ge 2 \, {J_{z,{\rm{rms}}}}$. Therefore, also in 3D, special locations of high electric current density are associated with particle injection.

The spatial locations with $|{J_z}| \ge 2\,{J_{z,{\rm{rms}}}}$ are associated with current ribbons that are predominantly elongated along the mean magnetic field ${\bm{B}}_0$. In Fig. \ref{fig_PartInj6}, we show the morphology of these regions for two representative planes perpendicular to ${\bm{B}}_0$ (taken at $ct/l=2.5$). These regions are sheet-like structures with a variety of length scales. We can see that the majority of the particles undergoing injection, whose location is shown by the red circles, resides at these current sheets. A large fraction of these current sheets are active reconnection layers, fragmenting into plasmoids. A typical example of such reconnecting current sheets is shown in Fig. \ref{fig_PartInj7}. We can see four flux ropes (3D plasmoids) that are formed within the current sheet (and elongated in the direction of the mean magnetic field), which is the typical signature of fast plasmoid-mediated reconnection. We will see in the next subsection that current sheets undergoing fast reconnection are important for having efficient particle injection, as they are capable to ``process'' a significant fraction of particles (from the thermal pool) during their lifetime in the turbulent plasma.

%%%%%%%%%%%%%%%%%%%%%%%%%%%%%%  
\begin{figure*}
 \centering 
  \includegraphics[width=0.99\textwidth]{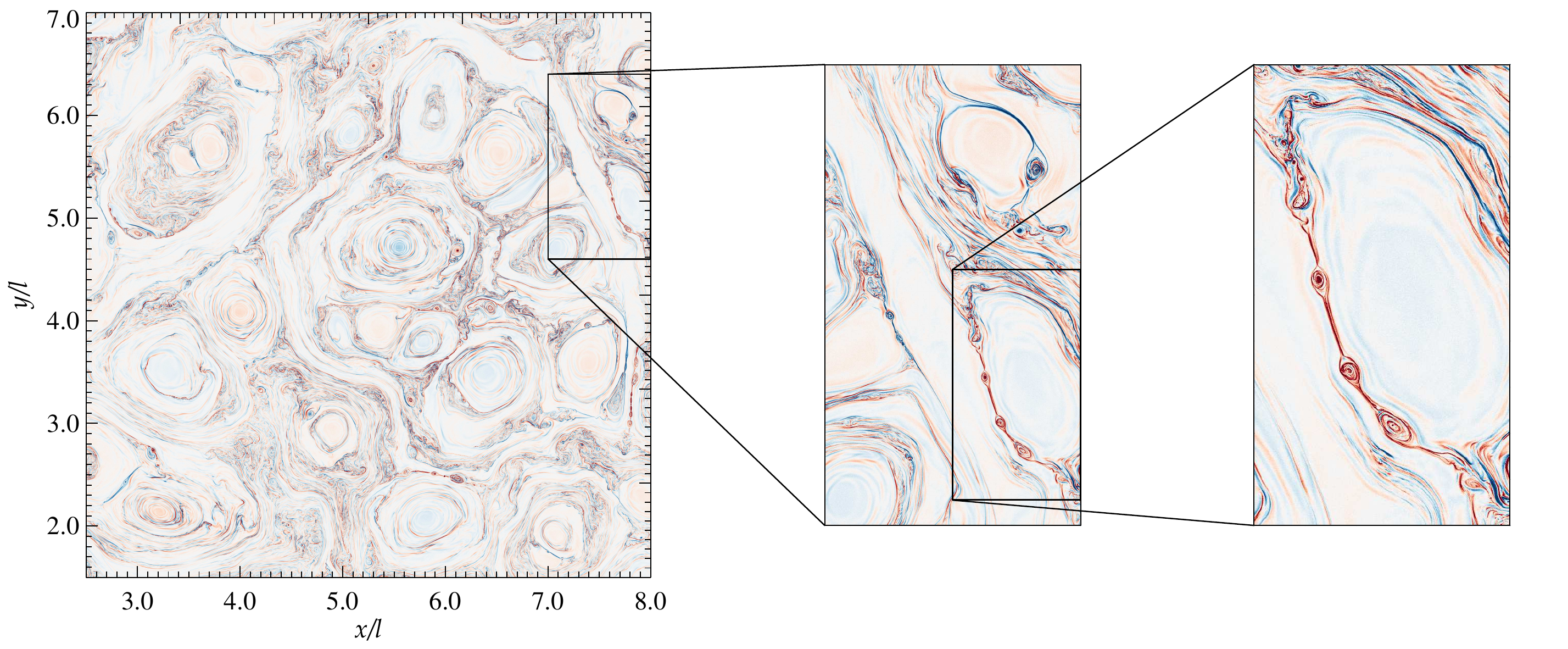}
 \caption{Chains of plasmoids in plasma turbulence from a 2D simulation with $L/d_{e0}=6560$ ($\sigma_0=10$, $\delta B_{{\rm{rms}}0}/ B_0=1$). The shaded isocontours represent the electric current density $J_z$ in a portion of the spatial domain given by $(x/l,y/l) \in [2.5,8.0] \times [1.5,7.0]$ at time $ct/l=4.5$. The color scheme is such that blue represents the most negative value, and red the most positive value. Zoomed-in subdomains are used to reveal one plasmoid chain.} 
\label{fig_PartInj3}
\end{figure*} 
%%%%%%%%%%%%%%%%%%%%%%%%%%%%%%

%%%%%%%%%%%%%%%%%%%%%%%%%%%%%%%
\begin{figure}
\begin{center}
\includegraphics[width=8.55cm]{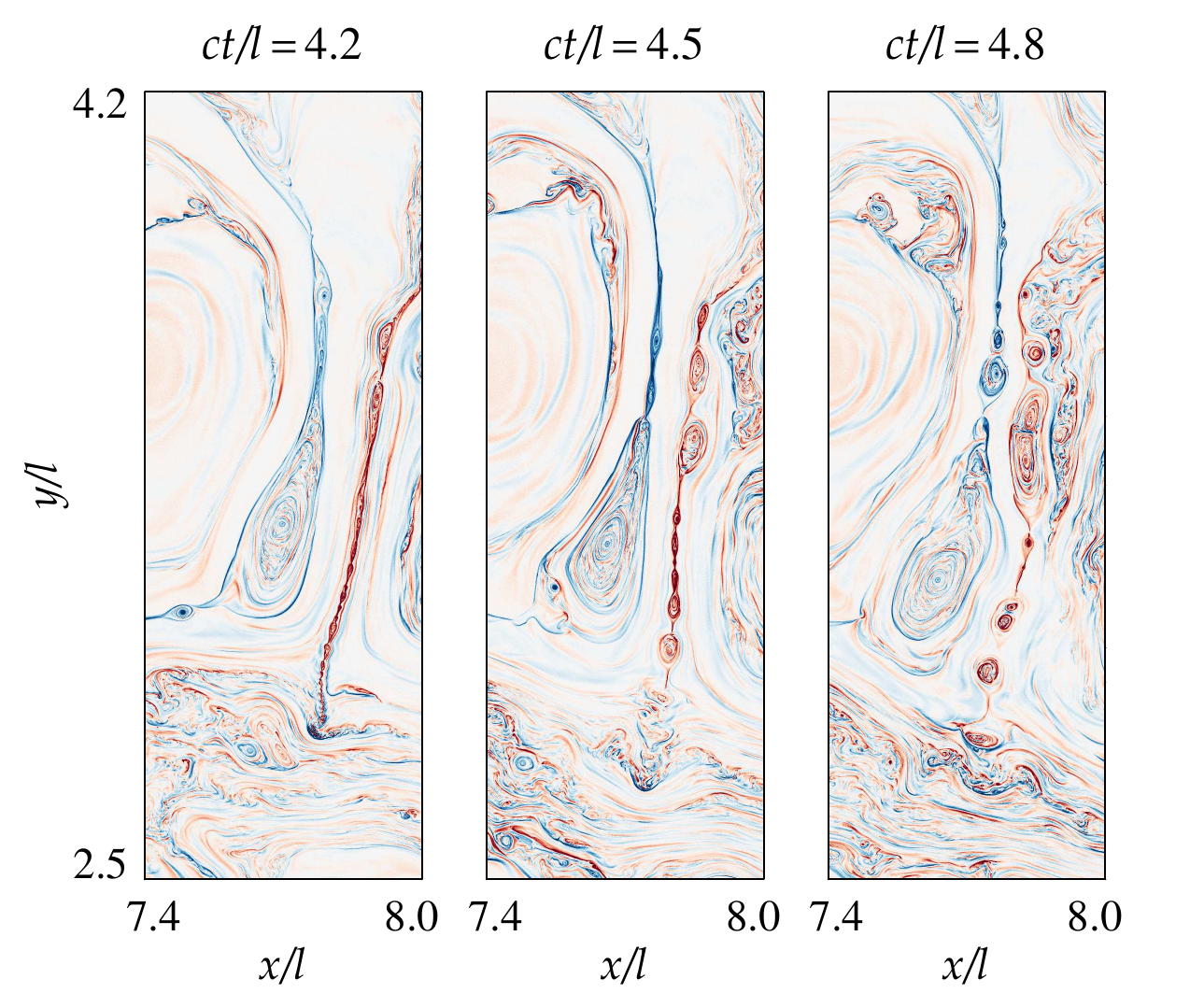}
\end{center}
\caption{Plasmoid formation and development from a 2D simulation with $L/d_{e0}=6560$ ($\sigma_0=10$, $\delta B_{{\rm{rms}}0}/ B_0=1$). The shaded isocontours represent the electric current density $J_z$ in a portion of the spatial domain given by $(x/l,y/l) \in [7.4,8.0] \times [2.5,4.2]$ at times $ct/l=4.2$ (left), $ct/l=4.5$ (center), and $ct/l=4.8$ (right). Colors range from blue ($J_z < 0$) to red ($J_z > 0$).}
\label{fig_PartInj4}
\end{figure}  
%%%%%%%%%%%%%%%%%%%%%%%%%%%%%%

\subsection{Plasmoid-mediated disruption of the current sheets and efficiency of reconnection-mediated injection}

Reconnecting current sheets are a viable source of particle injection in typical astrophysical systems ($\ell \ggg d_{e0}$) only if the injection efficiency (i.e., the fraction of particles going through the injection phase) is large and independent of  system size.  Here we show that this is indeed expected for our turbulence studies.

The rate at which a reconnecting current sheet can process particles is proportional to the normalized reconnection speed $\beta_R = v_R /c$, which essentially quantifies the speed of the reconnection process. This rate would be low for very elongated current sheets, as the large aspect ratio has the effect of throttling the reconnection rate. Indeed, a stable current sheet would be able to reach an asymptotic width determined by the microphysics of the plasma. For a relativistic pair plasma, the steady-state solution for the half-width of a reconnecting current sheet is \citep{ComiAsenjo14} 
\begin{equation} \label{} 
{\lambda_\infty} \simeq d_w = \sqrt {\frac{{m{c^2}}}{{4\pi n{e^2}}} w} \, ,
\end{equation}
where $w = K_3(1/\theta)/K_2(1/\theta)$ is the enthalpy per particle in units of $mc^2$. For a thinning current sheet, ${\lambda_\infty} $ is the asymptotic limit of its half-width. For $\theta = {k_B T}/{m c^2} \gg 1$, $d_w = \sqrt{{\gamma_{th} m c^2}/{3 \pi n{e^2}}} \sim d_e$. 
Then, for a current sheet of half-length $\xi \gg {\lambda_\infty}$, and a compression ratio between inflow and outflow of order unity, the steady-state reconnection rate is 
\begin{equation} \label{} 
\beta_R  \sim \frac{d_w}{\xi } \ll 1 \, .
\end{equation}
Since current sheets generated by  outer-scale eddies (which, as we discuss below, are the ones that dominate particle injection) have half-length $\xi \sim \ell$ larger than $d_w$ by many orders of magnitude, the reconnection rate, as well as the injection efficiency, would be extremely low in this scenario.

However, plasmoids (which form copiously in our simulations) can break the reconnection layer into shorter elements, consequently leading to a regime of fast nonlinear reconnection \citep{daughton_06,daughton_07,Daughton_09,Bhattacharjee09,Huang2010,uzdensky_10}. This can happen if the plasmoids disrupt the current sheet within its characteristic lifetime, i.e., within one nonlinear eddy turnover time  \citep{Carbone_1990,Mallet_2017,Loureiro2017,Boldyrev_2017,Comisso_2018,Dong_2018,Walker2018PhRvE}. Fast magnetic reconnection essentially begins when plasmoids become nonlinear, namely when the current density fluctuations caused by the growing plasmoids are of the same order of the current density of the reconnection layer (see Fig. 4 in \citet{Huang_2017}). 
Therefore, understanding the plasmoid formation in the context of a forming current sheet is essential to understand the onset of fast magnetic reconnection and ensuing particle injection.

In order to evaluate the conditions for plasmoid formation and current sheet disruption, we need to analyze the growth rate of tearing (or ``reconnecting'') modes in such current sheet. 
The tearing mode dispersion relation for a relativistic pair plasma can be obtained from the relativistic pair-plasma fluid equations \citep[e.g.][]{Koide2009} by applying the standard tearing mode analysis \citep{furth_63,Coppi76,AraBasuCoppi78}. In this way, one can obtain, for arbitrary values of the tearing stability parameter $\Delta '$ \citep{furth_63}, the dispersion relation 
\begin{equation} \label{general_disprelation} 
{\gamma ^{1/2}}\tau _H^{1/2}{\left( {\frac{\lambda}{{{d_w}}}} \right)^{3/2}}\frac{{\Gamma \left[ {(\Upsilon  - 1)/4} \right]}}{{\Gamma \left[ {(\Upsilon  + 5)/4} \right]}} =  - \frac{8}{\pi }\Delta ' \, ,
\end{equation}
where 
\begin{equation} \label{general_disprelation_small}
\tau_H = \frac{1}{k_{\xi} v_{A\lambda }} \, , \qquad  \Upsilon = \gamma \, {\tau_H} \, \frac{\lambda}{d_w}  \, ,
\end{equation}
$\gamma$ is the growth rate, $k_{\xi}$ is the wavenumber in the $\xi$-direction, $\lambda$ is the current sheet half-width, $v_{A \lambda}$ is the Alfv{\'e}n speed based on the reconnecting magnetic field, and $\Gamma(z)$ indicates the gamma function. This dispersion relation matches the non-relativistic one \citep{Porcelli91} when $w \to 1$, i.e., when the plasma is cold. \footnote{For the purpose of this study we have not considered oblique tearing modes, which can be included in a more general dispersion relation.} 
Eq. (\ref{general_disprelation}) can be further simplified for short-wavelength modes (small-$\Delta '$) and long-wavelength modes (large-$\Delta '$), which is convenient in order to derive analytically the conditions for current sheet disruption. For $\Upsilon \ll 1$, the small-${\Delta}'$ regime, the growth rate and the inner tearing layer half-width (where the ideal MHD approximation breaks down due to the finite electron and positron inertia) of the instability are 
\begin{equation} \label{general_disprelation_large}
\gamma_s  = {\left[ {\frac{{\Gamma (\frac{1}{4})}}{{2\pi \Gamma (\frac{3}{4})}}} \right]^2} \frac{{\left( {\Delta '} \lambda \right)^2}}{\tau_H} {\left( {\frac{d_w}{\lambda}} \right)^3} \, , \qquad \delta_{\rm{in}} = d_w^2 {\Delta '} \, .
\end{equation} 
On the other hand, for $\Upsilon \rightarrow 1^-$, in the large-${\Delta}'$ regime, the growth rate and the inner tearing layer half-width are
\begin{equation} \label{}
\gamma_l  = \frac{1}{\tau_H} \left( {\frac{d_w}{\lambda}} \right) \, , \qquad \delta_{\rm{in}} = d_w \, .
\end{equation} 
Using these relations, together with the tearing stability index \footnote{Here we assume a Harris-type current sheet \citep{Harris62}, which is a reasonably good approximation of current sheets occurring in magnetized turbulence \citep[e.g.][]{Servidio2010} and coalescing magnetic islands \citep[e.g.][]{Huang_2017}).}
\begin{equation} \label{}
\Delta ' = \frac{2}{\lambda }\left( {\frac{1}{{{k_\xi }\lambda }} - {k_\xi }\lambda } \right) \, , 
\end{equation} 
it can be shown that the dominant tearing mode at current sheet disruption scales like the fastest growing mode (see \citet{comisso_16,Comisso_ApJ2017,Huang_2017}), so that the instability wavenumber at current sheet disruption turns out to be simply
\begin{equation} \label{}
{k_{\xi,d}} \sim  \frac{d_w}{\lambda_d^2} \, ,
\end{equation}
where the subscript ``$d$'' denotes current sheet disruption. This implies also that the growth rate and the inner tearing layer half-width at current sheet disruption are
\begin{equation} \label{}
{{\gamma}_d} \frac{\xi }{v_{A \lambda}}  \sim \frac{d_w^2}{\lambda_d^3} \xi   \, ,  \qquad      {{\delta}_{{\rm in},d}}  \sim d_w \, .
\end{equation}
From these expressions, one still needs to determine the current sheet half-width at disruption, $\lambda_d$, in order to know the wavenumber $k_{\xi,d}$ and the growth rate ${{\gamma}_d}$.
We calculate the width of the current sheet at disruption by using the principle of least time introduced in \citet{comisso_16,Comisso_ApJ2017}, substituting the resistive tearing mode dispersion relation with the collisionless dispersion relation discussed above. Then, for a rapid current sheet that forms on the Alfv{\'e}nic timescale, the mode that becomes nonlinear in the shortest time disrupts the current sheet when 
\begin{equation} \label{Eq_aspectratio}
\frac{\lambda_d}{\xi} \, {\left[ \ln \left( {\frac{1}{{2{{\hat \epsilon }^{1/2}}}}{{\left( {\frac{{{d_w}}}{\xi }} \right)}^{1 + \alpha /2}}{{\left( {\frac{\xi }{{{\lambda _d}}}} \right)}^{1/2 + \alpha }}} \right) \right]^{1/3}}  \simeq  c_\lambda {\left( {\frac{d_w}{\xi}} \right)^{2/3}} \, ,
\end{equation} 
where $ c_\lambda$ is an $O(1)$ constant, ${\hat \epsilon} = \epsilon/(\delta {B_\lambda} \xi)$ is a normalized amplitude of the noise that seeds the instability (evaluated at the disruption scale), $\delta {B_\lambda}$ is the characteristic magnetic field fluctuation at scale $\lambda$, and $\alpha$ is an index that depends on the spectrum of the noise, which is related to the turbulence spectrum as $P_B({k_\xi }) \propto {{k_\xi}^{ 1- 2\alpha }}$ \citep{Comisso_2018}. Eq. (\ref{Eq_aspectratio}) can be solved exactly in terms of the Lambert $W$ function, but here we prefer to consider an asymptotic solution that yields more transparent results. Therefore, we solve Eq. (\ref{Eq_aspectratio}) by iteration obtaining, at the first order, the solution
\begin{equation} \label{Eq_aspectratio_esplicit}
\lambda_d  \sim  d_w^{2/3} \xi^{1/3} {\left[ {\ln \left( {\frac{1}{{2{{\hat \epsilon }^{1/2}}}}{\left( {\frac{{{d_w}}}{\xi }} \right)^{\frac{{4 - \alpha }}{6}}}} \right)} \right]^{ - 1/3}} \, ,
\end{equation}
which gives us the critical current sheet width that determines the layer disruption and the onset of fast reconnection.
Finally, the growth rate of the instability when the current sheet reaches this ratio is
\begin{equation} \label{gamma_exp_1fluid} 
\gamma_d  \sim  \frac{v_{A \lambda}}{\xi}   {\ln \left( {\frac{1}{{2{{\hat \epsilon }^{1/2}}}}{\left( {\frac{{{d_w}}}{\xi }} \right)^{\frac{{4 - \alpha }}{6}}}} \right)}   \, ,
\end{equation}
while the wavenumber of the dominant mode becomes
\begin{equation}\label{k_exp_1fluid} 
{k_{\xi,d}}  \sim d_w^{-1/3} \xi^{-2/3}  {\left[ {\ln \left( {\frac{1}{{2{{\hat \epsilon }^{1/2}}}}{\left( {\frac{{{d_w}}}{\xi }} \right)^{\frac{{4 - \alpha }}{6}}}} \right)} \right]^{2/3}} \, .
\end{equation}

We obtain from Eq. (\ref{Eq_aspectratio_esplicit}) that ${\lambda _d} \gg {\lambda_\infty} \simeq d_w$ for outer-scale current sheets with $\xi  \sim \ell \gg {d_w}$, as it is expected under typical astrophysical conditions. Therefore, an outer-scale current sheet disrupts in a chain of plasmoids before reaching the kinetic scale ${d_w}$, while inter-plasmoid layers, being shorter, can reach the thickness $d_w$. Eq. (\ref{Eq_aspectratio_esplicit}) tells us also that current sheets disrupt at a larger thickness for larger noise levels. However, the dependence is only logarithmic. 
From the other two relations, Eq. (\ref{gamma_exp_1fluid}) and Eq. (\ref{k_exp_1fluid}), we have that the growth rate of the plasmoid instability is $\gamma_d  {\xi}/{v_{A \lambda}} \gg 1$ at current sheet disruption, as it is required for the instability to amplify the perturbation to a significant level within the lifetime of the current sheet \citep{Comisso_2018}. Also, the number of plasmoids fragmenting the outer-scale current sheets, which is $\propto {k_{\xi,d}} \ell$, increases as $\ell/d_w$ increases and the noise of the system decreases. As an example, we show in Figs. \ref{fig_PartInj3} and \ref{fig_PartInj4} that a larger number of plasmoids forms when the  domain is increased by a factor $4$ with respect to the reference 2D simulation. In this simulation, as we argue below in this section, efficient plasmoid formation keeps the reconnection speed and the injection efficiency high when increasing system size. As a result, the fraction of nonthermal particles remains about the same when moving from the reference box size $L/d_{e0}=1640$ up to $L/d_{e0}=6560$ (see Fig. 2(b) in \citet{ComissoSironi18}). 

When the reconnection layer becomes dominated by the presence of plasmoids, soon after the condition $\lambda \sim \lambda_d$ is met, the complexity of the dynamics gives rise to a strongly time-dependent process. Nevertheless, in a statistical steady-state, we may expect that the reconnection layer containing  the main $X$-point, which is the one that determines the global reconnection rate, has a bounded aspect ratio $\xi_X/\lambda_X$. If $\xi_X/\lambda_X \sim 1$, the reconnection process would choke itself off, since this  would imply $\beta_R \sim 0$ \citep{ComiBhatta2016}. This means that $\xi_X/\lambda_X \gg 1$  in a steady reconnection process. On the other hand, the reconnection layer at the main $X$-point cannot be longer than the marginally stable sheet. Indeed, the fractal-like process of current sheet disruption due to the plasmoid instability terminates when the length of the innermost local current layer of the chain is shorter than the critical length $\xi_c$ \citep{Huang2010,uzdensky_10,ComissoPoP2015,ComissoGrasso2016}. Therefore, $\xi_X \lesssim \xi_c$ is also expected. At present there are no analytical estimates for the aspect ratio $\xi_c/\lambda_X$, which might also depend on the noise level \citep[e.g.][]{LeiNi2010,Huang_2017,Shi2018}. However, numerical simulations have found $\xi_c/\lambda_X \sim 50$ in the collisionless regime \citep[e.g.][]{daughton_06,daughton_07,Ji2011}.
As a consequence, for a compression ratio between inflow and outflow of order unity, the reconnection rate is bounded from above and below as 
\begin{equation} \label{} 
1/50 \lesssim \beta_R  \ll 1 \, ,
\end{equation}
which classify it as a fast reconnection rate. More precisely, numerical simulations consistently indicates that $\beta_R$ is an $O(0.1)$ quantity (for relativistic pair plasmas, see, e.g, \citet{zenitani_hesse_09,bessho_12,Cerutti2012ApJL,guo_14,kagan_15,liu_15,sironi_16,werner_17,Liu2017}).

The aforementioned properties of reconnecting current sheets are important in regulating the particle injection efficiency. Here we show that the fraction of particles processed by  reconnecting current sheets is independent of the system size and is quite large (despite the small filling fraction of current sheets) as long as the reconnection rate is high. To this purpose, let us consider a generic current sheet of characteristic length $2\xi$ and thickness $2\lambda$, whose lifetime is approximately given by the local eddy turnover time $\tau_{\rm{nl}} \sim {\tau_{A\xi}} = \xi /{v_{A\lambda }}$, assuming critical balance \citep{gs95,Boldyrev2006}. If fast reconnection occurs for a time close to the eddy turnover time (see, e.g, Fig. \ref{fig_PartInj4}), a single reconnecting current sheet can ``process'' the upstream plasma up to a distance 
\begin{equation} \label{}
{\lambda_{{R},j}} = {\beta _{{R},j}} \, c \, {\tau_{{\rm{nl}},j}} \sim {\beta _{{R},j}} \, c \frac{\xi_j}{ {v_{{A\lambda}, j}}} \, ,
\end{equation}
where $j$ labels the $j$-th current sheet among the population of current sheets present at a given time, and the subscript ``$R$'' stands for reconnection. Since the surface processed by the entire population of reconnecting current sheets is in good approximation the one processed by the largest-scale ones, whose length scale corresponds to the turbulence integral length $\ell$ \citep[e.g.][]{Servidio2009}, we have that magnetic reconnection can process a plasma surface  
\begin{equation} \label{2Dprocessed}
{\mathcal{A}}_R =  \sum \nolimits_j {{\lambda_{{R},j}} \, {\xi_j} }   \sim \beta_R  L^2 
\end{equation}
in one large-eddy turnover time. Here,  we have used $n_{\rm{cs}} \sim (L/\ell)^2$ as an estimate for the number of outer-scale current sheets, and $\beta_R$ is the average reconnection rate.
Furthermore, if we consider that current sheets in 3D are sheet-like structures with $2 \lambda \ll 2\xi \lesssim 2 l_{\parallel}$, with $2 l_{\parallel}$ indicating the direction along the magnetic field, we can obtain that, in one large-eddy turnover time, the reconnecting current sheets process a plasma volume 
\begin{equation} \label{3Dprocessed}
{\mathcal{V}}_R =  \sum \nolimits_j {{\lambda_{{R},j}} \, {\xi_j} \, l_{{\parallel},j}}   \sim  \beta_R  L^3  \, .
\end{equation}
Therefore, according to Eqs. (\ref{2Dprocessed})/(\ref{3Dprocessed}), the plasma surface/volume processed by the reconnecting current sheets is a fixed fraction of the domain if $\beta_R$ is independent of the system size, as discussed above. Moreover, since $\beta_R$ is an $O(0.1)$ quantity, magnetic reconnection can process large volumes of magnetic energy in few outer-scale eddy turnover times.

In the next sections, we will address how particles are energized both in the injection phase and in the subsequent stochastic acceleration phase, and we will analyze the signatures of the acceleration process on the particle distribution.

\section{Mechanisms of particle energization}  \label{SecEnergiz}

\begin{figure*}
 \centering 
  \includegraphics[width=8.75cm]{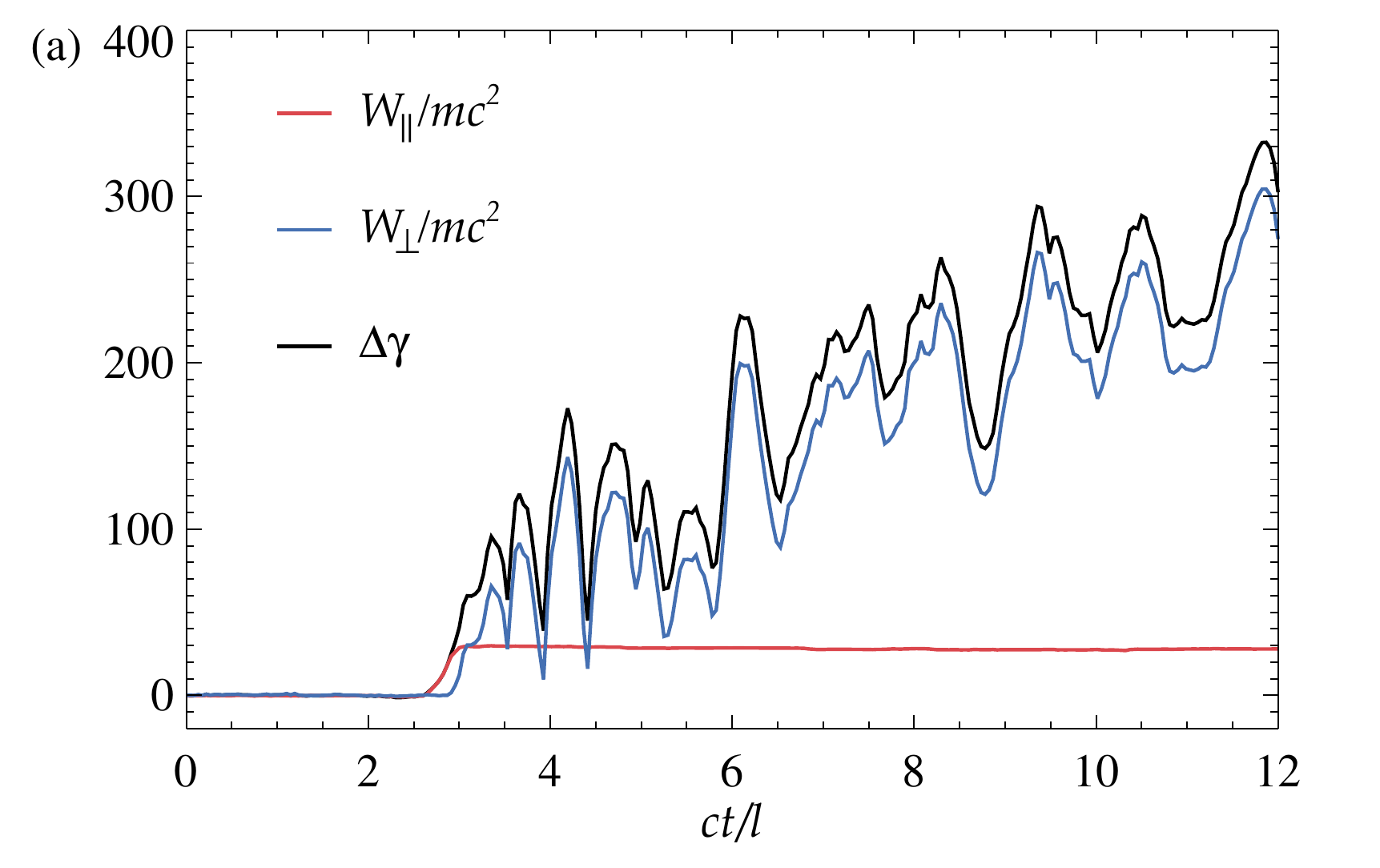}
  \includegraphics[width=8.75cm]{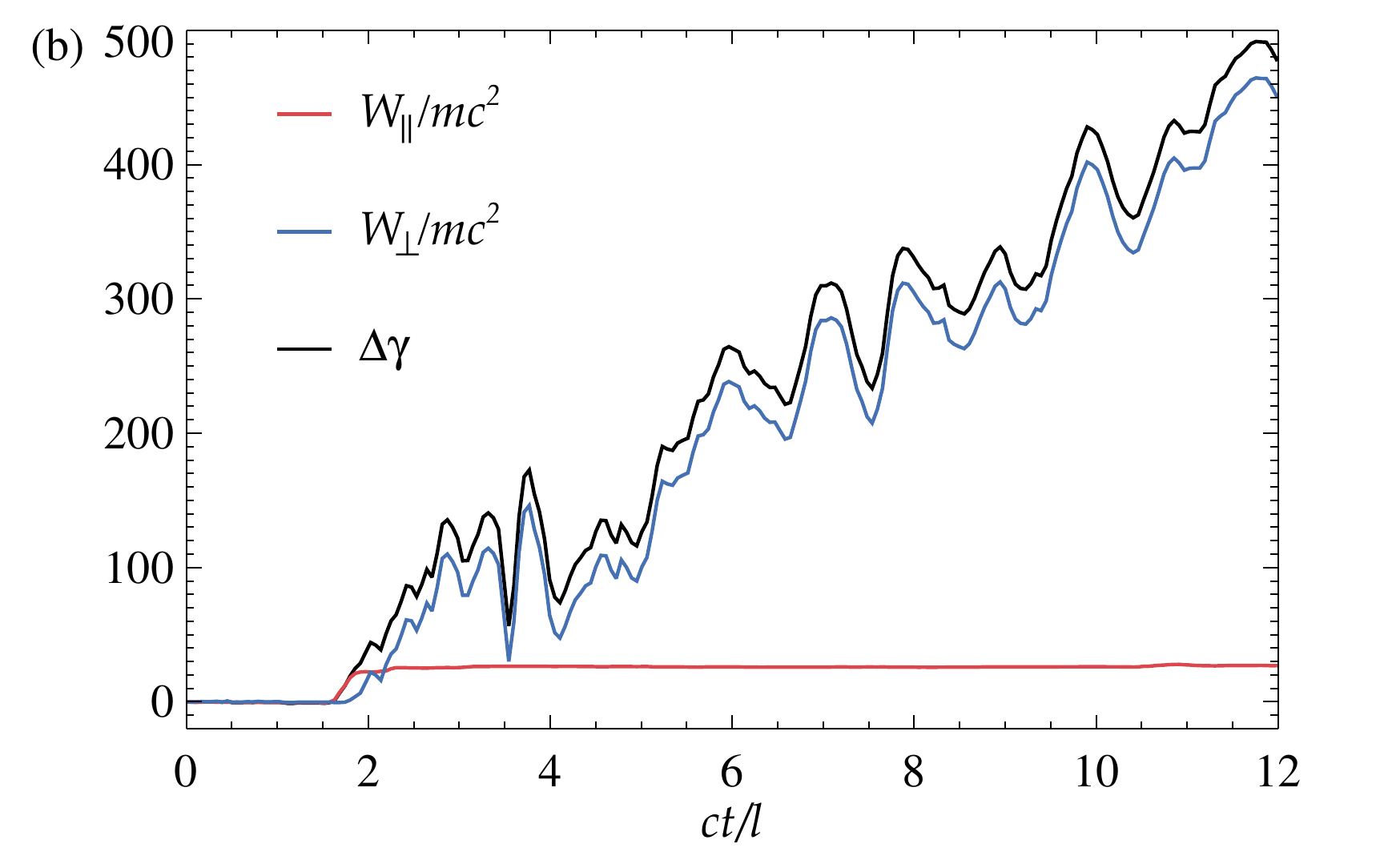}
  \hfill
  \includegraphics[width=8.75cm]{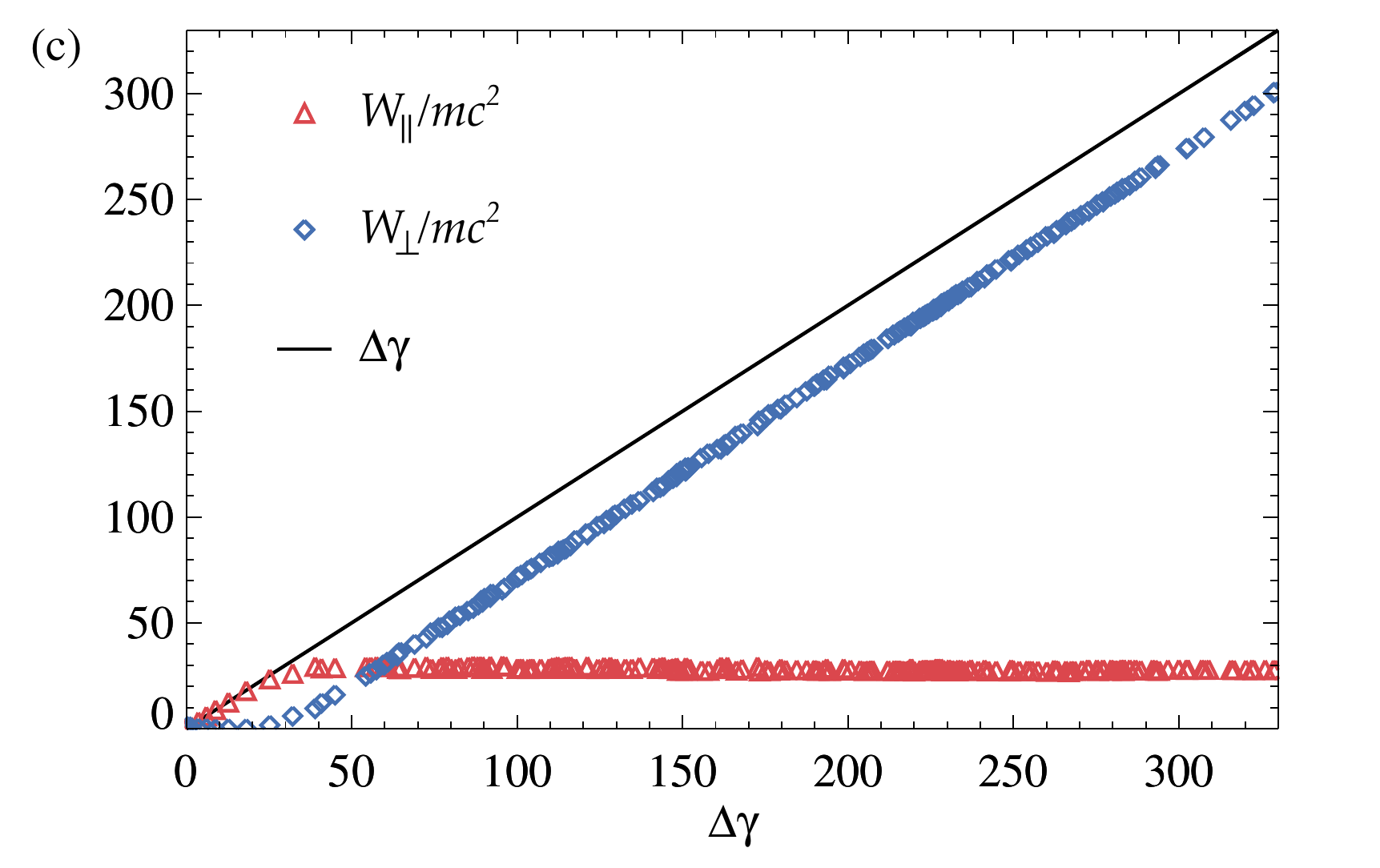}
  \includegraphics[width=8.75cm]{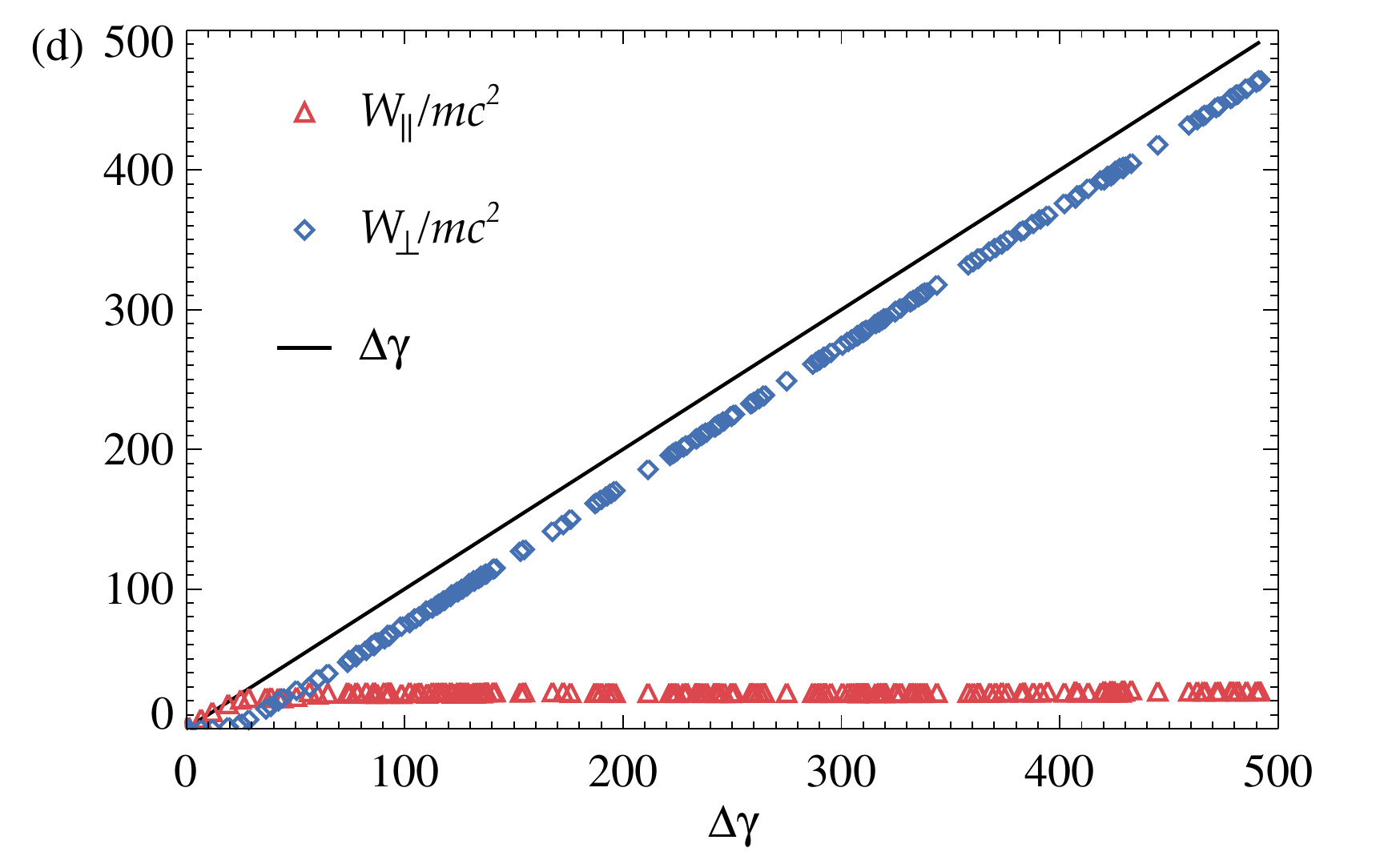}
  \hfill
  \includegraphics[width=8.75cm]{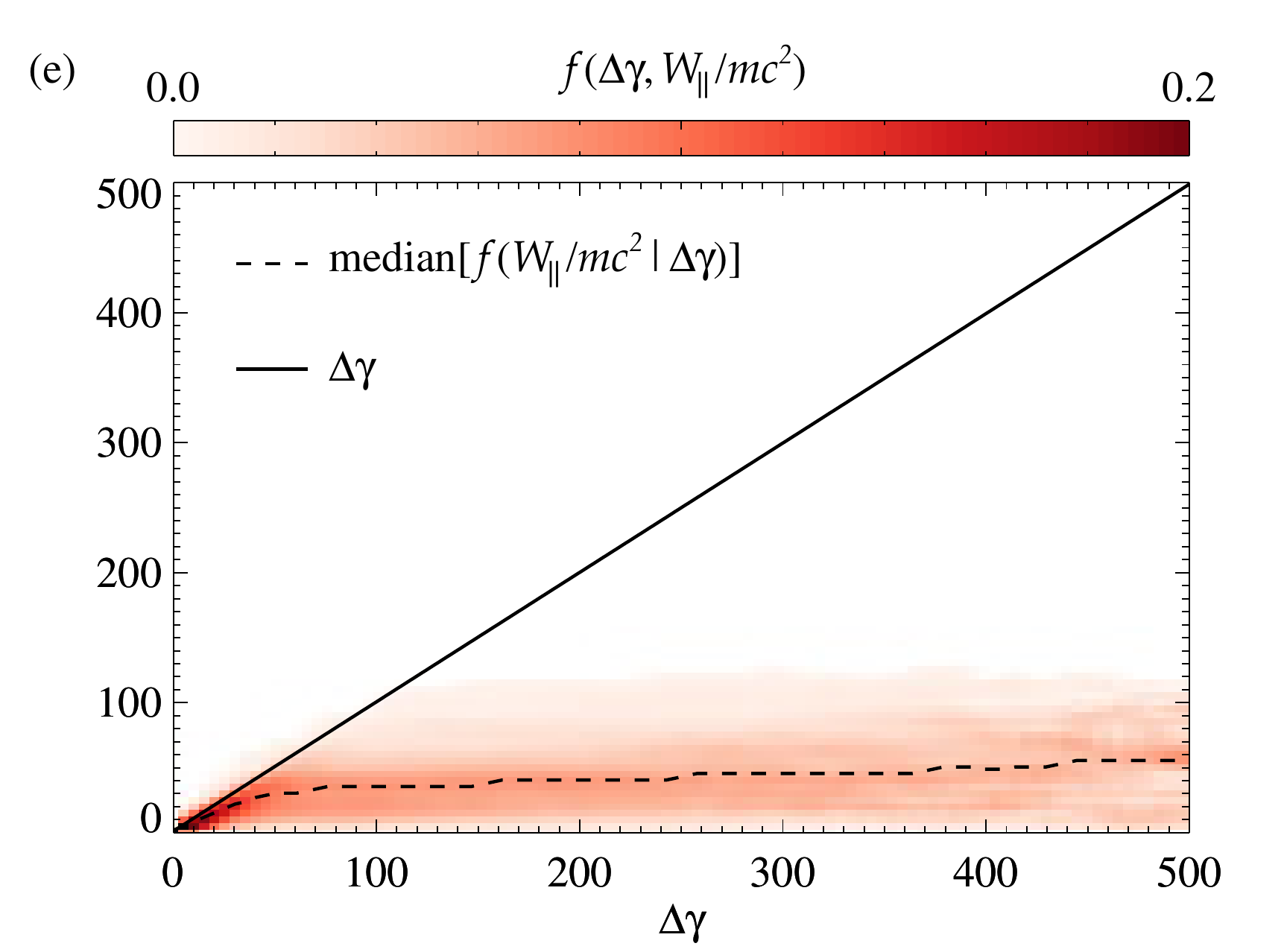}
  \includegraphics[width=8.75cm]{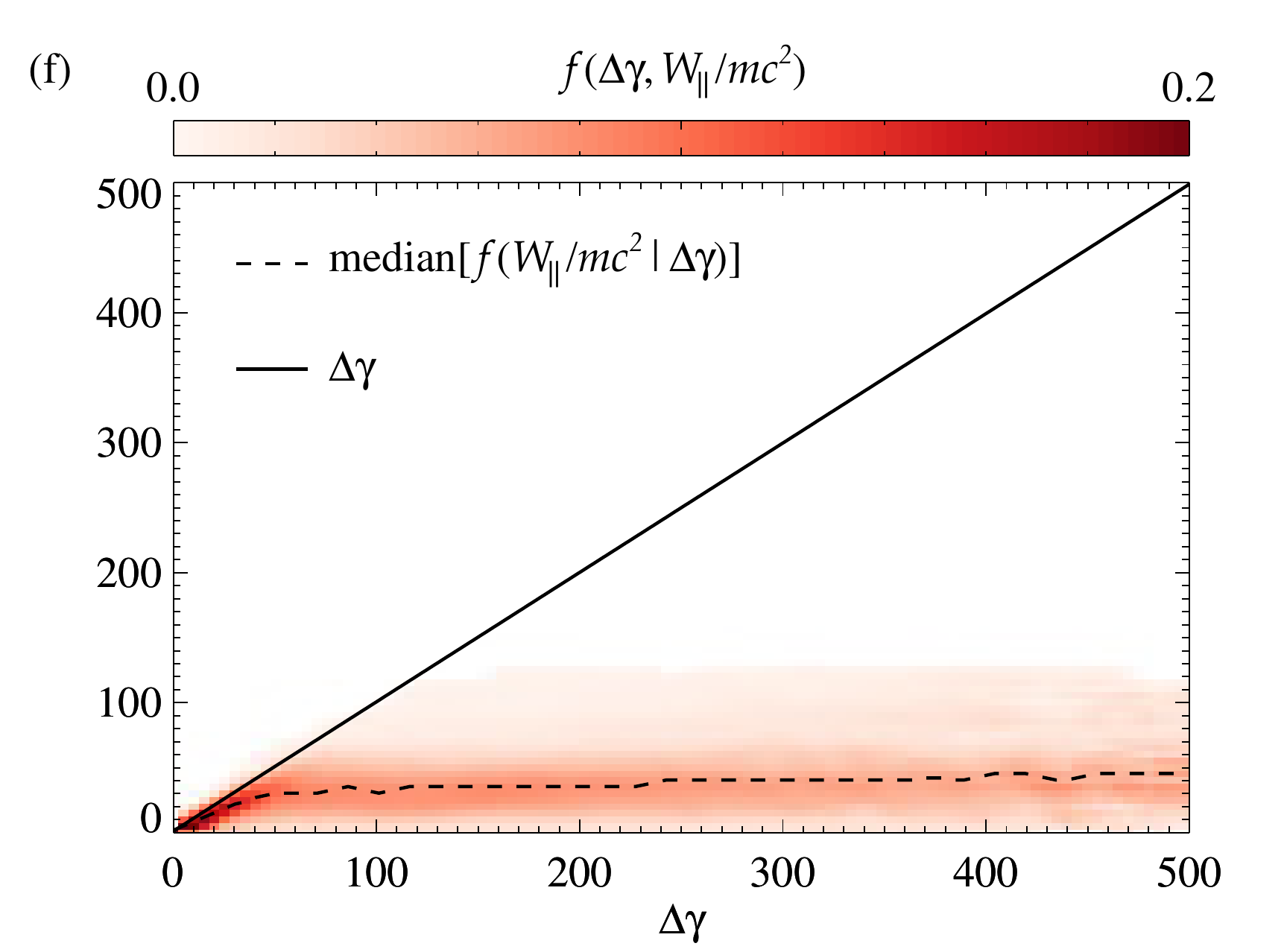}
\caption{Relative contributions of ${\bm{E}}_\parallel = ({\bm{E}} \cdot {\bm{B}}) {\bm{B}}/B^2$ and ${\bm{E}}_\bot = {\bm{E}} - {\bm{E}}_\parallel$ to the particle energization in 2D (left) and 3D (right) simulations with $\sigma_0=10$ and $\delta B_{{\rm{rms}}0}/ B_0=1$. The 2D simulation has domain size $L/d_{e0}=1640$ (with $l=L/8$), while the 3D simulation has domain size $L/d_{e0}=820$ (with $l=L/4$). Top row: for a typical high-energy particle, time evolution of the normalized particle energy gain, $\Delta \gamma$ (black solid line), normalized work done by the parallel electric field, ${W_\parallel}/mc^2$ (red solid line), and normalized work done by the perpendicular electric field, ${W_\bot} /mc^2$ (blue solid line). Middle row: scatter plot of ${W_\parallel}/mc^2$ versus $\Delta \gamma$ (red triangles) and ${W_\bot} /mc^2$ versus $\Delta \gamma$ (blue diamonds), for the same particle displayed in the top frame. The solid black line indicates the expected sum ${W_\parallel}/mc^2 + {W_\bot}/mc^2=\Delta \gamma$. Bottom row: distribution of particles with respect to $\Delta \gamma$ and ${W_\parallel}/mc^2$, for particles ending up with $\gamma \geq 18 \sigma_0$ at $ct/l=12$. The median of the conditional PDF at given $\Delta \gamma$,  $f \left( {{W_\parallel }/m{c^2} | \Delta \gamma} \right)$, is shown with a dashed black line. Again, the solid black line indicates the expected sum ${W_\parallel}/mc^2 + {W_\bot}/mc^2 = \Delta \gamma$.}
\label{fig_parallelvsperpelectricfield}
\end{figure*}

In order to distinguish the relative roles of  different energization mechanisms, it is convenient to compute the work done by the parallel electric field, ${W_\parallel}(t) = q \int_0^t {{{\bm{E}}_\parallel}(t') \cdot {\bm{v}}(t') \, dt'}$, as well as the work done by the perpendicular electric field, ${W_\bot}(t) = q \int_0^t {{{\bm{E}}_\bot}(t') \cdot {\bm{v}}(t') \, dt'}$, for a statistically significant sample of particles (here, as usual, $q$ is the electric charge, ${\bm{E}}$ is the electric field, and ${\bm{v}}$ is the particle velocity). To this aim, we tracked a sample of $\sim10^7$ particles randomly selected from each of our PIC simulations.\footnote{${W_\parallel}(t)$ and ${W_\bot}(t)$ are computed on the fly in order to achieve high accuracy, regardless of the time sampling of particle outputs.}  Note that in this section, parallel ($\parallel$) and perpendicular ($\bot$) components are defined with respect to the \emph{local} magnetic field, i.e. ${\bm{E}}_\parallel = ({\bm{E}} \cdot {\bm{B}}) {\bm{B}}/B^2$ and ${\bm{E}}_\bot = {\bm{E}} - {\bm{E}}_\parallel$. The main results of our analysis, for the reference 2D and 3D simulations (see Table \ref{tab:param}), are presented in Fig. \ref{fig_parallelvsperpelectricfield} (left column for 2D and right column for 3D). 
We first discuss the energization process of representative particles that end up in the high-energy tail, and then we present a statistical analysis that allows us to quantify the contributions of parallel and perpendicular electric fields for the overall acceleration of nonthermal particles.

%%%%%%%%%%%%%%%%%%%%%%%%%%%%%%  
%%%%%%%%%%%%%%%%%%%%%%%%%%%%%%

Figs. \ref{fig_parallelvsperpelectricfield}(a) and \ref{fig_parallelvsperpelectricfield}(b) show the particle energy gain normalized to rest mass energy, $\Delta \gamma (t) = \gamma (t) - \gamma (0)$, as well as the relative contributions ${W_\parallel}(t)/m c^2$ and ${W_\bot}(t)/m c^2$, for representative high-energy particles in 2D and 3D turbulence. The total work done by the electric field is not plotted here, since $q \int_0^t {{\bm{E}}(t') \cdot {\bm{v}}(t') \, dt'} = m c^2 \Delta \gamma (t)$ is satisfied to high accuracy and is essentially indistinguishable from the black solid line representing $\Delta \gamma (t)$. Both figures indicate that the work done by ${\bm{E}}_\parallel$  is responsible for the initial energy gain, while the work done by ${\bm{E}}_\bot$ takes over at relatively low energies and propels the particle to the highest energies. Alternative plots that provide similar information, but can be more easily generalized to analyze a large population of particles (as we do below), are shown in Figs. \ref{fig_parallelvsperpelectricfield}(c) and \ref{fig_parallelvsperpelectricfield}(d), for 2D and 3D, respectively. In this case, the relative contributions ${W_\parallel}/m c^2$ and ${W_\bot}/m c^2$ are plotted as a function of $\Delta \gamma$, and the black solid line indicates the expected sum of the two terms. The plots  show that the low $\Delta \gamma$-range is dominated by  $W_\parallel$, while $W_\bot \gg W_\parallel$ when  particles reach high energies.

Figs. \ref{fig_parallelvsperpelectricfield}(c) and \ref{fig_parallelvsperpelectricfield}(d) are generalized in Figs. \ref{fig_parallelvsperpelectricfield}(e) and \ref{fig_parallelvsperpelectricfield}(f), respectively, to account for a statistical assessment of the energization of a sample of particles. We consider all tracked particles that end up well into the nonthermal tail at late times, more precisely all tracked particles for which $\gamma \geq 18 \sigma_0$ at $ct/l=12$. The figures show the distribution $f\left( {\Delta \gamma , {W_\parallel }/m{c^2}} \right)$ of particles with respect to $\Delta \gamma$ and  $W_\parallel/m{c^2}$. We normalize $f\left( {\Delta \gamma , {W_\parallel }/m{c^2}} \right)$ such that
\begin{equation} \label{} 
\int\limits_{ - \infty }^\infty  {f\left( {\Delta \gamma ,\frac{W_\parallel}{m c^2}} \right)d(\Delta \gamma )}  = 1  \, .
\end{equation}
 The distribution $f\left( {\Delta \gamma , {W_\bot}/m{c^2}} \right)$  is not plotted here since it conveys the same message.
We can see that the peak of the distribution for a given $\Delta \gamma$ is around ${W_\parallel }/m{c^2} \sim 40$, for all $\Delta \gamma > 50$. This occurs both in 2D and 3D simulations. We also calculated the median of the histogram as a function of $\Delta \gamma$, which is shown as a black dashed line in  Figs. \ref{fig_parallelvsperpelectricfield}(e) and \ref{fig_parallelvsperpelectricfield}(f). The median also approaches a constant value ${W_\parallel }/m{c^2} \sim 40$ at $\Delta \gamma>50$. \footnote{For low values of $\Delta \gamma$, the mode and the median of $f \left( {{W_\parallel }/m{c^2} | \Delta \gamma} \right)$ are independent of the final particle energy only if the selected threshold satisfies $\gamma \gg (\sigma_0/2) \gamma_{th0}$ at late times, i.e., for particles that end up well into the nonthermal tail (see also Sec. \ref{SecAnisotropy}).} 
This confirms for a statistically-significant sample of particles the same conclusions presented above: high-energy particles are first energized via ${\bm{v}} \cdot {\bm{E}}_\parallel$, which brings them up to $\Delta \gamma \sim W_\parallel /m{c^2}  \sim 40$, and then further energization is provided by perpendicular electric fields, with $W_\bot \gg W_\parallel$ for the highest energy particles.

In summary, we find both for individual particles and for a large sample of particles, that the initial stages of acceleration are controlled by parallel electric fields. This is consistent with the fact that strong parallel electric fields are expected at active reconnection layers, where we have indeed shown that particle injection (i.e., the first stage of acceleration) occurs.

%%%%%%%%%%%%%%%%%%%%%%%%%%%%%%  
\begin{figure}
\begin{center}
\hspace*{-0.085cm}\includegraphics[width=8.75cm]{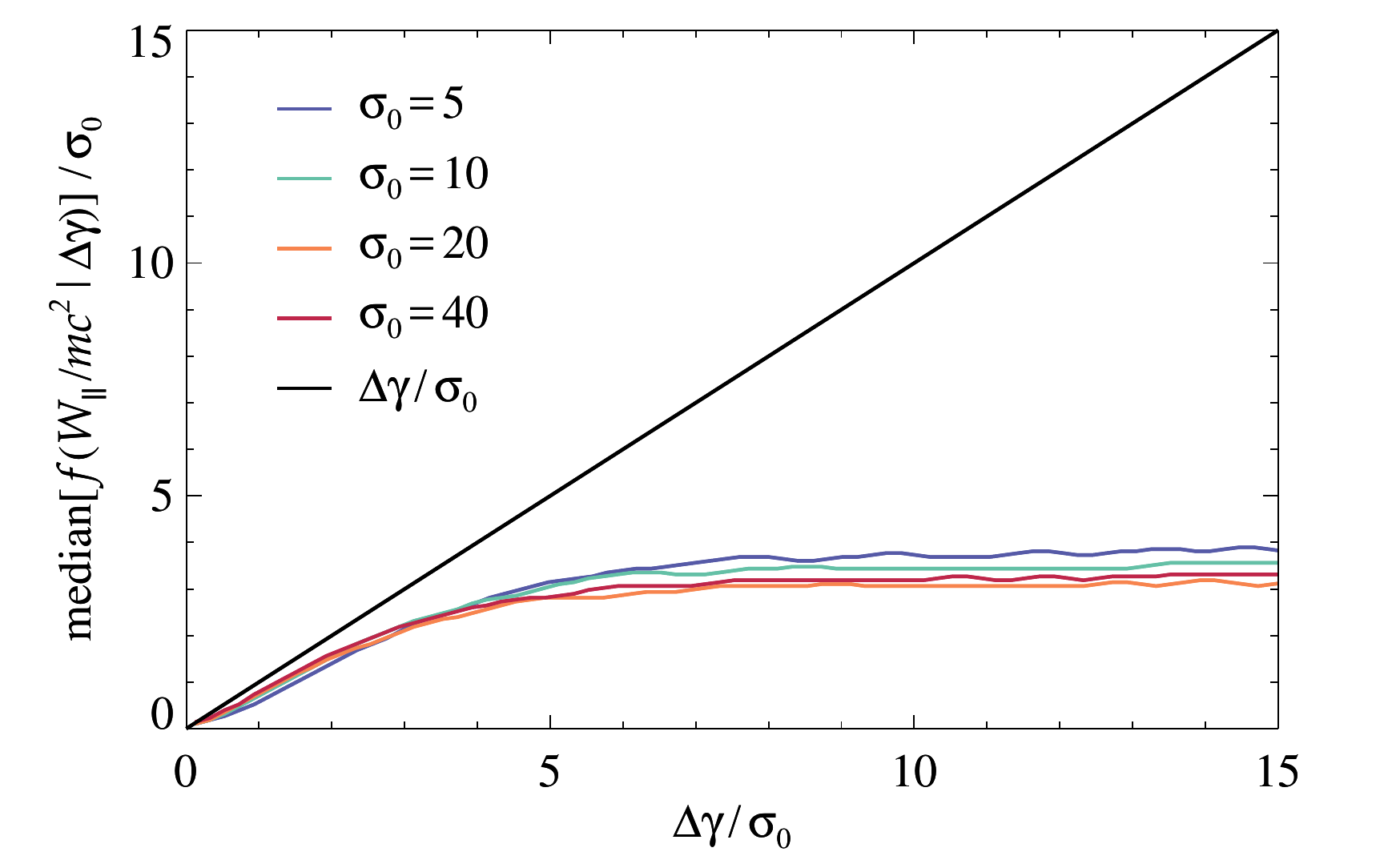}
\end{center}
\caption{Median of $f \left( {{W_\parallel }/m{c^2} | \Delta \gamma} \right)$, divided by $\sigma_0$, for high-energy particles from different 3D simulations having $\sigma_0=5$ (blue), $\sigma_0=10$ (green), $\sigma_0=20$ (orange), and $\sigma_0=40$ (red).  We considered all tracked particles with $\gamma \geq 18 \sigma_0$ at $ct/l=12$. All the simulations have $\delta B_{{\rm{rms}}0}/ B_0=1$ and $L/d_{e0}=820$. The solid black line indicates the expected sum ${W_\parallel}/mc^2 + {W_\bot}/mc^2 = \Delta \gamma$.}
\label{fig_energy_jump_magnetizations}
\end{figure}  
%%%%%%%%%%%%%%%%%%%%%%%%%%%%%%  

The initial energy gain due to  ${\bm{v}} \cdot {\bm{E}}_\parallel$ is dependent on magnetization. 
It can be seen from Fig. \ref{fig_energy_jump_magnetizations} that the typical energy gain provided by ${\bm{v}} \cdot {\bm{E}}_\parallel$ increases with $\sigma_0$. From our simulations we find that the typical increase in Lorentz factor during the injection process (which is governed by parallel fields at reconnection layers) is 
\begin{equation} \label{}
\Delta \gamma_{\rm{inj}}  \sim W_\parallel/m{c^2} \sim  \kappa \sigma_0 \gamma_{th0} \, , \quad \sigma_0 \gg 1
\end{equation}
where $\kappa$ is a numerical factor of order unity ($\kappa \sim 2$ from Fig. \ref{fig_energy_jump_magnetizations}).
In general, the time-dependent magnetization $\sigma  = \delta B_{{\rm{rms}}}^2/4\pi n_0 w m c^2 $ decreases with time in decaying turbulence, implying that the time-dependent $\Delta \gamma_{\rm{inj}} = \kappa \sigma \gamma_{th}$ also decreases with time ($\gamma_{th}$ is the instantaneous mean Lorentz factor).

The length $l_\parallel$ along ${\bm{B}}$ (which in reconnection layers is dominated by the mean field ${\bm{B}}_0$)  required to attain the energy gain $\Delta \gamma_{\rm{inj}}$ can be obtained from the reconnection electric field $E_R$ by assuming particles moving along ${\bm{E}}_\parallel$ at the speed $|{\bm{v}}| \sim c$. If $E_\parallel\sim E_R \approx {\rm{const}}$ during the acceleration time, then 
\begin{equation} \label{}
\Delta \gamma_{\rm{inj}} =   \beta_R \delta B_{{\rm{rms}}} \frac{e \, l_\parallel}{m c^2}  \, .
\end{equation}
Here we have used the typical reconnection electric field $E_R = \beta_R \delta B_{{\rm{rms}}} v_A/c \sim \beta_R \delta B_{{\rm{rms}}}$. Therefore, the length scale $l_{\rm{inj}}$ required to attain the increase $\Delta \gamma_{\rm{inj}}$ can be expressed as 
\begin{equation} \label{}
l_{\rm{inj}}  = \frac{\kappa}{\beta_R} \sqrt {\frac{\sigma}{w}} \, \gamma_{th} d_e \, ,
\end{equation}
where the different physical quantities have to be evaluated at the injection time. This expression indicates that a sufficient length for particle injection is always guaranteed for a large enough system, i.e., $l \gg  l_{\rm{inj}}$. Similarly, as most of injection happens at outer-scale current sheets, the time $\tau_{\rm{inj}}$ required for accelerating particles up to this energy is always granted if  $\tau_{\rm{nl}} = l/\delta V_{\rm{rms}} \gg \tau_{\rm{inj}} =  l_{\rm{inj}}/c$, where $\tau_{\rm{nl}}$ is the outer-scale nonlinear time and $\delta V_{\rm{rms}} = \langle {\delta {V^2}} \rangle^{1/2}$ is the space-averaged root-mean-square value of the velocity field fluctuations. The two requirements coincide for $\delta {V_{{\rm{rms}}}} \to c$.

%%%%%%%%%%%%%%%%%%%%%%%%%%%%%%  
\begin{figure}
\begin{center}
\hspace*{-0.185cm}\includegraphics[width=8.75cm]{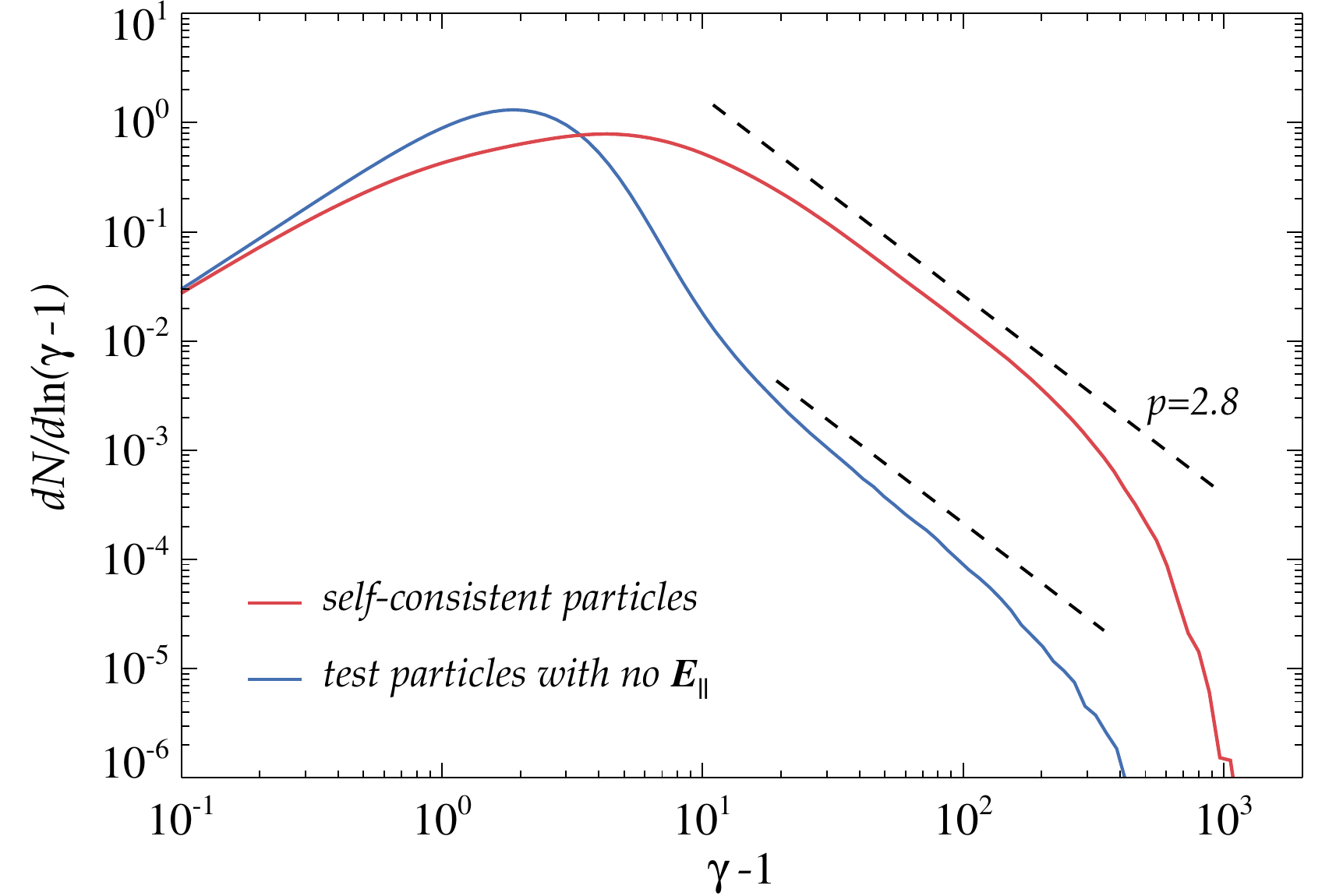}
\end{center}
\caption{Particle spectra $dN/d\ln(\gamma-1)$ at $ct/l=12$ for normal particles (red solid line) and test-particles that are evolved with ${\bm{E}} \rightarrow {\bm{E}} - {\bm{E}}_\parallel$ (blue solid line) from a 2D simulation with $\sigma_0=10$, $\delta B_{{\rm{rms}}0}/ B_0=1$, and $L/d_{e0}=1640$. The two spectra display similar power-law index but different power-law normalization. The high-energy tail with $\gamma \geq 12$ contains $17\%$ of the normal particles, while only $0.2 \%$ of the test-particles are contained in the tail with $\gamma \geq 12$. }
\label{fig_test_particles}
\end{figure}  
%%%%%%%%%%%%%%%%%%%%%%%%%%%%%%  

Even though ${W_\bot} \gg {W_\parallel}$ for high-energy particles, the initial ${\bm{v}} \cdot {\bm{E}}_\parallel$ energization process is important to promote the particles to energies large enough such that they can experience the subsequent ${\bm{v}} \cdot {\bm{E}}_\bot$ acceleration. Hence,  energization  by ${\bm{v}} \cdot {\bm{E}}_\parallel$ controls the number of particles that have the possibility to reach nonthermal energies. This point, which was already discussed in Section \ref{SecInjection}, can be probed in a direct way by comparing the self-consistent PIC particles with a population of test-particles for which we artificially exclude acceleration by ${\bm{E}}_\parallel$, assuming that the electric field they feel is ${\bm{E}} \rightarrow {\bm{E}} - {\bm{E}}_\parallel$.  
To this aim, we performed a 2D PIC simulation where we added a population of $\sim 5 \times 10^9$ such test-particles. The resulting particle spectra at late time ($ct/l=12$) are shown in Fig. \ref{fig_test_particles}. The particle spectrum of normal particles has a much larger fraction of particles contained in the high-energy tail ($17\%$ vs $0.2\%$). \footnote{In both cases we consider a nonthermal tail starting at $\gamma \geq 12$, since the power-law range starts at $\gamma \sim 12$ for both particle spectra. Note that this value is quite larger than the termal peak of the test-particle population, which is consistent with the low normalization $N_0$ of its nonthermal power-law tail.} Equivalently, the normalization of the power-law tail in the test-particle spectrum is much lower than for self-consistent particles. On the other hand, the index $p = - d\log N/d\log (\gamma -1)$ of the power-law tail is similar (see dashed black lines), indicating that the ${\bm{v}} \cdot {\bm{E}}_\bot$ energization is the crucial process responsible for setting the power-law slope. Also, the cutoff energy is about the same for the two population of particles, indicating that it is not controlled by parallel electric fields.

In the next section we will see that the two different energization processes, which dominate in different energy ranges (${\bm{v}} \cdot {\bm{E}}_\parallel$ for $\Delta \gamma_{\rm{inj}}  \lesssim  \kappa \sigma \gamma_{th}$ and ${\bm{v}} \cdot {\bm{E}}_\bot$ for $\Delta \gamma_{\rm{inj}}  \gtrsim  \kappa \sigma \gamma_{th}$), also affect the anisotropy of the particle distribution.

\section{Anisotropy and particle mixing} \label{SecAnisotropy}

Anisotropic features of the particle distribution can have a significant impact on the observed synchrotron emission \citep[e.g.][]{Tavecchio2019}. Here we show that even if the initial velocity distribution is isotropic, the particle energization process drives a significant energy-dependent anisotropy, as the pitch angle scattering rate is not sufficient to keep the particle distribution close to isotropy. 
In order to characterize the anisotropy of the particle distribution, we first examine the pitch-angle distribution of the particles, namely the statistics of the pitch-angle cosine $\cos \alpha  = {\bm{v}} \cdot {\bm{B}} /({\left| {\bm{v}} \right|\left| {\bm{B}} \right|})$. Then, we analyze the anisotropy of the four-velocity distribution of the particles. We perform these analyses on a statistically significant sample of $\sim 10^7$ particles, both in 2D and 3D. These particles were randomly selected and tracked over time for each of the  simulations. Finally, we also look at the spatial mixing of particles.

\subsection{Pitch-angle distribution}

%%%%%%%%%%%%%%%%%%%%%%%%%%%%%%  
\begin{figure*}
 \centering 
  \includegraphics[width=8.75cm]{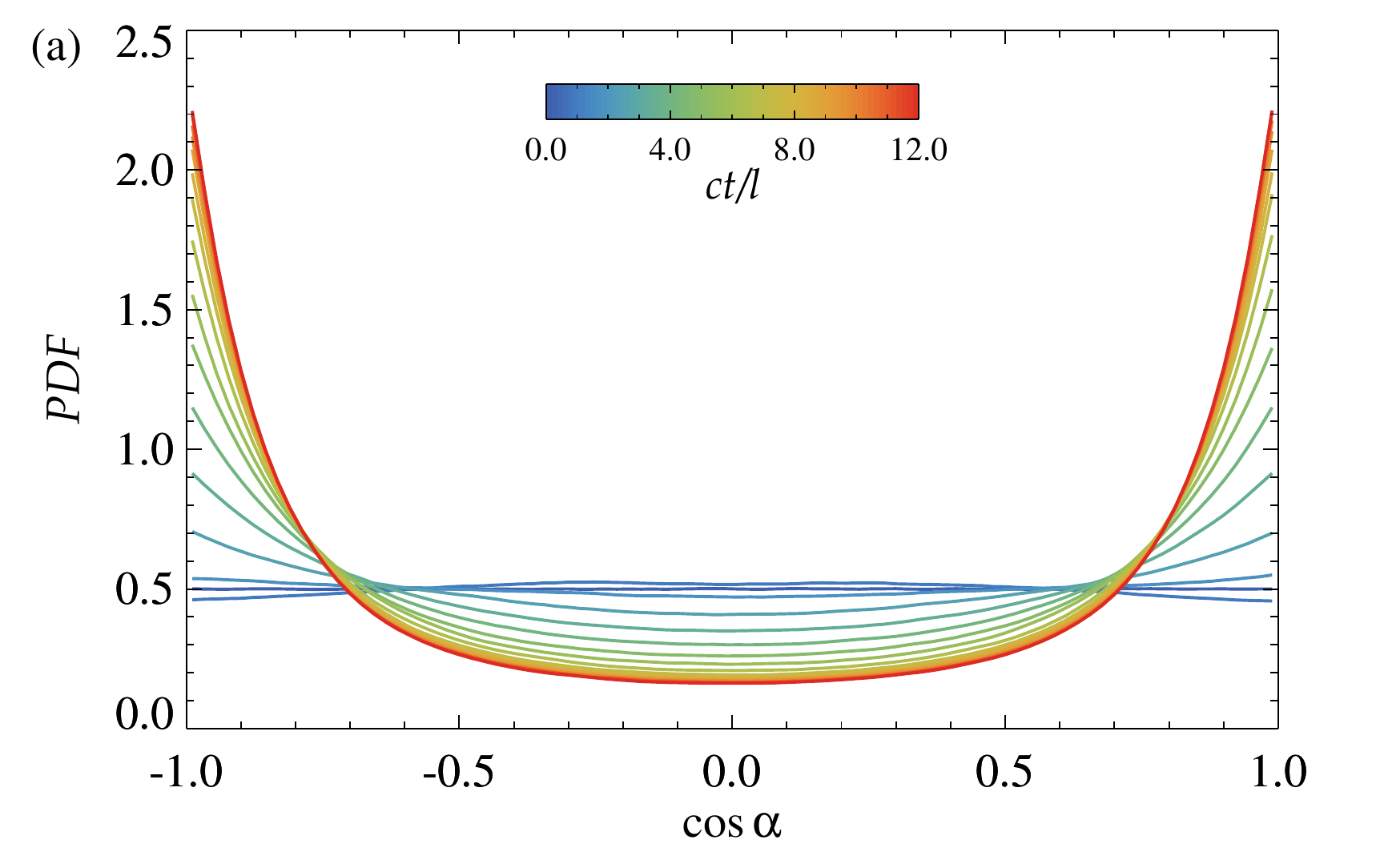}
  \includegraphics[width=8.75cm]{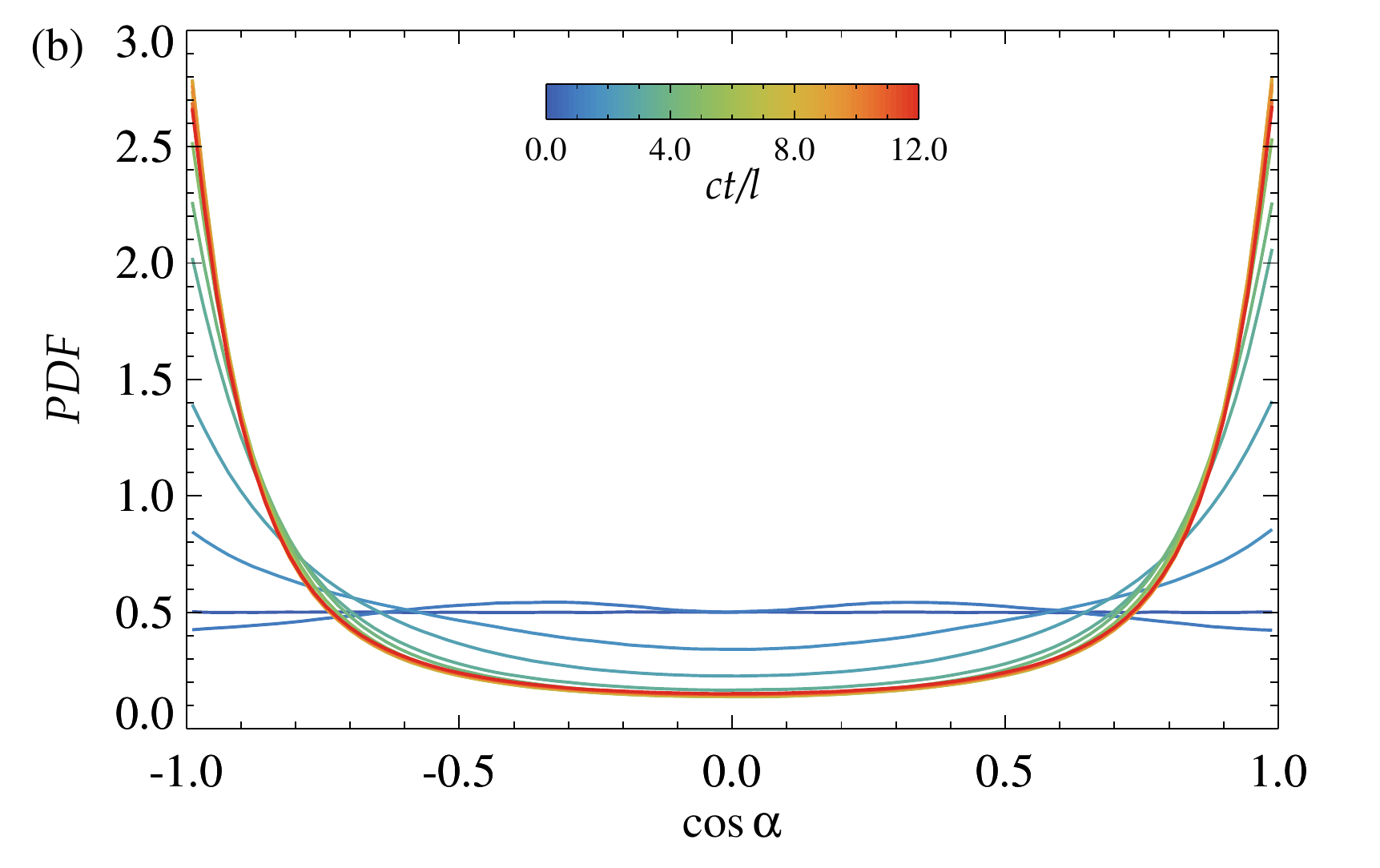}
\caption{Probability density functions of the pitch-angle cosine $\cos \alpha  = {\bm{v}} \cdot {\bm{B}} /({\left| {\bm{v}} \right|\left| {\bm{B}} \right|})$ at different times, obtained from 2D (left) and 3D (right) simulations. Both simulations have $\sigma_0=10$ and $\delta B_{{\rm{rms}}0}/ B_0=1$. The 2D simulation has domain size $L/d_{e0}=1640$ (with $l=L/8$), while the 3D simulation has domain size $L/d_{e0}=820$ (with $l=L/4$).}
\label{Fig_pitchA_time}
\end{figure*} 
%%%%%%%%%%%%%%%%%%%%%%%%%%%%%%

%%%%%%%%%%%%%%%%%%%%%%%%%%%%%%  
\begin{figure*}
 \centering 
  \includegraphics[width=8.75cm]{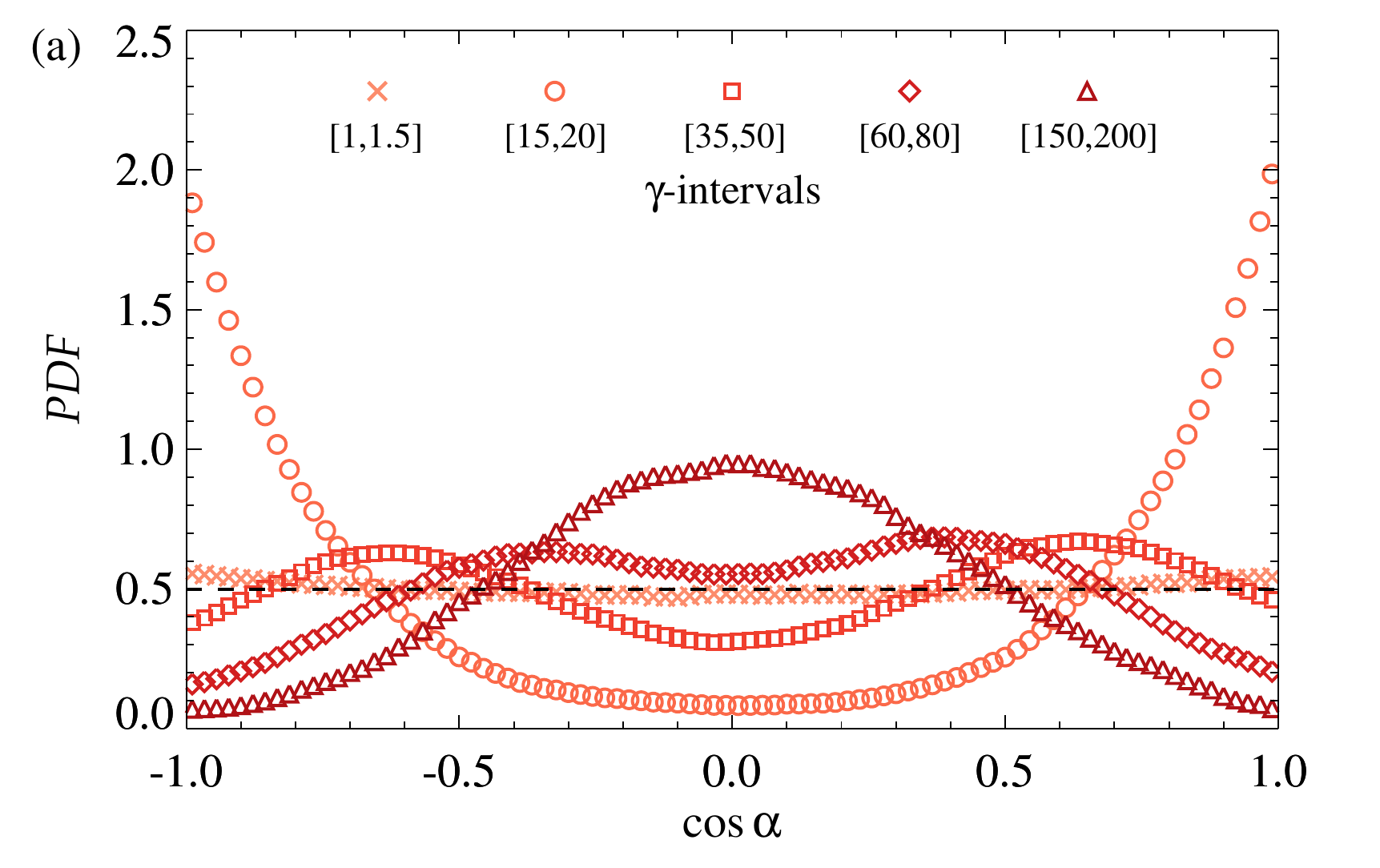}
  \includegraphics[width=8.75cm]{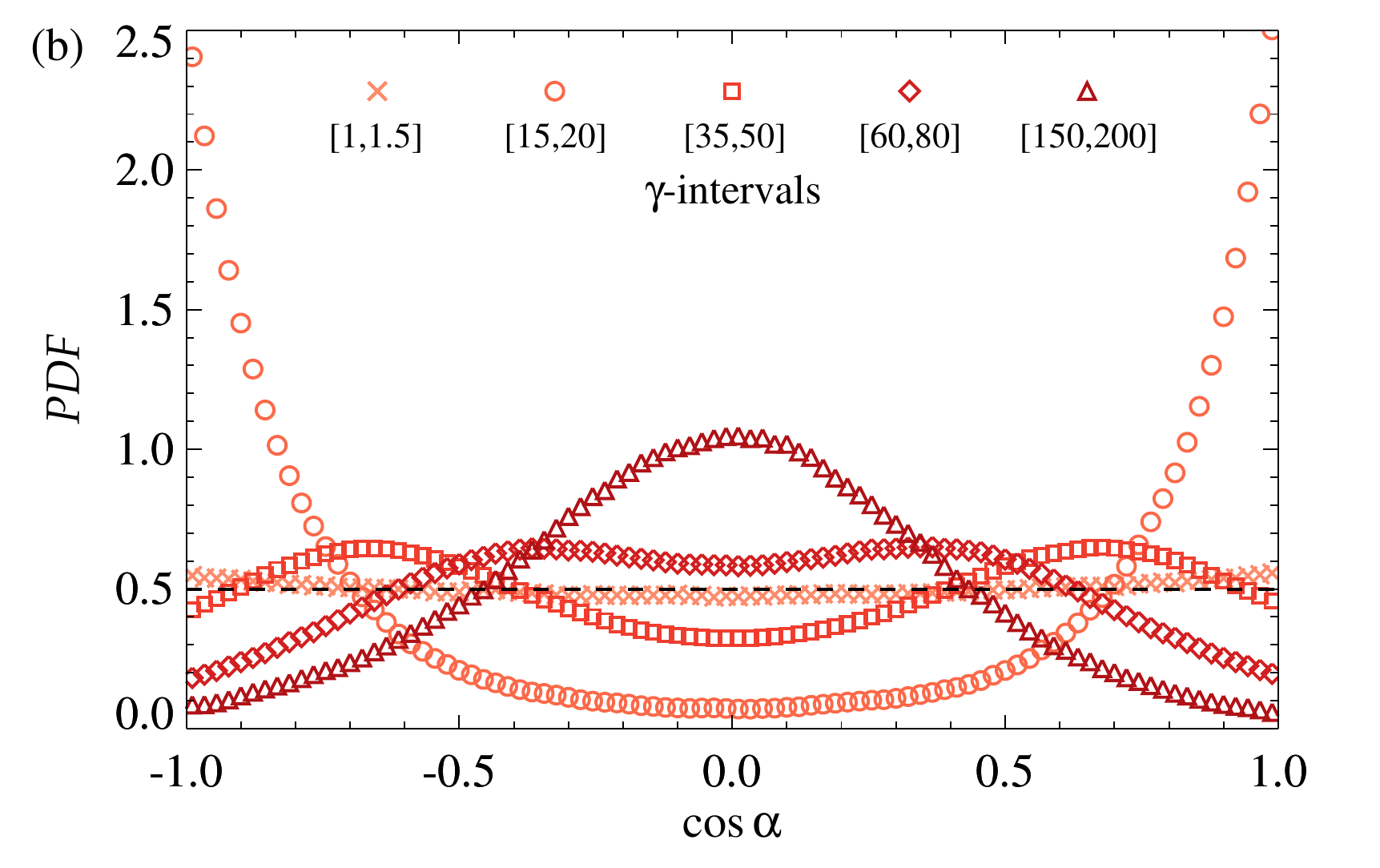}
    \hfill
  \includegraphics[width=8.75cm]{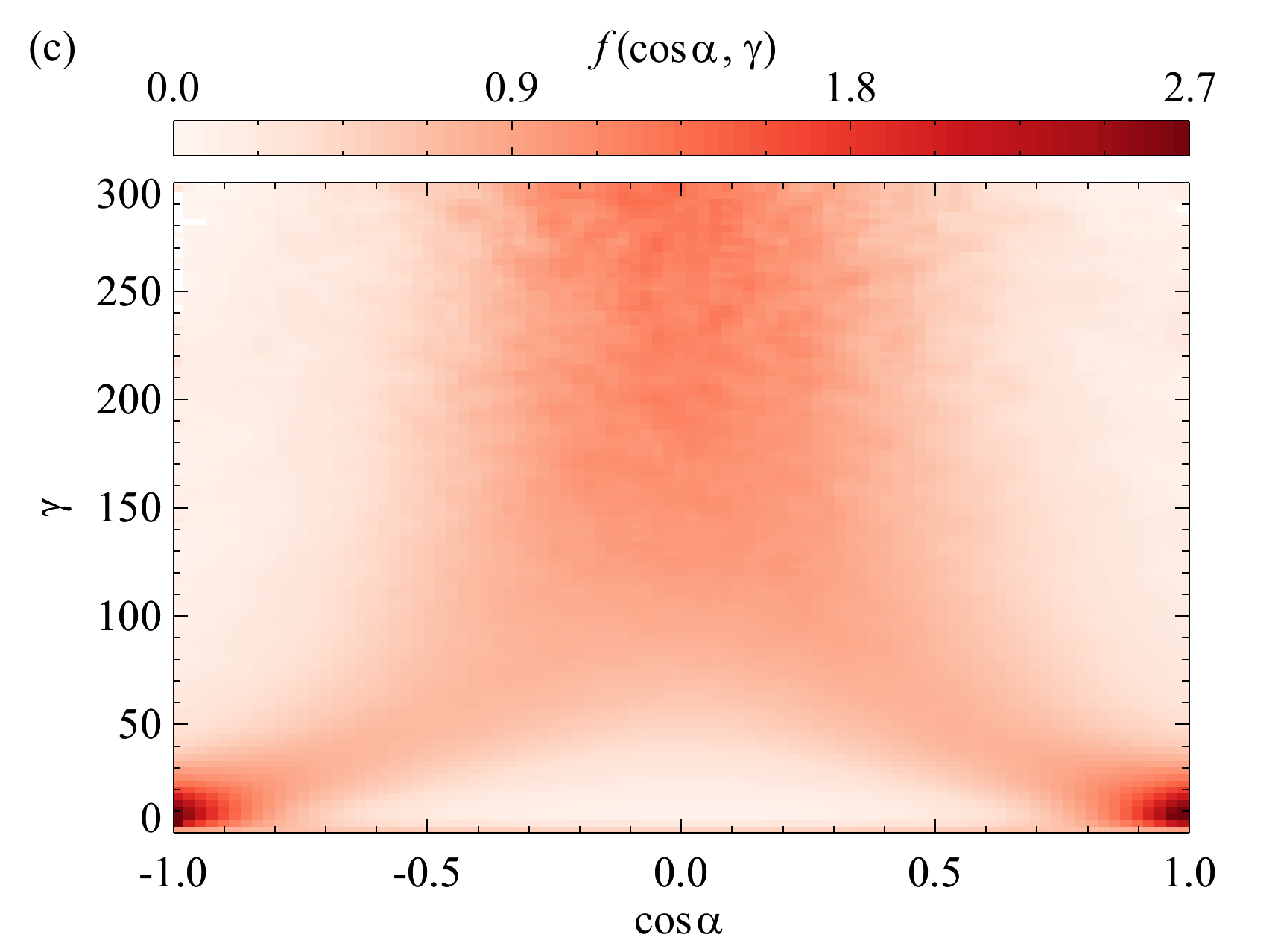}
  \includegraphics[width=8.75cm]{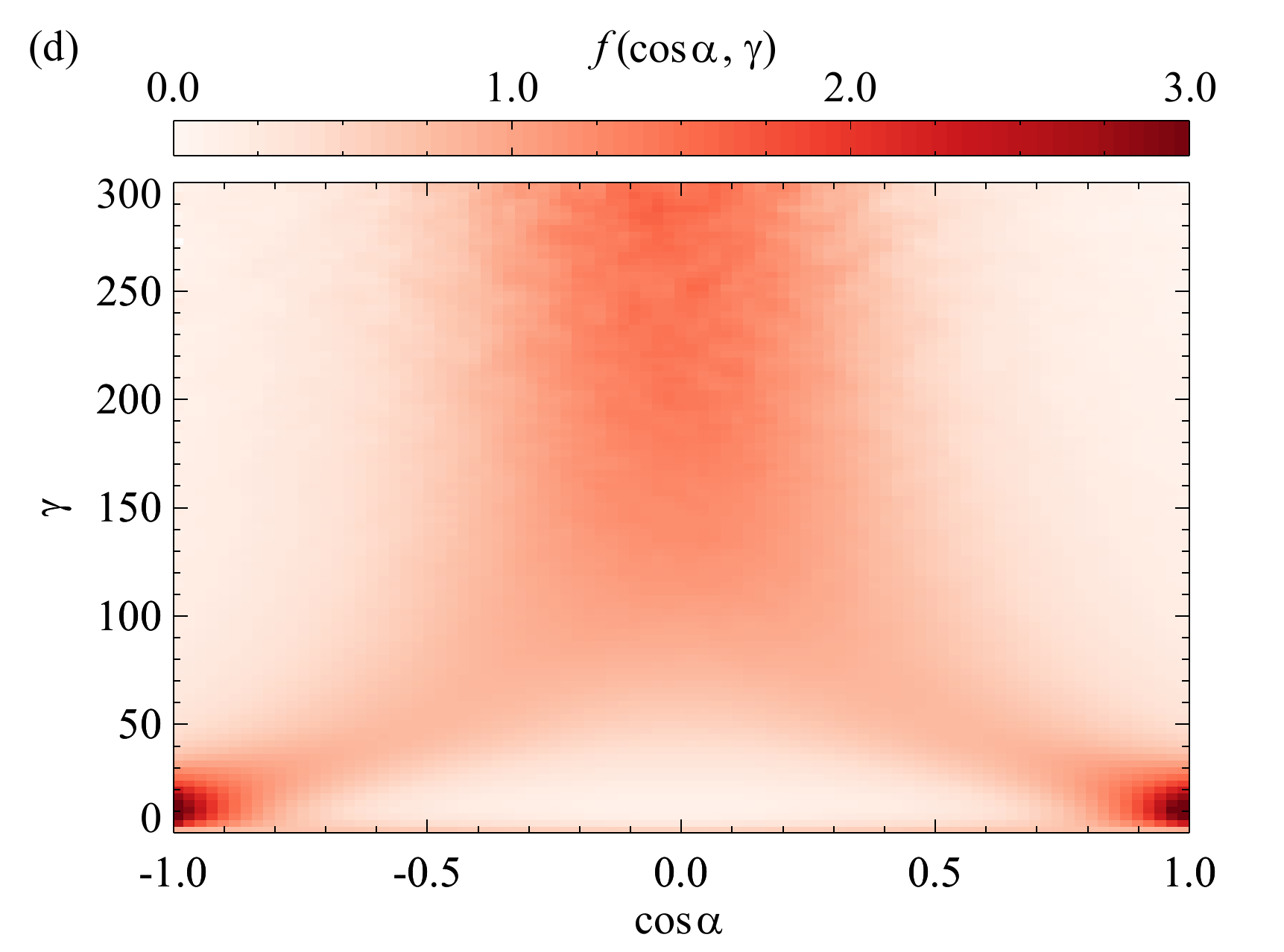}
\caption{Particle distributions obtained from 2D (left) and 3D (right) simulations with $\sigma_0=10$ and $\delta B_{{\rm{rms}}0}/ B_0=1$. The 2D simulation has domain size $L/d_{e0}=1640$ (with $l=L/8$), while the 3D simulation has domain size $L/d_{e0}=820$ (with $l=L/4$). Top row: probability density functions of the pitch-angle cosine $\cos \alpha  = {\bm{v}} \cdot {\bm{B}} /({\left| {\bm{v}} \right|\left| {\bm{B}} \right|})$  for particle Lorentz factors in the intervals $\gamma  \in [1,1.5]$ ($\times$ symbol), $\gamma  \in [15,20]$ ({\large$\circ$} symbol), $\gamma  \in [35,50]$ ($\square$ symbol),  $\gamma  \in [60,80]$ ({\large$\diamond$} symbol), and $\gamma  \in [150,200]$ ($\vartriangle$ symbol).  Bottom row: particle distribution with respect to the pitch-angle cosine $\cos \alpha$ and the Lorentz factor $\gamma$. The plots are obtained from data in the time range $ct/l \in [3,12]$ to increase  statistics.}
\label{Fig_pitchA}
\end{figure*} 
%%%%%%%%%%%%%%%%%%%%%%%%%%%%%%

The time evolution of the overall particle distribution with respect to $\cos \alpha$ is shown in Fig. \ref{Fig_pitchA_time}(a) for the reference 2D simulation  and in Fig. \ref{Fig_pitchA_time}(b) for the reference 3D simulation. As turbulence evolves, the pitch-angle distribution becomes anisotropic with strong peaks at $\cos \alpha = \pm 1$, i.e., for particles moving along the magnetic field lines. Pronounced peaks of the pitch-angle distribution near $\cos \alpha = \pm 1$ have also been found in nonrelativistic plasma turbulence at low-$\beta_p$ \citep[e.g.][]{PecoraJPP18}, with $\beta_p = n_0 k_B T/({\langle B^2 \rangle}/8 \pi)$ indicating the plasma beta, i.e., the ratio of  thermal pressure to  magnetic pressure. Indeed, the low-$\beta_p$ regime is similar to the high-$\sigma$ regime investigated here, in the sense that in both cases the magnetic energy density dominates over the thermal energy density (in our simulations the initial plasma beta is $\beta_p = 2 \theta_0/[w_0(\sigma_z + \sigma_0)]$). The  PDFs of  2D and 3D simulations are similar; nevertheless, the 3D case exhibits higher probability peaks near $\cos \alpha = \pm 1$. Furthermore, the pitch-angle distribution evolves more rapidly in  3D  as a consequence of the faster conversion of magnetic energy into particle energy. The large fraction of particles having velocity strongly aligned/antialigned with the local magnetic field is a natural expectation of injection mediated by magnetic reconnection, which can efficiently energize particles through the work done by ${\bm{E}}_\parallel$. As we have shown, reconnecting current sheets can process a large fraction of particles in just a few $c/l$ (see Eqs. (\ref{2Dprocessed}) and (\ref{3Dprocessed})).

The PDFs illustrated in Fig. \ref{Fig_pitchA_time} are dominated by low-energy particles (i.e., near the spectral peak), since they control the number census (see Figs. \ref{fig3} and \ref{fig8}). In order to characterize the anisotropy of  particles of higher energy, we construct PDFs of $\cos \alpha$ for different populations of particles depending on their Lorentz factor. We collected  particle data from a time range $ct/l \in [3,12]$ in order to increase statistics. However, we have verified that decreasing this range (up to a single time snapshot taken at late times) does not modify the results. These results are shown in Fig. \ref{Fig_pitchA}(a) and Fig. \ref{Fig_pitchA}(b) for 2D and 3D turbulence, respectively. At very low energies ($\gamma \sim 1$), the particle distribution remains nearly isotropic. These are the particles of our initial Maxwellian, which have not been energized. At moderate Lorentz factors ($\gamma \sim 15$), the particle distribution displays stong peaks close to $\cos \alpha = \pm 1$, in analogy with the results shown in Fig. \ref{Fig_pitchA_time}. At higher Lorentz factors, the pitch-angle distribution evolves into a ``butterfly distribution'' with minima at both $\cos \alpha = \pm 1$ and $\cos \alpha = 0$. This phenomenon occurs at Lorentz factors $\gamma \sim 50$ for the simulations with $\sigma_0=10$ shown in Fig. \ref{Fig_pitchA_time}. At even higher energies ($\gamma \gg 50$), the pitch-angle distribution become eventually peaked at $\cos \alpha = 0$, i.e. for particles moving in the plane perpendicular to the local magnetic field. This  trend can be displayed using a distribution $f \left( {\cos \alpha , \gamma} \right)$ of particles with respect to $\cos \alpha$ and $\gamma$. This distribution, shown in  Fig. \ref{Fig_pitchA}(c) and Fig. \ref{Fig_pitchA}(d) for 2D and 3D turbulence, respectively, has been normalized such that
\begin{equation} \label{} 
\int\limits_{ - 1 }^1  {f\left( {\cos \alpha ,\gamma} \right) d(\cos \alpha)}  = 1  \, .
\end{equation}
In these plots, the peaks of $f\left( {\cos \alpha ,\gamma} \right)$ are located   at $\cos \alpha = \pm 1$ for low energies,  and then they move towards  $\cos \alpha = 0$ until $\gamma \sim 80$. At higher energies, the peak of the distribution remains located at $\cos \alpha = 0$, with particles that lie progressively more perpendicular to the local magnetic field as their energy increases.

The energy-dependent anisotropy illustrated in Fig. \ref{Fig_pitchA} reflects the different acceleration mechanisms that operate at different energies (see Section \ref{SecEnergiz}). At low energies, the contribution of the ${\bm{v}}_\parallel \cdot {\bm{E}}_\parallel$ energization is dominant, so that particles end up being strongly aligned/antialigned with the magnetic field ($\cos \alpha \sim \pm 1$). On the other hand, as the energy increases, the ${\bm{v}}_\bot \cdot {\bm{E}}_\bot$ energization takes over and propels the particles in the direction perpendicular to the local magnetic field. The time scale of this acceleration is fast compared to the pitch-angle scattering timescale,  so that particles retain their orientation $\cos \alpha \sim 0$ for long times.

%%%%%%%%%%%%%%%%%%%%%%%%%%%%%%  
\begin{figure}
\begin{center}
\hspace*{-0.085cm}\includegraphics[width=8.75cm]{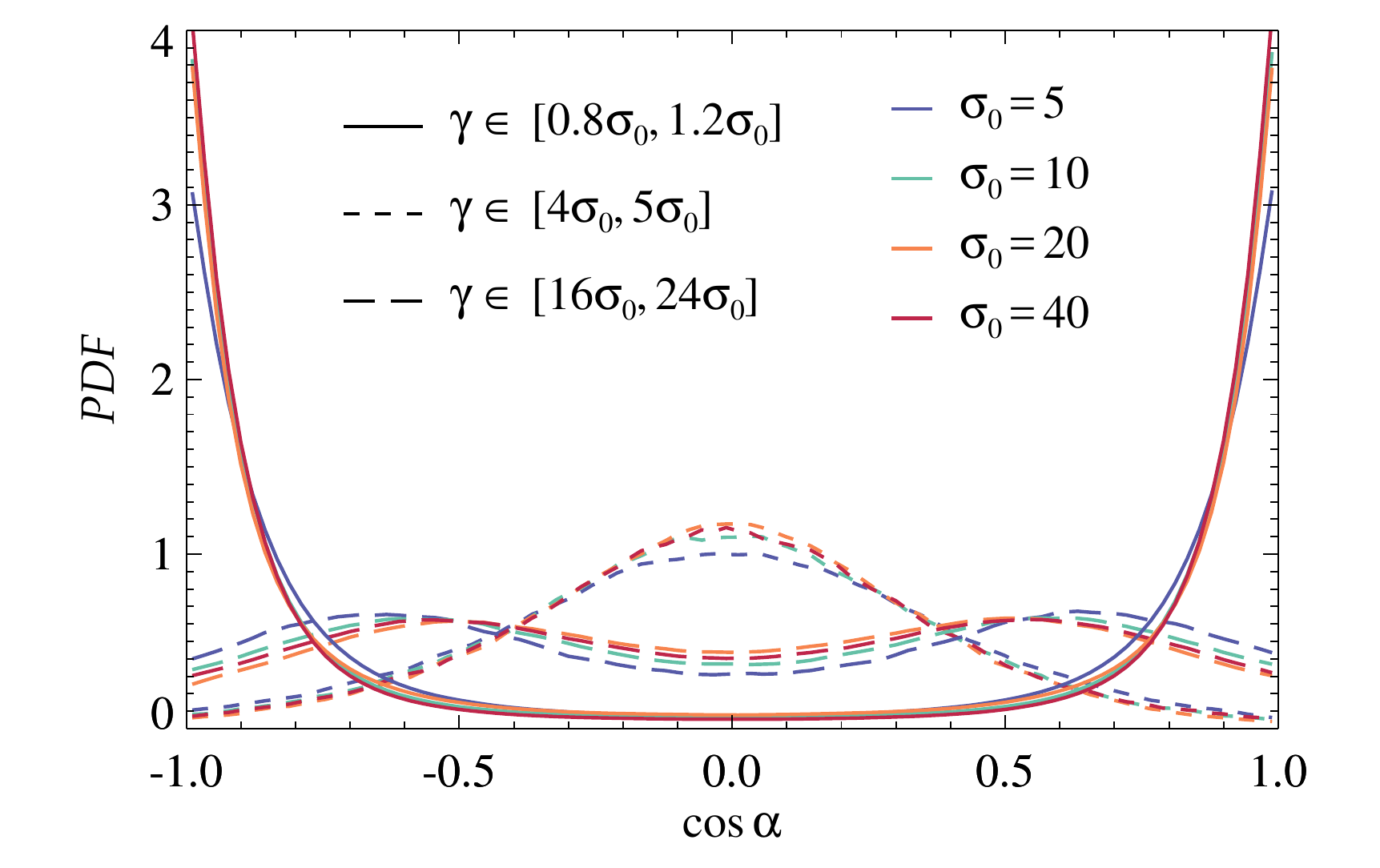}
\end{center}
\caption{Probability density functions of the pitch-angle cosine $\cos \alpha  = {\bm{v}} \cdot {\bm{B}} /({\left| {\bm{v}} \right|\left| {\bm{B}} \right|})$ for particles with Lorentz factors $\gamma  \in [0.8 \sigma_0, 1.2 \sigma_0]$ (solid lines), $\gamma  \in [4 \sigma_0, 5 \sigma_0]$ (long-dashed lines), and $\gamma  \in [16 \sigma_0, 24 \sigma_0]$ (dashed lines). Different colors refer to different 3D simulations having $\sigma_0=5$ (blue), $\sigma_0=10$ (green), $\sigma_0=20$ (orange), and $\sigma_0=40$ (red). All 3D simulations have $\delta B_{{\rm{rms}}0}/ B_0=1$ and $L/d_{e0}=820$. We also recall that $\gamma_{th0} \approx 1.58$. Data is collected from a time range $ct/l \in [3,12]$.}
\label{Fig_pitchA_sigma}
\end{figure} 
%%%%%%%%%%%%%%%%%%%%%%%%%%%%%%  

The results shown in Fig. \ref{Fig_pitchA} for magnetization $\sigma_0 = 10$ hold also for the other magnetizations we investigate. In Fig. \ref{Fig_pitchA_sigma},  we present the results from four simulations that differ in magnetization $\sigma_0 \in \left\{ {5,10,20,40} \right\}$. Here we show only the results from  3D simulations, since those from  2D simulations are analogous. The ranges in $\gamma$ are scaled with $\sigma_0$, which provides the typical energy scale (e.g., the starting point of the high-energy nonthermal tail, $\gamma_{st}$, increases linearly with $\sigma_0$, as illustrated by Eq.  (\ref{gamma_st})). For  $\gamma \sim (\sigma_0/2) \gamma_{th0}$ (solid lines), we have a pitch-angle distribution peaked at $\cos \alpha \sim \pm 1$ (the only difference is that the percentage of particles aligned/antialigned with the local magnetic field slightly increases with $\sigma_0$). The butterfly distribution with minima at $\cos \alpha = \pm 1,0$ appears for $\gamma \sim 5 (\sigma_0/2) \gamma_{th0}$ (long-dashed lines). Finally, for $\gamma \gg 5 (\sigma_0/2) \gamma_{th0}$, well into the nonthermal tail, the particle velocities become mostly perpendicular to the magnetic field and we can  see that all the distributions are peaked at $\cos \alpha = 0$ (see dashed lines).

\subsection{Particle four-velocity distribution}

%%%%%%%%%%%%%%%%%%%%%%%%%%%%%%  
\begin{figure}
\begin{center}
\includegraphics[width=8.55cm]{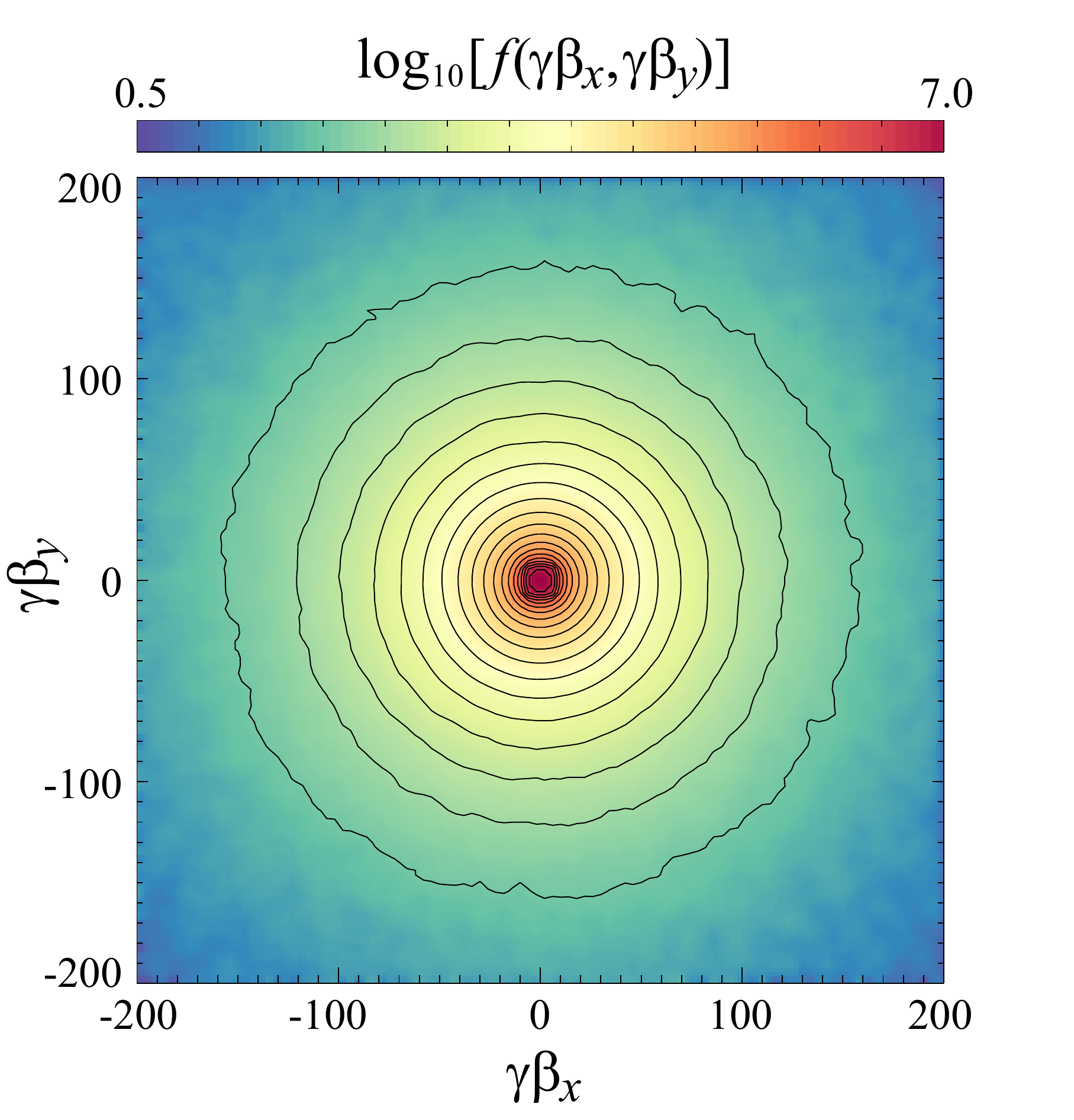}
\includegraphics[width=8.55cm]{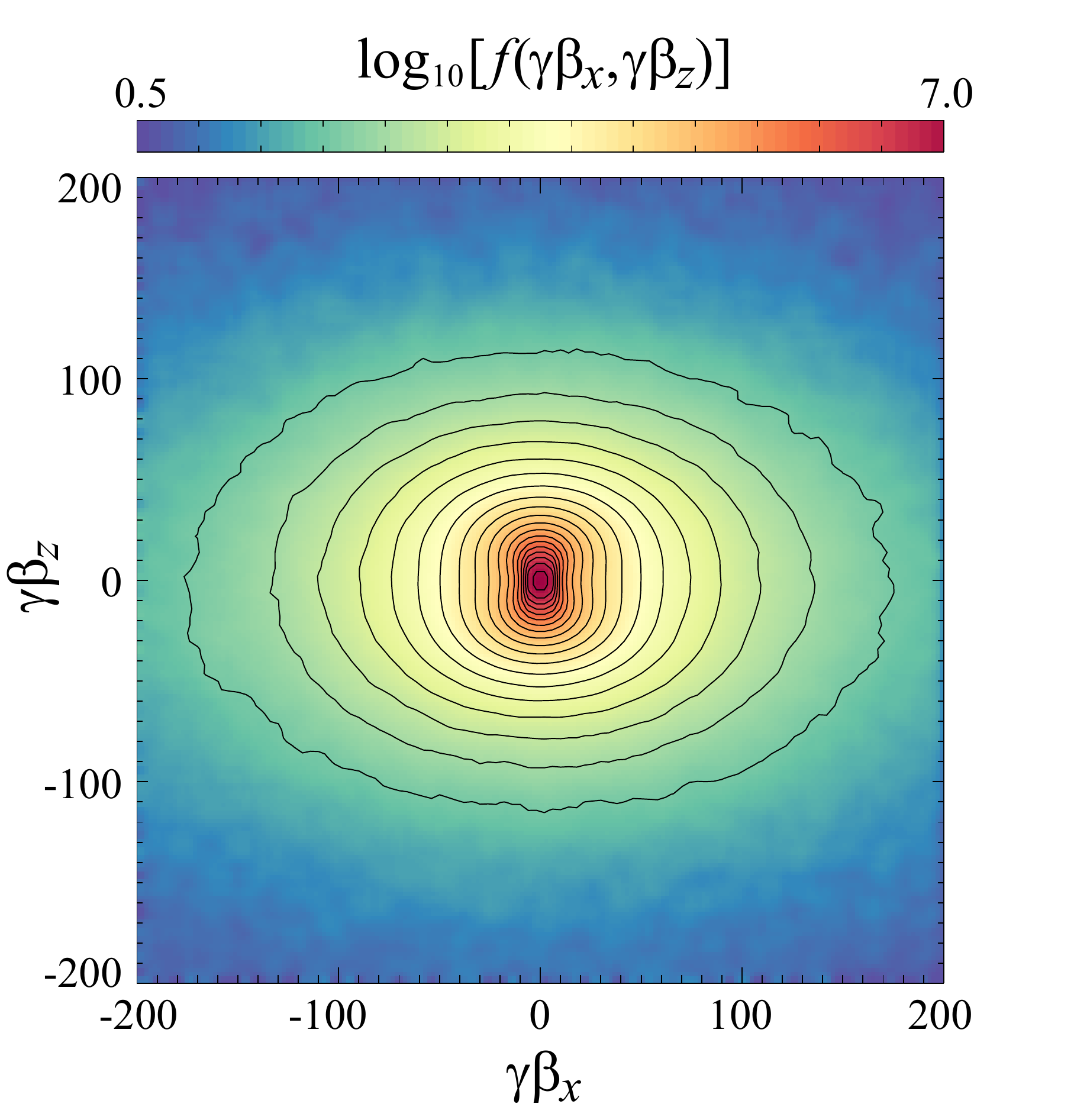}
\end{center}
\caption{Top frame: box-averaged four-velocity distribution function $f(\gamma \beta_x, \gamma \beta_y)$ for a 3D simulation with $\sigma_0=10$, $\delta B_{{\rm{rms}}0}/ B_0=1$, and $L/d_{e0}=820$. Bottom frame: from the same simulation, box-averaged four-velocity distribution function $f(\gamma \beta_x, \gamma \beta_z)$. The plots are obtained from data in the time range $ct/l \in [3,12]$. Normalization is arbitrary.}
\label{Fig_VelocityDistributionFunction}
\end{figure} 
%%%%%%%%%%%%%%%%%%%%%%%%%%%%%%  

%%%%%%%%%%%%%%%%%%%%%%%%%%%%%%  
\begin{figure}
\begin{center}
\includegraphics[width=8.55cm]{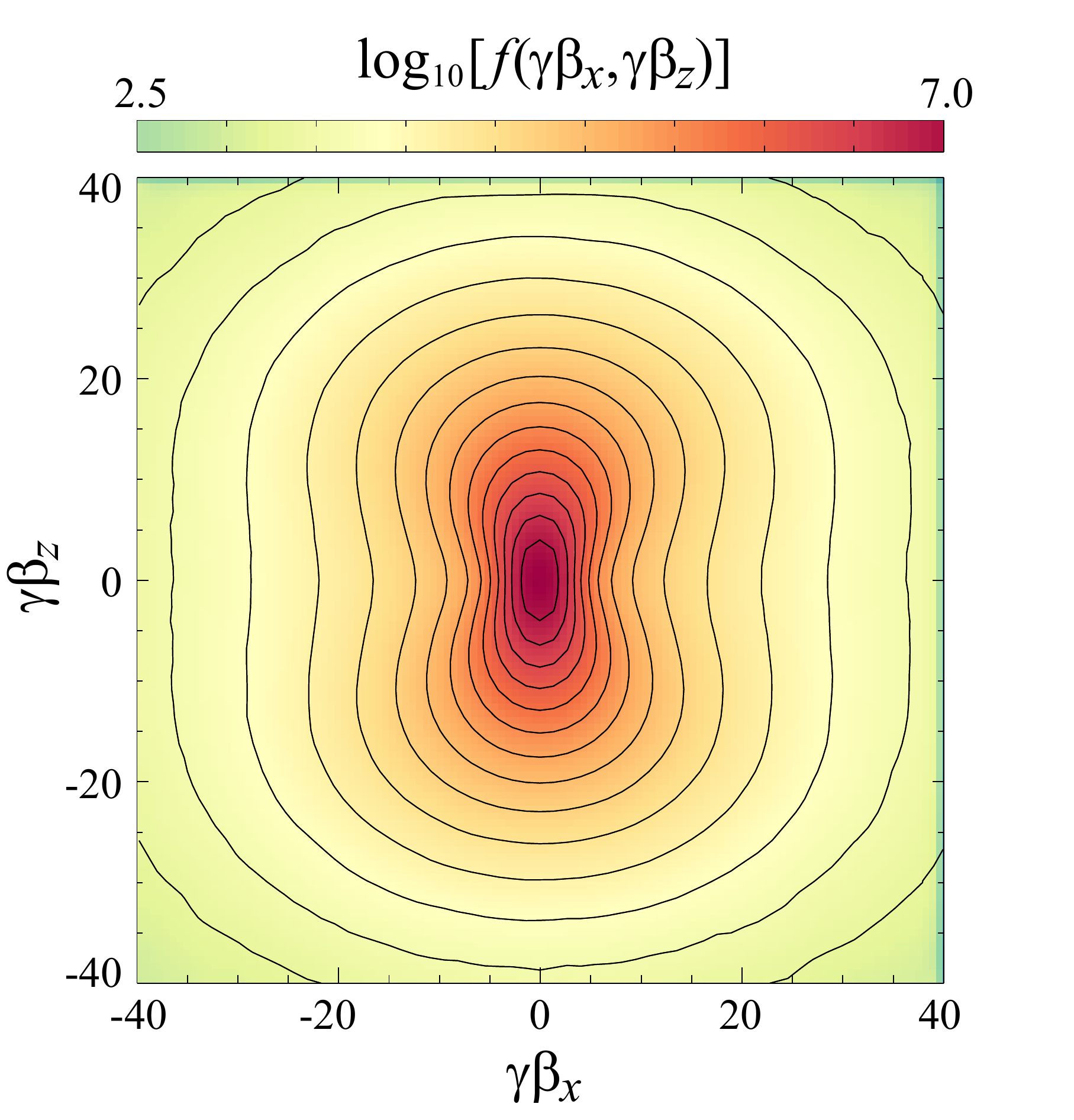}
\end{center}
\caption{Zoom around the intermediate-energy region of the four-velocity distribution function $f(\gamma \beta_x, \gamma \beta_z)$ shown in Fig. \ref{Fig_VelocityDistributionFunction}.}
\label{Fig_VelocityDistributionFunction_Zoom}
\end{figure} 
%%%%%%%%%%%%%%%%%%%%%%%%%%%%%%  

The results on the anisotropy of the pitch-angle distributions, computed with respect to the \emph{local} magnetic field ${\bm{B}} = {{\bm{B}}_0} +  \delta {\bm{B}}$, suggest that the four-velocity distribution function, with respect to the  \emph{mean} magnetic field ${\bm{B}}_0  = B_0 {\bm{\hat z}}$, should also display significant anisotropy. Indeed, as the turbulence fluctuations decay, the local magnetic field becomes progressively more aligned with the direction of the mean magnetic field. 

We calculated the domain-averaged four-velocity distributions in the $xy$ plane, $f\big(\gamma \beta_x, \gamma \beta_y \big)$, and in the $xz$ plane, $f\big(\gamma \beta_x, \gamma \beta_z \big)$, from the same samples of particles used to analyze the local pitch-angle distributions. The results, for our reference 3D simulation (the 2D case is analogous) are shown in Fig. \ref{Fig_VelocityDistributionFunction} (and a zoom in Fig. \ref{Fig_VelocityDistributionFunction_Zoom}). As for Fig. \ref{Fig_pitchA}, we collected particle data in the time range $ct/l \in [3,12]$ in order to increase  statistics, but we have also verified that decreasing this range (up to a single time snapshot taken at late times) does not modify the results. As we expected, from Fig. \ref{Fig_VelocityDistributionFunction} we find that the four-velocity distribution is isotropic in the plane perpendicular to ${\bm{B}}_0$ (top panel), while it develops more complex features with respect to planes that contain ${\bm{B}}_0$, as for the case of $f\big(\gamma \beta_x, \gamma \beta_z \big)$ (bottom panel). The results are analogous when considering $f\big(\gamma \beta_y, \gamma \beta_z \big)$ or $f\big((\gamma^2 \beta_x^2 + \gamma^2 \beta_y^2)^{1/2} , \gamma \beta_z \big)$. The distribution $f\big(\gamma \beta_x, \gamma \beta_z \big)$ displays a core region elongated in the $\gamma \beta_z$ direction, as particle velocities are mostly aligned/antialigned with the magnetic field at low energies. Furthermore, a close inspection shows that there is an intermediate-energy region with the majority of  particles residing in a double cone  whose axis is the direction of the mean magnetic field ${\bm{B}}_0$ (see Fig. \ref{Fig_VelocityDistributionFunction_Zoom}). This is the intermediate-energy range in which the peak of pitch-angle distribution moves from $\cos \alpha = \pm 1$ towards $\cos \alpha =0$. At even higher energies, Fig. \ref{Fig_VelocityDistributionFunction} shows that $f\big(\gamma \beta_y, \gamma \beta_z \big)$ becomes elongated in the direction perpendicular to ${\bm{B}}_0$, consistently with the dominance of  ${\bm{v}}_\bot \cdot {\bm{E}}_\bot$ energization at higher energies and the resulting anisotropy of the pitch angle cosine.

\subsection{Particle mixing}

%%%%%%%%%%%%%%%%%%%%%%%%%%%%%%  
\begin{figure*}
 \centering 
  \includegraphics[width=0.47\textwidth]{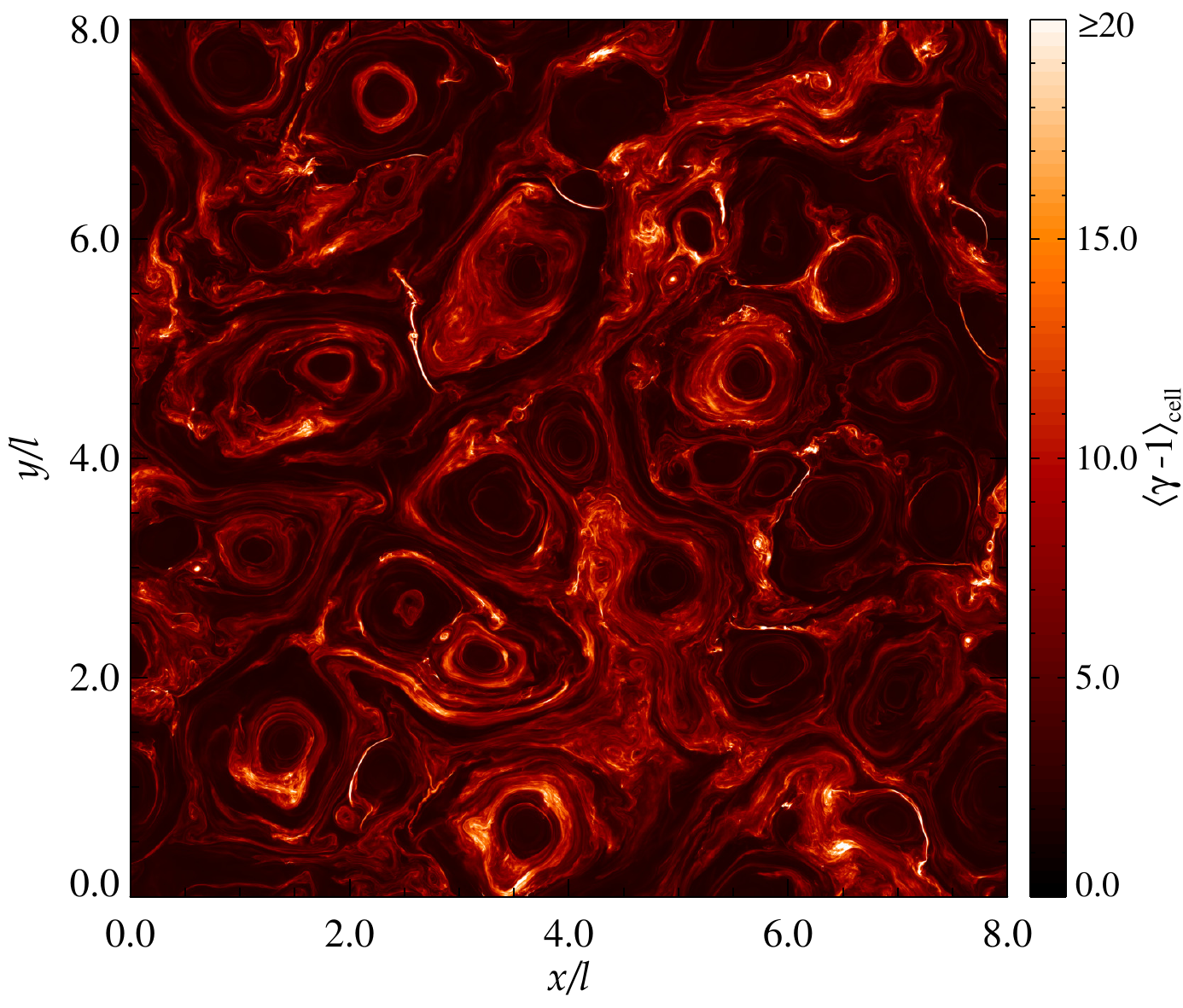}
  \includegraphics[width=0.47\textwidth]{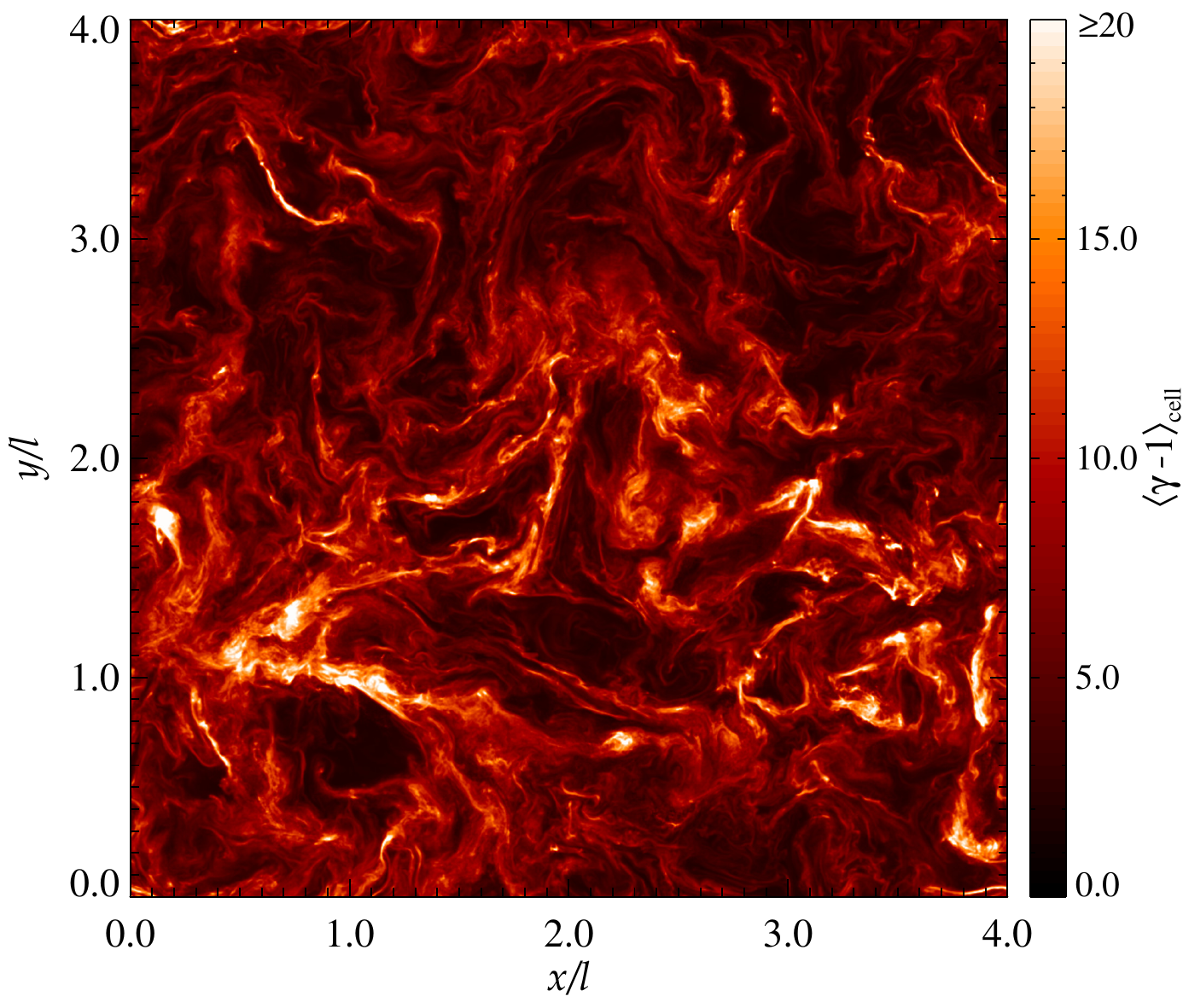} 
  \hfill
  \includegraphics[width=0.47\textwidth]{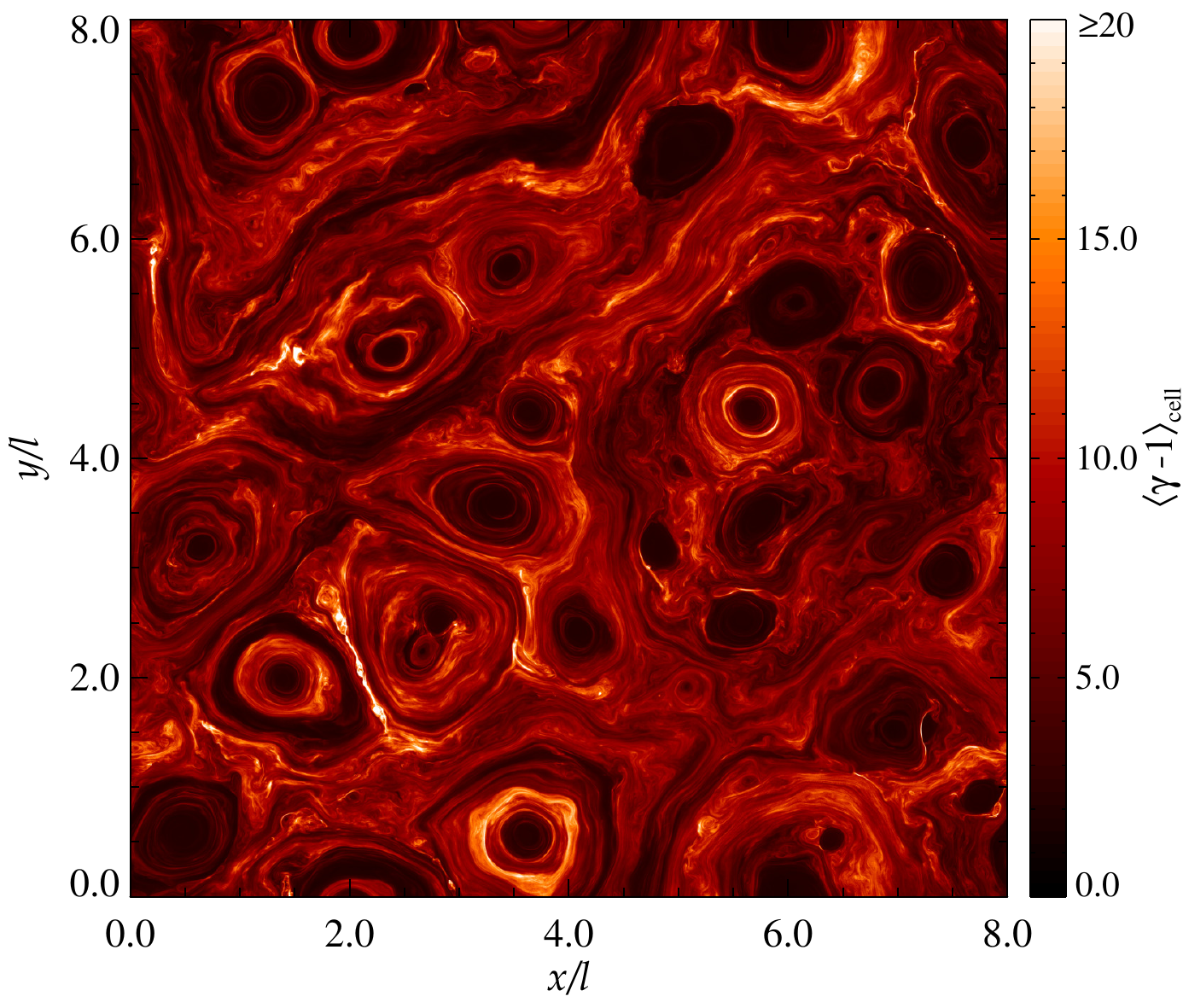}
  \includegraphics[width=0.47\textwidth]{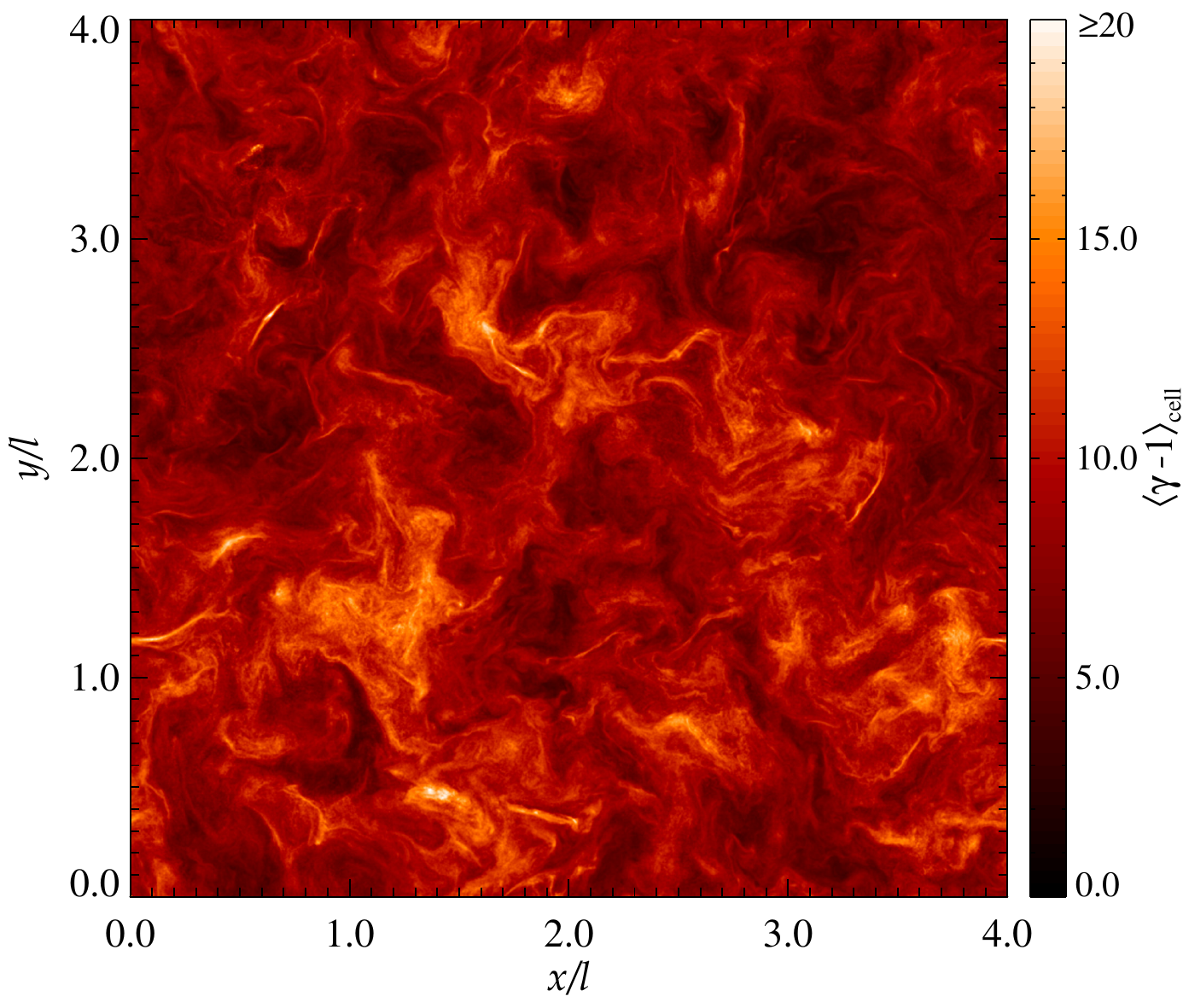}
    \hfill
  \includegraphics[width=0.47\textwidth]{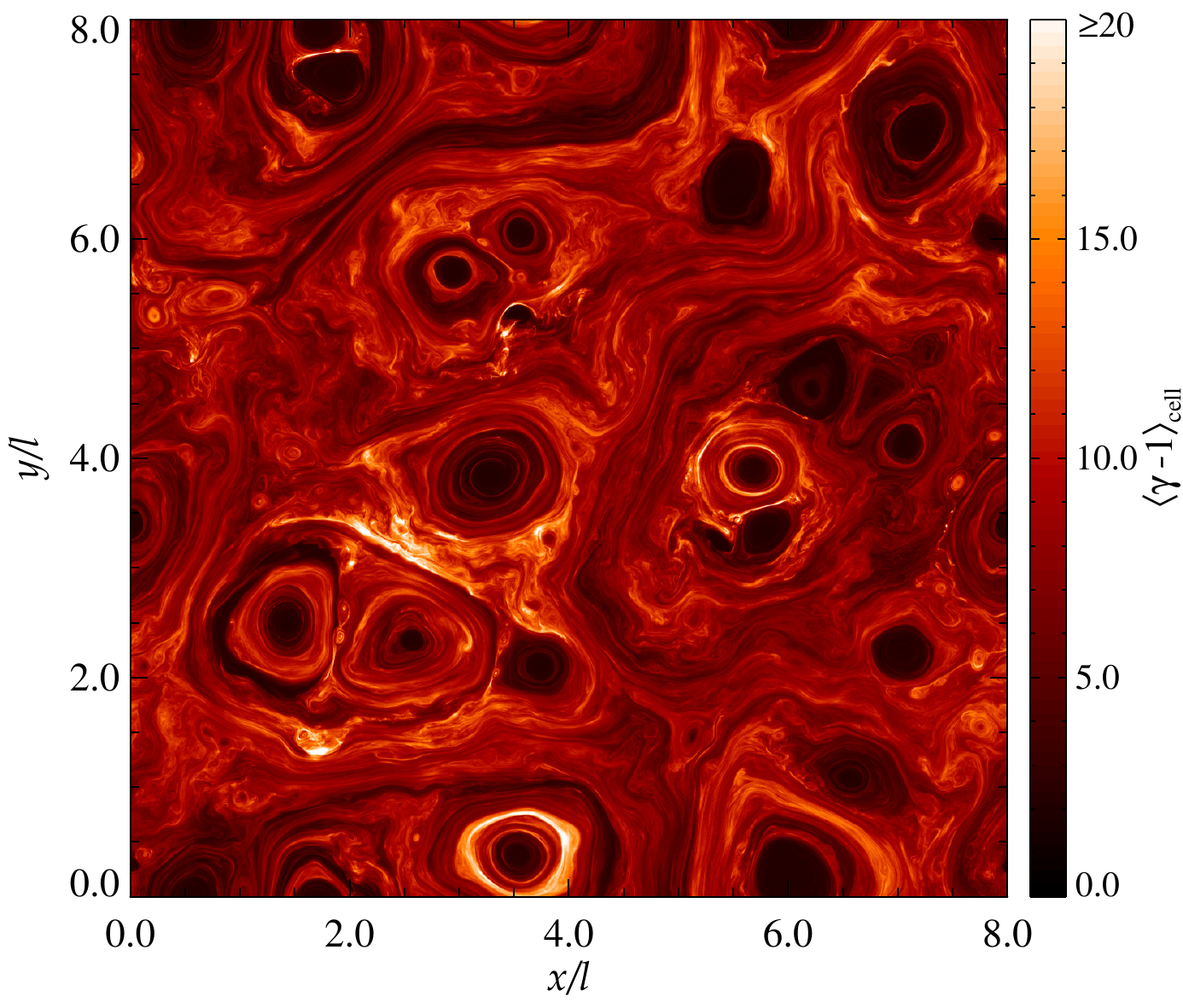}
  \includegraphics[width=0.47\textwidth]{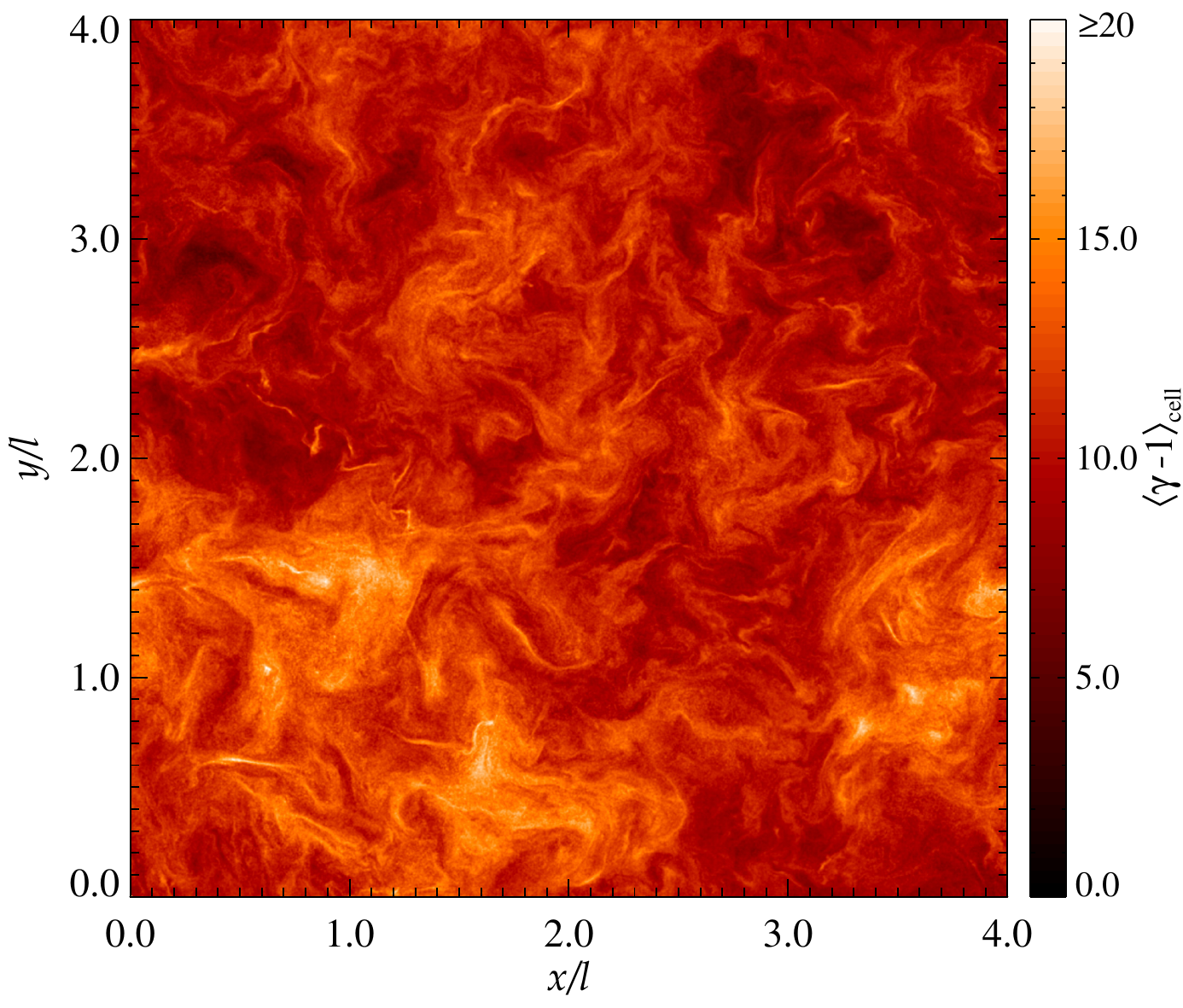}
 \caption{2D plots of the cell-averaged mean kinetic energy per particle normalized by $mc^2$, $\langle \gamma-1 \rangle_{\rm{cell}}$, for 2D turbulence (left column) and 3D turbulence (right column). For 3D turbulence, the 2D plots refer to a slice of the domain at constant $z/l = 0$. The normalized times $ct/l$ for the plots in the left column are (from top to bottom): $ct/l=4.6$, $ct/l=7.7$ and $ct/l=10.8$, while those for the plots in the right column are (from top to bottom): $ct/l=2.7$, $ct/l=4.5$ and $ct/l=6.3$. The 2D simulation has a domain size $L/d_{e0}=1640$ (with $l=L/8$), while for the 3D simulation $L/d_{e0}=820$ (with $l=L/4$). Both simulations have $\sigma_0=10$ and $\delta B_{{\rm{rms}}0}/ B_0=1$. An animation showing $\langle \gamma-1 \rangle_{\rm{cell}}$ at $ct/l=2.7$ in different $xy$ slices can be found at \url{https://doi.org/10.7916/d8-prt9-kn88}.} 
\label{fig_Mixing}
\end{figure*} 
%%%%%%%%%%%%%%%%%%%%%%%%%%%%%%
We show that while the qualitative and quantitative features of the pitch-angle distributions are similar in our 2D and 3D simulations, the turbulent mixing (in space) of the energized particles is quite different. Particle mixing in 2D is expected to be less efficient than in 3D, since the translation-invariant symmetry along ${\bm{\hat z}}$ seriously constrains the 2D dynamics. As a consequence, regions of space devoid of high-energy particles can be retained for  a larger number of outer-scale eddy turnover times in  2D simulations.

In Fig. \ref{fig_Mixing}, we show how the energized particles are distributed in the spatial domain in 2D (left column) and 3D (right column). For both cases, we plot the cell-averaged  kinetic energy per particle, $\langle \gamma-1 \rangle_{\rm{cell}}$, at three different times (from top to bottom). The mean kinetic energy is normalized by $m c^2$. The time snapshots are different for 2D ($ct/l$=4.6,7.7,10.8) and 3D ($ct/l$=2.7,4.5,6.3), to account for the faster turbulence decay in 3D. In both cases, the initial energization occurs at current sheets, which display high values of $\langle \gamma-1 \rangle_{\rm{cell}}$, and then particles propagate outside current sheets in other regions of the domain (see top frames in Fig. \ref{fig_Mixing}). As time progresses, energized particles diffuse in the spatial domain, and the mean kinetic energy per particle becomes more uniform  (middle frames in Fig. \ref{fig_Mixing}). However, in 2D the cores of the large-scale flux tubes remain essentially unaffected, as these overdense regions with $n \gg n_0$ and with higher fluctuation magnetic energy density $\delta {B^2}/8\pi$  (see Fig. \ref{fig1}) are mainly populated by low energy particles that have not been processed by reconnecting current sheets. On the other hand, the 3D domain does not present such isolated regions of low $\langle \gamma-1 \rangle_{\rm{cell}}$. At quite early times, the mean kinetic energy per particle becomes fairly homogeneous across the entire 3D domain, whereas the 2D simulation preserved regions of low $\langle \gamma-1 \rangle_{\rm{cell}}$ for much longer (bottom frames in Fig. \ref{fig_Mixing}).  This different behavior between 2D and 3D turbulence is also reflected in the particle energy spectrum, with 2D turbulence retaining more particles with $\gamma \lesssim \gamma_{th0}$ until late times (see Figs.~\ref{fig3} and \ref{fig8}).

\section{Particle Energy Diffusion and Stochastic Acceleration} \label{SecEnergyDiff}

We have seen that after the injection phase, the subsequent energy gain is dominated by perpendicular electric fields via stochastic scattering off the turbulent fluctuations \citep{ComissoSironi18}. Here, in order to elucidate the properties of the stochastic acceleration phase, we evaluate the energy diffusion coefficient directly from the self-consistent particle evolution of our PIC simulations. This allows us to determine the acceleration timescale associated with stochastic acceleration. Then we show that the two-stage process that accelerates particles is well modeled by an initial injection phase powered by reconnection electric fields, followed by a second acceleration phase modeled with the measured  energy diffusion coefficient.

\subsection{Particle energy diffusion}

Particles that are stochastically scattered off the turbulent fluctuations experience a biased random walk in momentum space, which can be modeled with a Fokker-Planck approach \citep[e.g.][]{BlandfordEichler87}, provided that the fractional momentum change in a single scattering is sufficiently small. In this case, one could describe the process of stochastic acceleration from the point of view of a Fokker-Planck equation in energy space \citep[e.g.][]{Ramaty1979}
\begin{equation} \label{eqFP}
\frac{{\partial N}}{{\partial t}} = - \frac{\partial }{{\partial \gamma }}\left( {{A_\gamma }N} \right) + \frac{{{\partial ^2}}}{{\partial {\gamma ^2}}}\left( {{D_\gamma }N} \right)  \, .
\end{equation}
Here, as usual, $N$ is the particle spectrum differential in energy, $A_\gamma$ is the energy convection coefficient, and $D_\gamma$ is the energy diffusion coefficient. 
%Terms describing escape, injection and losses may also be added to this equation. 
Note that in general, the convection and diffusion coefficients are time dependent (this is indeed the case for the turbulence simulations performed here). The convection coefficient $A_\gamma$ represents the mean energy gain due to stochastic acceleration, and is related to the diffusion coefficient in energy space as 
\begin{equation} \label{}
A_\gamma = \frac{d \langle \gamma \rangle}{dt} = \frac{1}{\gamma^2} \frac{\partial }{\partial \gamma} \left( {\gamma^2 D_\gamma } \right) \, , 
\end{equation} 
The diffusion coefficient in energy space $D_\gamma$ is also related to the diffusion coefficient in momentum space $D_{p}$, with $D_\gamma\simeq D_{p}$ for the ultra-relativistic particles considered here.
Given the fact that high-energy particles preferentially lie in the plane perpendicular to the mean field (Sec. \ref{SecAnisotropy}), and that their energization is mostly contributed by perpendicular electric fields (Sec. \ref{SecEnergiz}), the momentum diffusion coefficient  $D_{p}$ is essentially identical to $D_{p_\perp}$, i.e., to the diffusion coefficient of momenta perpendicular to the mean field. The determination of this coefficient, or equivalently of the energy diffusion coefficient,  establishes the properties of the stochastic acceleration phase.

\begin{figure}
\begin{center}
\hspace*{-0.085cm}\includegraphics[width=8.75cm]{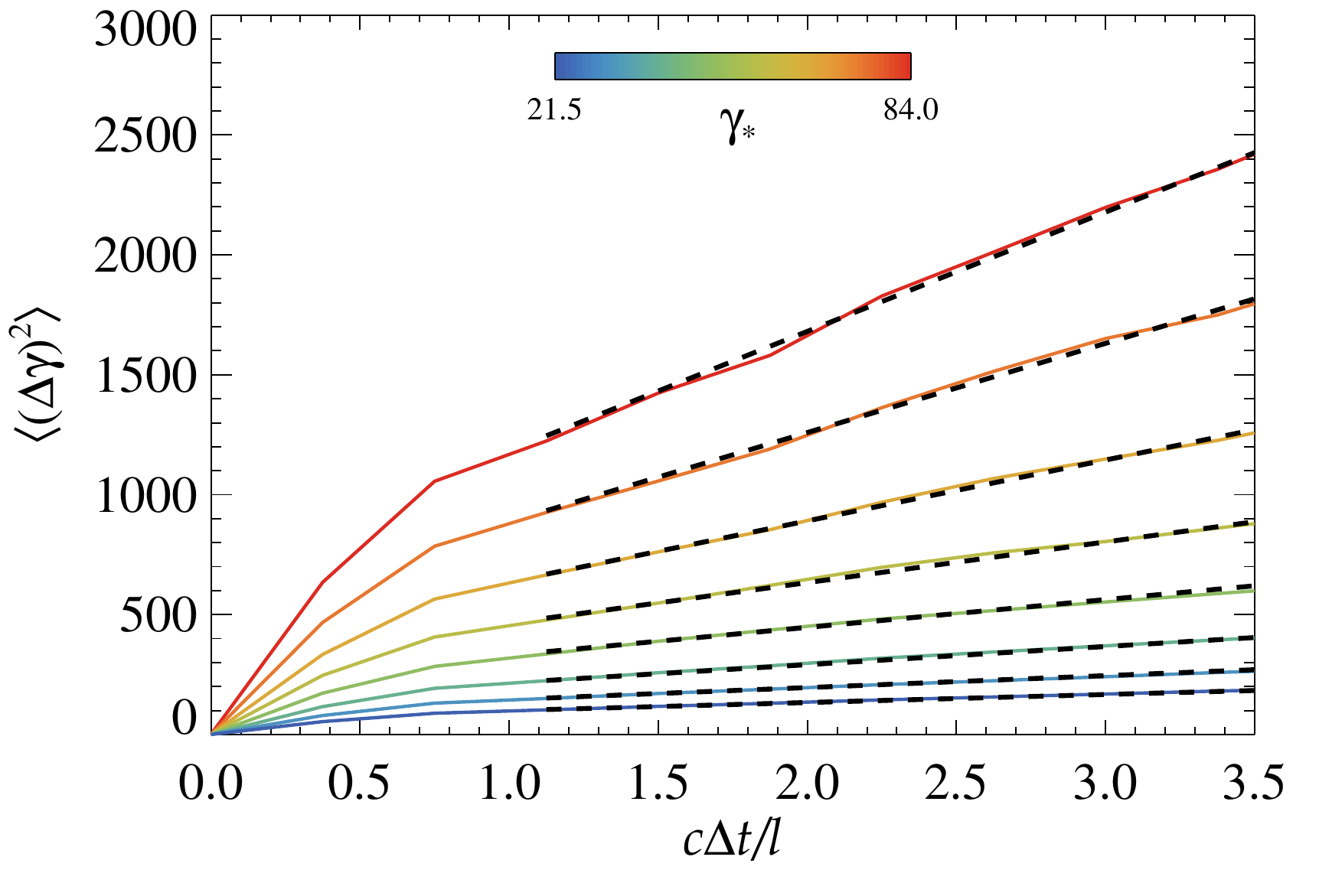}

\vspace{0.415cm}

\hspace*{-0.085cm}\includegraphics[width=8.75cm]{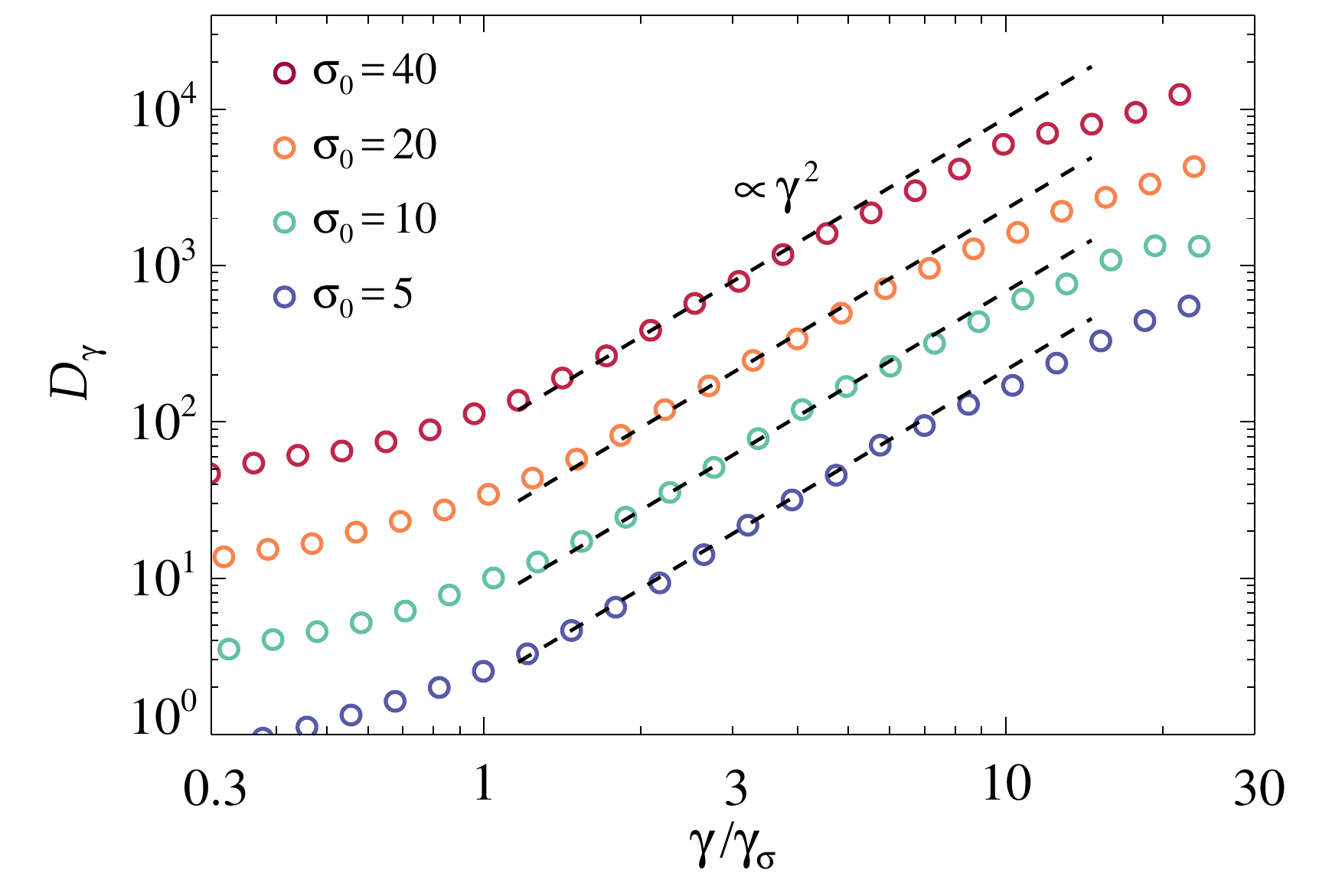}
\end{center}
\caption{Diffusion in energy space from 2D simulations with $\delta B_{{\rm{rms}}0}/ B_0=1$ and different initial magnetizations $\sigma_0$. Top panel: mean square variation of the Lorentz factor for particles binned in logarithmic intervals $[\gamma_*/\nu , \, \gamma_* \nu]$ with $\nu=1.1$ and $\gamma_* = 21.5 \to 84$ (from blue to red) at time $ct_*/l=5.25$ for the reference 2D simulation. The dashed black lines indicate linear fits. Bottom panel: energy diffusion coefficient $D_{\gamma}$ (in units of $c/l$), as a function of the Lorentz factor $\gamma$ (divided by ${\gamma _\sigma}$ to align cases with different magnetization), measured at the time interval $c \Delta t /l  = 1.875$ from four simulations having initial magnetization $\sigma_0 \in \left\{ {5,10,20,40} \right\}$.}
\label{fig_Diff_2D}
\end{figure}

\begin{figure}
\begin{center}
\hspace*{-0.085cm}\includegraphics[width=8.75cm]{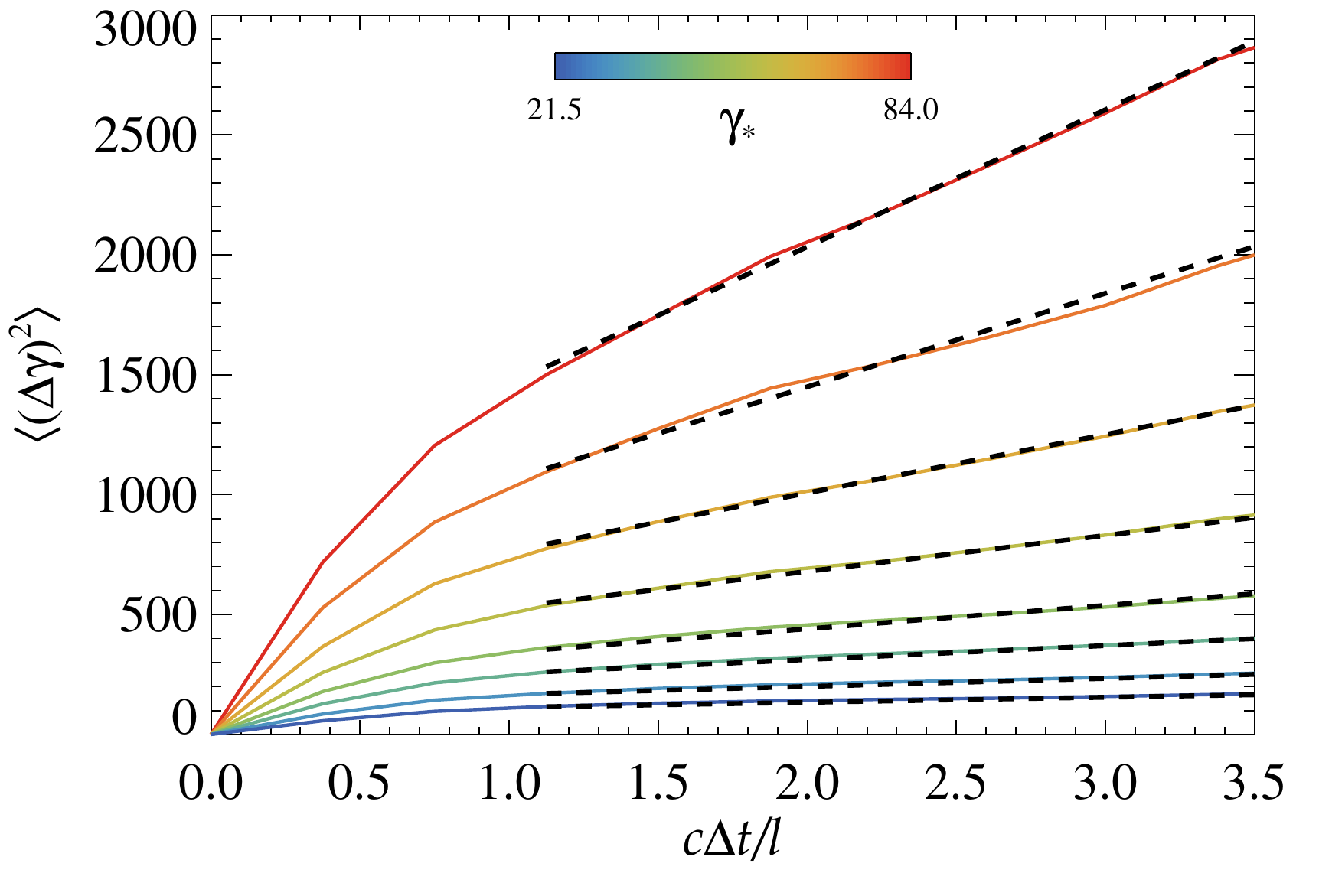}

\vspace{0.415cm}

\hspace*{-0.085cm}\includegraphics[width=8.75cm]{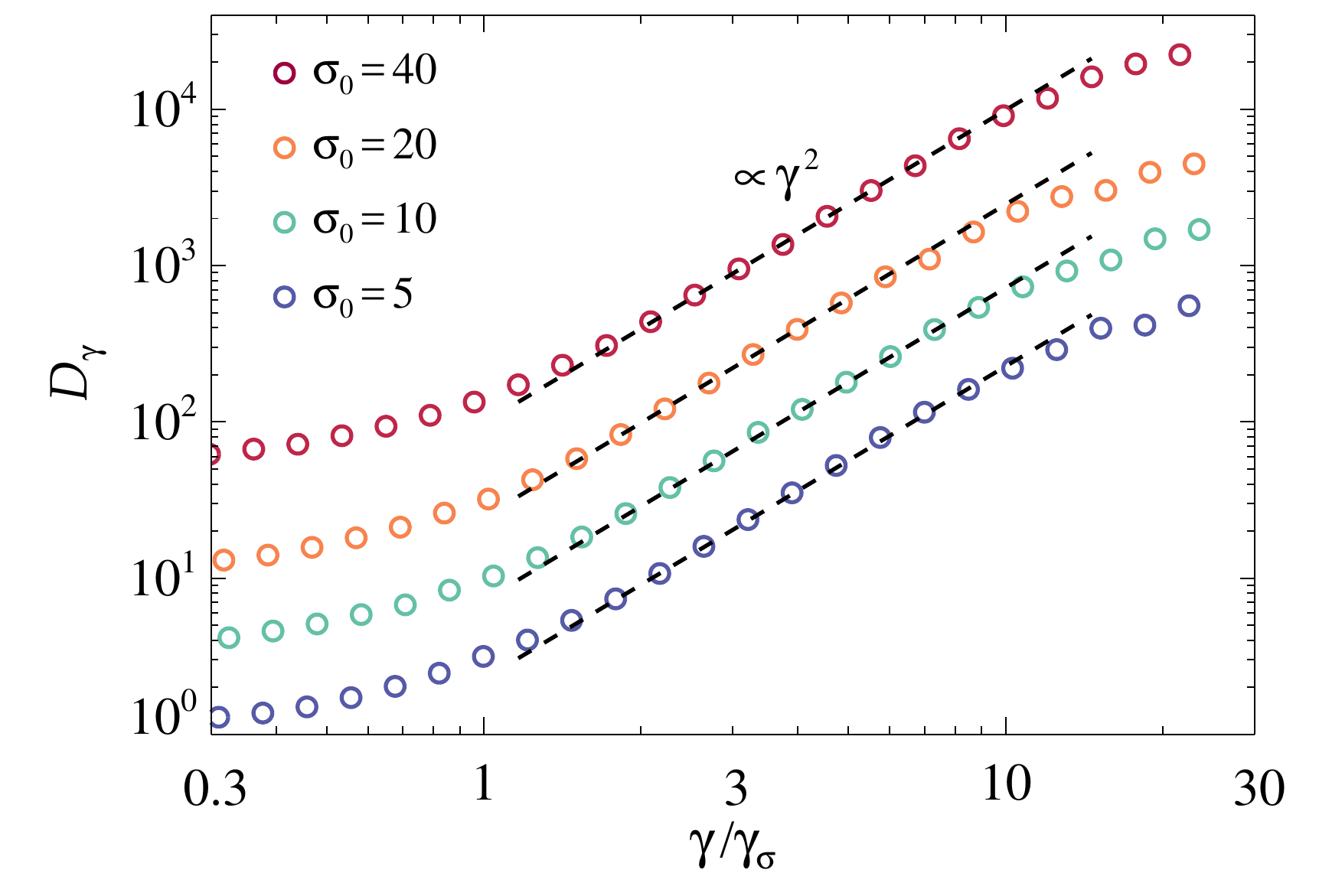}
\end{center}
\caption{Diffusion in energy space from 3D simulations with $\delta B_{{\rm{rms}}0}/ B_0=1$ and different initial magnetizations $\sigma_0$. Top panel: mean square variation of the Lorentz factor for particles binned in logarithmic intervals $[\gamma_*/\nu , \, \gamma_* \nu]$ with $\nu=1.1$ and $\gamma_* = 21.5 \to 84$ (from blue to red) at time $ct_*/l=3$ for the reference 3D simulation. The dashed black lines indicate linear fits. Bottom panel: energy diffusion coefficient $D_{\gamma}$ (in units of $c/l$), as a function of the Lorentz factor $\gamma$ (divided by ${\gamma _\sigma}$ to align cases with different magnetization), measured at the time interval $c \Delta t /l  = 1.875$ from four simulations having initial magnetization $\sigma_0 \in \left\{ {5,10,20,40} \right\}$.}
\label{fig_Diff_3D}
\end{figure}

We evaluate the energy diffusion coefficient directly from PIC simulations (see also \citet[]{Wong2019arXiv}). To this aim, from each of the 2D and 3D simulations employed for this analysis, we tracked in time the positions, four-velocities, and electromagnetic field values of about $10^7$ particles that were randomly selected at the beginning of the simulation.
From the time history of the particles evolution, we calculate the mean square $\gamma$-variation 
\begin{equation} \label{mean_square_variation_PIC}
\langle {{{(\Delta \gamma)}^2}} \rangle  = \frac{1}{N_p}\sum\limits_{n = 1}^{N_p} {{{\left( {{\gamma_n}(t) - {\gamma_n}({t_*})} \right)}^2}}
\end{equation}
for particles grouped in such a way that at an initial time $t_*$, they belong to the same energy bin ($N_p$ is the number of particles in the selected bin). The energy bin at $t_*$ is chosen accordingly to the particle energy calculated in the frame comoving with the drift velocity ${\bm{v}}_D = c{\bm{E}} \times {\bm{B}} /B^2$. For each particle, we perform a Lorentz boost from the observer/simulation frame to the local ${\bm{E}} \times {\bm{B}}$ frame, which results in the boosted Lorentz factor ${\gamma}'  = {\gamma_D} \gamma \left( {1 - {{{\bm{v}}_D \cdot {\bm{v}}}}/{c^2}} \right)$, where $\gamma_D = 1/\sqrt {1 - {({\bm{v}}_D/c)}^2}$ is the Lorentz factor for the drift velocity. Then, we evaluate Eq. (\ref{mean_square_variation_PIC}) by selecting particles in a small energy bin with ${\gamma}'  \in [\gamma_*/\nu , \, \gamma_* \nu]$, where $\gamma_*$ is the characteristic Lorentz factor of the energy bin, and $\nu$ is a constant factor that should be close to unity (we choose $\nu=1.1$). Finally, the diffusion coefficient in energy space can be calculated as
\begin{equation} \label{diffcomp}
{D_{\gamma} }= \frac{{ \langle {{{(\Delta \gamma)}^2}} \rangle }}{{2 \Delta t }} \, ,
\end{equation}
where $\Delta t  = t - t_*$ is a time interval that should be (i) long enough that the initial conditions become insignificant, and particles are in the diffusive regime; and (ii) short enough that the turbulence properties have not significantly changed. By using a large sample of particles in each energy bin, non-secular variations of the particle energy are averaged out and the mean energy gain can be obtained.

The results of our analysis of the particle energy diffusion are reported in Figs. \ref{fig_Diff_2D} and \ref{fig_Diff_3D}, for 2D and 3D simulations, respectively. The top frames show the mean square variation $\langle {{{(\Delta \gamma)}^2}} \rangle$ for particles binned according to their initial energy at $ct_*/l = 5.25$ for the reference 2D simulation and $ct_*/l = 3$ for the reference 3D simulation. In both cases, at the selected time $t_*$, turbulence is well developed and the time dependent magnetization calculated with the magnetic energy in  turbulent  fields is $\sigma(t_*) \sim 1$. The plots indicate that a diffusive behavior in energy space, $\langle {{{(\Delta \gamma)}^2}} \rangle \propto \Delta t$ (compare with dashed black lines), is achieved after $c \Delta t/l \sim 1$, in both 2D and 3D reference simulations. For shorter time intervals, particles preserve memory of the initial conditions and their motion is not diffusive. The slope at late times (dashed lines) depends on  particle energy, and it allows to quantify the energy dependence of the diffusion coefficient.

The bottom frames on Figs. \ref{fig_Diff_2D} and \ref{fig_Diff_3D} show the particle energy dependence of the energy diffusion coefficient from simulations with different initial magnetization $\sigma_0$ (indicated with different colors in the figures). The diffusion coefficient is evaluated using Eq.~(\ref{diffcomp}) in the time interval $c \Delta t /l = 1.875$, starting from $c t_*/l =5.25$ for  2D simulations and $c t_*/l = 3$ for 3D. We verified that the energy dependence remains the same when taking different time intervals, or by fitting the slopes of $\langle {{{(\Delta \gamma)}^2}} \rangle$ as a function of time in the diffusive regime (as done with the dashed lines in the top panels). In order to properly compare different $\sigma_0$, we display the energy diffusion coefficient as a function of the Lorentz factor normalized by $\gamma _\sigma$ (see Eq.~(\ref{gamma_st})).
The energy range where stochastic acceleration occurs starts at the beginning of the power-law high-energy tail of the particle spectrum, i.e. for ${\gamma} / {\gamma _\sigma} \gtrsim 1$.  In the stochastic acceleration range, the energy diffusion coefficient scales as $D_{\gamma} \propto \gamma^2$ (compare with the dashed black lines in the bottom panels). A similar dependence on the particle energy was also found in \citet{Lynn2014,KimuraApJ16,KimuraMNRAS19,Wong2019arXiv}, and is consistent with particle acceleration by non-resonant and/or broadened resonant interactions with the turbulent fluctuations \citep[e.g.][]{Skilling75,BlandfordEichler87,Schlickeiser89,ChandranPRL2000,ChoLazarian2006,Lemoine19}. Then, at higher energies, near the  high-energy cutoff of the power law, the energy dependence of $D_{\gamma}$ becomes weaker  as the particle Larmor radius gets closer to the energy-containing scale of the turbulence.

\begin{figure}
\begin{center}
\hspace*{-0.085cm}\includegraphics[width=8.75cm]{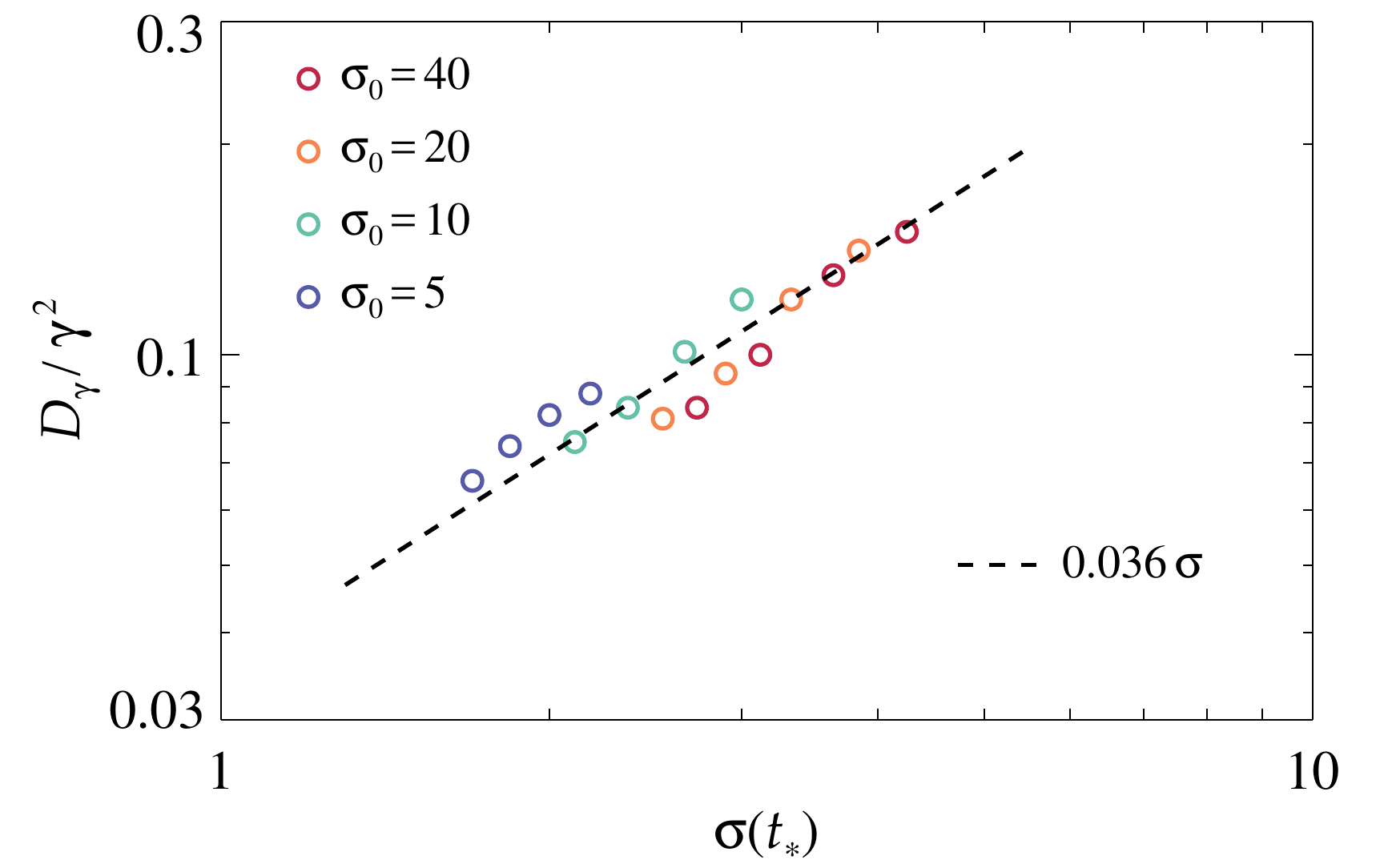}

\vspace{0.415cm}

\hspace*{-0.085cm}\includegraphics[width=8.75cm]{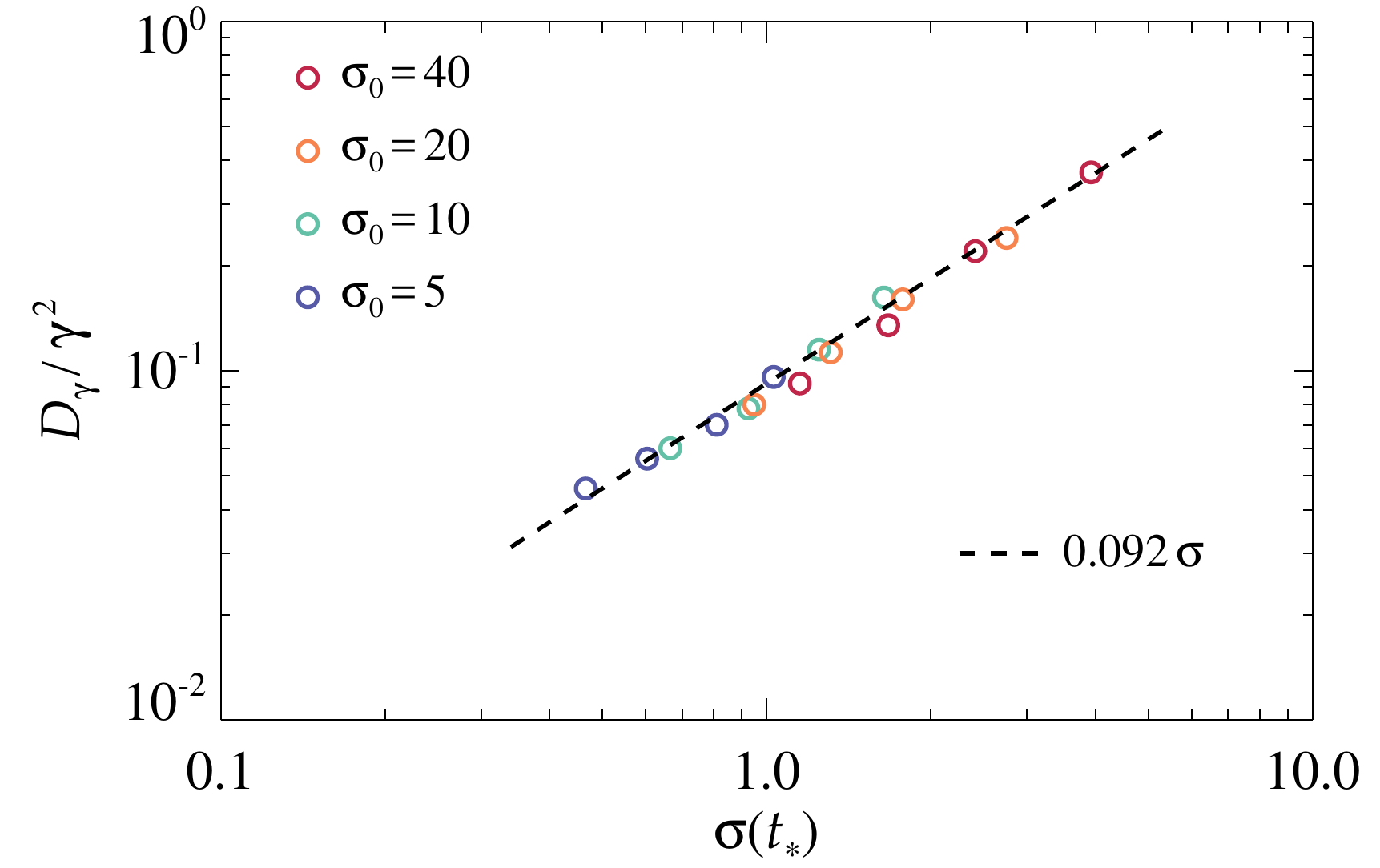}
\end{center}
\caption{Diffusion coefficient in energy space as a function of the actual magnetization $\sigma(t_*)$ from 2D simulations (top) and 3D simulations (bottom) with same $\delta B_{{\rm{rms}}0}/ B_0=1$ but different initial magnetization $\sigma_0 \in \left\{ {5,10,20,40} \right\}$. We employed $c \Delta t /l = 1.875$ for all measurements of the energy diffusion coefficient $D_{\gamma}$. Note that here $D_{\gamma}$ is in units of $c/l$. A linear fit is shown with a dashed black line. }
\label{fig_Diff_sigma}
\end{figure}

\begin{figure}
\begin{center}
\hspace*{-0.185cm}\includegraphics[width=8.75cm]{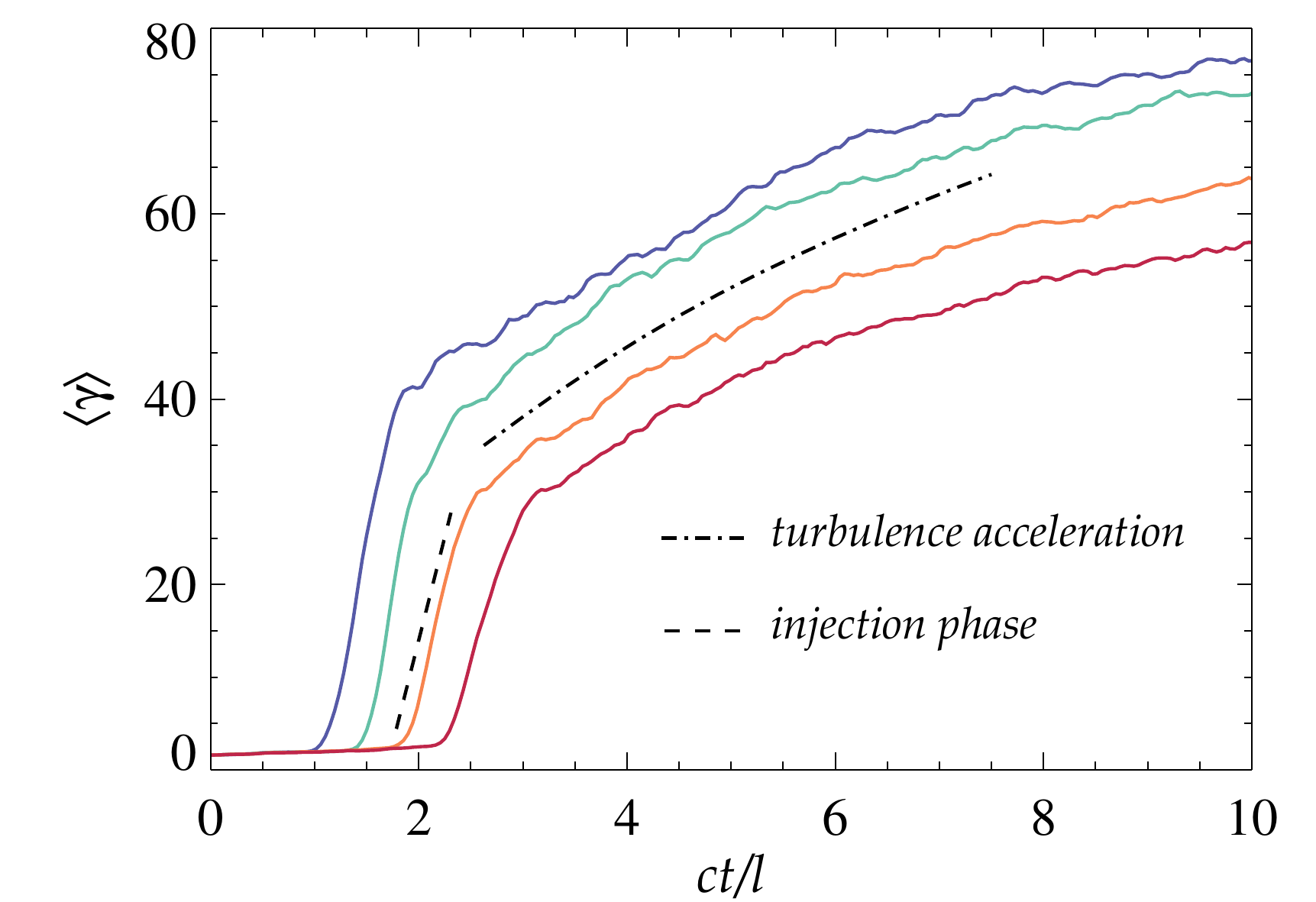}

\vspace{0.415cm}

\hspace*{-0.185cm}\includegraphics[width=8.75cm]{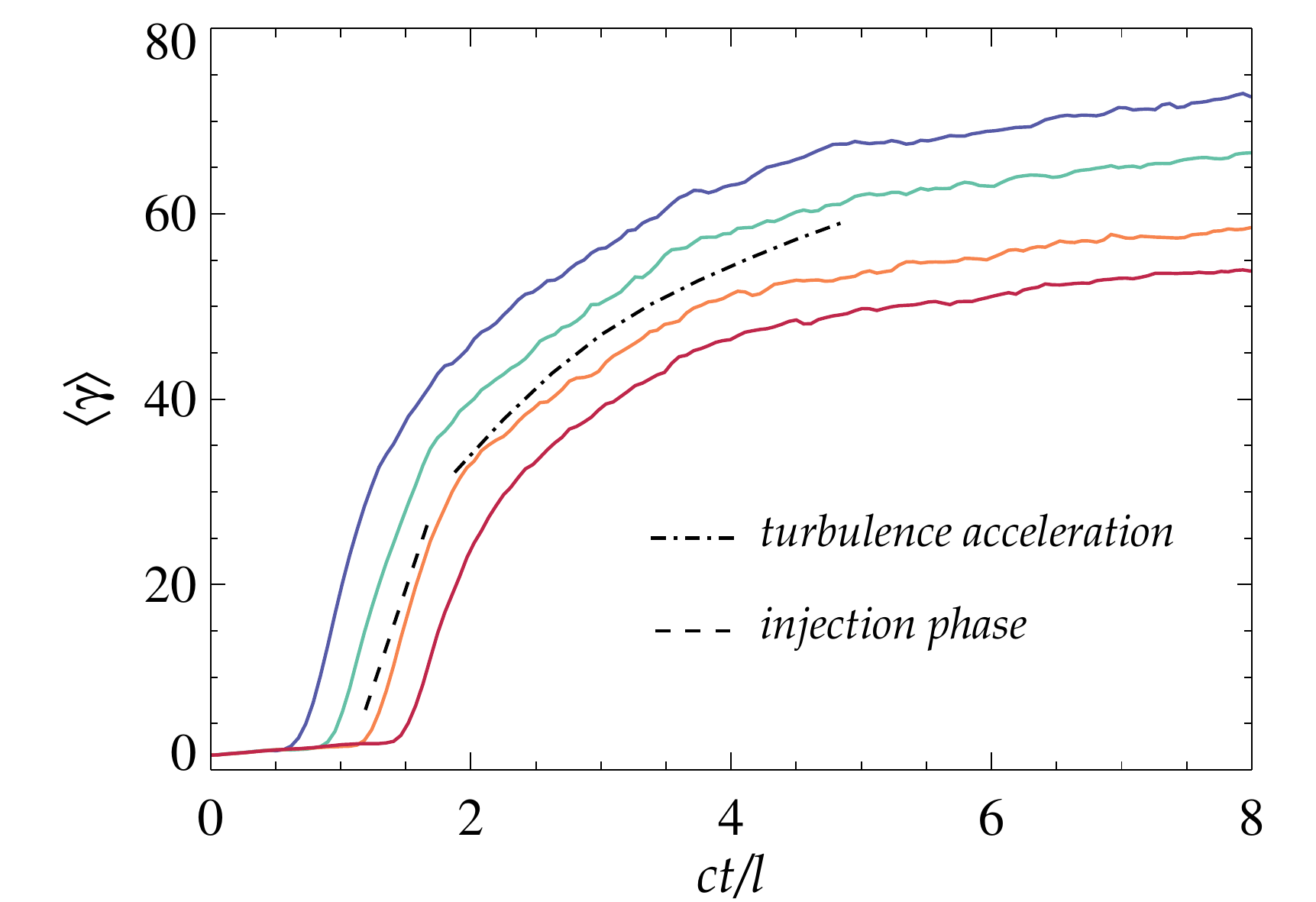}
\end{center}
\caption{Evolution of the mean Lorentz factor of different generations of particles undergoing injection at early times ($ct_{in\!j}/l\lesssim 2$) for 2D turbulence (top) and 3D turbulence (bottom). Both simulations have $\sigma_0=10$ and $\delta B_{{\rm{rms}}0}/ B_0=1$.  
The initial energy gain, due to the reconnection electric field, can be modeled as in Eq.~(\ref{eqRECONN}) with $\beta_R=0.05$ (dashed lines), while the subsequent evolution, governed by stochastic interactions with the turbulent fluctuations, follows Eq.~(\ref{eqSTOCevol}) (dot-dashed line).}
\label{fig_mean_gamma_evol}
\end{figure}

The energy diffusion coefficient depends also on the actual magnetization $\sigma(t_*) = {\langle {\delta B^2} \rangle }/{{4\pi n_0 w m c^2}}$. In order to better understand this dependence, in Fig.~\ref{fig_Diff_sigma} we plot the energy diffusion coefficient as a function the magnetization $\sigma$ at four different times $t_*$ (in the range $c t_*/l \in [4,6]$ for 2D, and $c t_*/l \in [2,4]$ for 3D) for the four simulations having different initial magnetization. Both 2D and 3D simulations show a clear trend of increasing diffusion coefficient with increasing magnetization. The 3D simulations are well fitted by a linear relation in $\sigma$ (compare with dashed black line),
\begin{equation} \label{Dgamma_numerical}
D_{\gamma}  \sim  0.1 \sigma  \, \left( {\frac{c}{l}} \right) \, \gamma^2  \, .
\end{equation}
This scaling can be understood by noting that for a stochastic process akin to the original Fermi mechanism \citep[e.g.][]{BlandfordEichler87,Lemoine19}, the energy diffusion coefficient is
\begin{equation} \label{Dgamma_analytical}
D_{\gamma}  = \frac{1}{3} \langle {\gamma_V^2 \beta_V^2} \rangle  \frac{c}{\lambda_{\rm mfp}} \gamma^2 \, ,
\end{equation}
where ${\langle {\gamma_V^2 \beta_V^2} \rangle}^{1/2}$ is the typical four-velocity of the scatterers, and $\lambda_{\rm mfp}$ is the particle scattering mean-free-path. Therefore, if we estimate the scattering mean-free-path as $\lambda_{\rm mfp} \sim ( B_0/\delta B_{\rm{rms}})^2 l$ and identify $\gamma_V \beta_V$ with the dimensionless Alfv\'enic four-velocity,  ${\langle {\gamma_V^2 \beta_V^2} \rangle}  \sim \langle {B^2} \rangle /4\pi nwm{c^2}$, from Eq. (\ref{Dgamma_analytical}) we obtain $D_{\gamma} \propto \sigma$ for $\langle {B^2} \rangle /B_0^2 \sim 1$, in agreement with Eq. (\ref{Dgamma_numerical}).
Then, from these results we can also estimate the stochastic acceleration timescale 
\begin{equation} \label{t_acc_stoc}
t_{acc} = {\left| {\frac{1}{\gamma }\frac{{d \langle \gamma \rangle }}{{dt}}} \right|^{ - 1}} \sim \frac{3}{\sigma }\frac{l}{c} \, .
\end{equation}
In our simulations, the acceleration timescale increases in time since $\sigma$ decreases in time as a combined effect of the decaying turbulent fluctuations $\delta B_{\rm{rms}}(t)$ and the increase of the enthalpy per particle $m c^2 w(t)$.

\subsection{Injection and turbulence acceleration}

As discussed in Sections  \ref{SecInjection} and \ref{SecEnergiz}, a large fraction of particles is preaccelerated by magnetic reconnection before being accelerated by scattering off the turbulent fluctuations. This two-stages acceleration process is shown in Fig.~\ref{fig_mean_gamma_evol} for both 2D and 3D simulations. Here, each colored curve represents the average Lorentz factor of particles having the same injection time $t_{in\!j}$ (within $\Delta t_{in\!j} = 0.32 c/l$ for 2D and $\Delta t_{in\!j} = 0.22 c/l$ for 3D). The linear growth from $\langle \gamma \rangle \sim 1$ up to $\langle \gamma \rangle \sim 30$ (i.e., the injection phase) is powered by field-aligned electric fields, whose magnitude is $|E_{\parallel}|\simeq \beta_R \delta B_{\rm rms}$, via
\begin{equation} \label{eqRECONN}
\frac{d \langle \gamma \rangle}{dt} =  \frac{e}{m c} \beta_R {\delta B_{\rm rms}} \, .
\end{equation} 
The dashed black lines in Fig.~\ref{fig_mean_gamma_evol} show Eq.~(\ref{eqRECONN}) taking a reconnection rate $\beta_R \simeq 0.05$, as appropriate for relativistic reconnection with guide field comparable to the alternating fields \citep{werner_17}. After this first acceleration phase, stochastic acceleration takes place, and, as discussed above, we can estimate
\begin{equation} \label{eqSTOCevol}
\frac{d \langle \gamma \rangle}{dt} = 4 \kappa_{\rm{stoc}} \sigma  \, \left( {\frac{c}{l}} \right) \, \gamma \, .
\end{equation}
with $\kappa_{\rm{stoc}} \sim 0.03$ from the 2D simulations and $\kappa_{\rm{stoc}} \sim 0.1$ from the 3D ones. Taking the temporal decay of the magnetic fluctuations, as well as the temporal increase of the relativistic enthalpy, directly from our simulations, we obtain the dot-dashed lines shown in Fig.~\ref{eqSTOCevol}. For the 3D case, the decrease in time of the stochastic acceleration rate is more pronounced than the 2D case as a consequence of the faster magnetic energy decay, and the corresponding decrease of the magnetization $\sigma$.

A final remark concerns the acceleration timescales associated with magnetic reconnection and turbulence fluctuations. Fast magnetic reconnection leads to the acceleration timescale $t_{acc} = \beta_R^{-1} (\rho_L/c)$, where $\rho_L$ is the particle Larmor radius. On the other hand, we have seen that stochastic acceleration by turbulent fluctuation yields $t_{acc} = (3/\sigma) (l/c)$. Therefore, for the hypothetical case in which reconnection could drive particles up to the highest energies ($\rho_L \sim l$), the acceleration timescale of fast magnetic reconnection could actually be longer than the one associated with the turbulence fluctuations for $\sigma \gtrsim 1$. Indeed, in this magnetically-dominated regime, turbulence provides an exceptionally fast acceleration mechanism that can potentially explain the most extreme astrophysical accelerators.

\section{Summary} \label{SecConclusions} 

In this article, we have presented the results of a series of first-principles kinetic PIC simulations of decaying turbulence in magnetically-dominated plasmas, with the goal of understanding how plasma turbulence, and its interplay with magnetic reconnection, can accelerate charged particles. 
We considered a pair (electron-positron) plasma, which is relevant for various astrophysical systems, such as jets from supermassive black holes, pulsar and magnetar magnetospheres, winds, and wind nebulae. 
In this regime, our computational domain ($2460^3$ cells in 3D; from $16400^2$ to $65600^2$ cells in 2D) is large enough to capture the turbulence cascade from large (MHD) scales to small (kinetic) scales.

In the following, we itemize the main points of this paper.
\vspace{0.1in}

1. The generation of a large population of nonthermal particles is a self-consistent by-product of both 2D and 3D magnetically-dominated turbulence. In particular, the late time particle energy spectrum displays a power-law high-energy range whose slope $p$, high-energy cutoff $\gamma_c$, and fraction of particles in the power-law tail $\zeta_{\rm{nt}}$ are markedly similar in 2D and 3D, even though the time development of the particle energy spectrum is different.

2. The power-law slope decreases (i.e., becomes harder) with increasing initial values of  magnetization and fractional strength of the turbulence fluctuations, with slopes that can be as hard as $p \lesssim 2$. In contrast, the initial plasma temperature does not affect the power-law slope, but only yields an overall energy shift to larger energies for higher initial plasma temperatures. For power-law energy tails with $p >2$ (i.e., not limited by energy budget constraints), the wider the MHD inertial range ${2 \pi}/{k_I d_e}$, the larger the high-energy cutoff, which can extend up to $\gamma_c \sim (e/m c^2) \sqrt{\langle B^2 \rangle} 2 \pi /k_I $, if turbulence survives long enough to allow the particles to reach this upper limit. The fact that the power-law starts close to the peak of the distribution yields a large fraction of particles in the nonthermal tail. For the physical parameters explored in this work, we obtain a number fraction $\zeta_{\rm{nt}} \sim 15 \%$ - $31 \%$ of particles in the nonthermal tail.

3. The majority of particles are injected into acceleration at regions of high electric curent density. More specifically, a large fraction of particles is extracted from the thermal pool and injected into the acceleration process by reconnecting current sheets. These reconnecting current sheets are strongly unstable to the formation of plasmoids, which allows fast magnetic reconnection to occur. We observe the development of plasmoids in current sheets formed as a self-consistent result of magnetized turbulence, both in 2D and in 3D. In 3D, they appear as a chain of flux ropes elongated in the direction of the mean magnetic field.

4. Reconnecting current sheets are efficient in injecting particles (i.e., they promote a large fraction of particles in the nonthermal tail) in spite of their small filling fraction, as they can process a large fraction of particles within the sheet lifetime.  
The efficiency remains high also when increasing system size, as we have shown that the plasmoid instability (whose properties are obtained from a tearing mode dispersion relation generalized for relativistically hot plasmas) ensures the triggering of fast magnetic reconnection within the lifetime of the large-scale current sheets, which are the ones that dominate the particle injection census. As a consequence, magnetic reconnection can process a large volume of plasma in few large (outer-scale) eddy turnover times (a volume $ {\mathcal{V}}_R \sim  \beta_R  L^3$ in one outer-scale eddy turnover time).

5. Particle acceleration at reconnecting current sheets can propel particles up to a typical Lorentz factor gain $\Delta \gamma_{in\!j} = \kappa \sigma \gamma_{th}$, after which the acceleration is continued by means of stochastic scattering off turbulent fluctuations. It is the stochastic acceleration process that allows particles to reach the highest energies, up to a Larmor radius roughly equal to the energy-containing scale of the turbulence. The work done by the electric field parallel to the magnetic field (which is expected at reconnecting current sheets), $W_\parallel$, is responsible for most of the early particle energy gain (injection). On the other hand, the second acceleration phase is powered by  perpendicular electric fields. For high-energy particles, i.e., such that $\Delta \gamma \gg \kappa \sigma \gamma_{th}$, we find $W_\bot \gg W_\parallel$, i.e., the work done by  perpendicular electric fields dominates the overall energy gain.

6. An additional confirmation of the fact that the parallel electric field controls the injection physics but not the subsequent acceleration process comes from a numerical simulation with extra (test) particles that do not feel parallel electric fields. This shows that the injection fraction is strongly suppressed. In fact, only a small fraction of these test particles participate in the acceleration process ($\zeta_{\rm{nt}}$ decreases by almost two orders of magnitude). On the other hand, for those test particles that can participate in the acceleration process, the power-law slope $p$ is very similar to that of the regular particles. This  indicates that  acceleration  by the perpendicular electric field controls the slope of the power-law high-energy tail.

7. The fact that different energization mechanisms dominate at different energy ranges affects the particle pitch-angle distribution, $f \left( {\cos \alpha , \gamma} \right)$. 
We find that the pitch-angle distribution develops distinguishing features at \emph{low}, \emph{intermediate}, and \emph{high} values of $\gamma$. These values depend on the initial mean Lorentz factor and magnetization. 
For $\gamma \sim (\sigma_0/2) \gamma_{th0}$, particles velocities are strongly aligned/antialigned with the local magnetic field ${\bm{B}}$, while at $\gamma \gg 5 (\sigma_0/2) \gamma_{th0}$, particles velocities are mostly perpendicular to ${\bm{B}}$. At intermediate energies such that $\gamma \sim 5 (\sigma_0/2) \gamma_{th0}$,  particles follow a distribution which has minima for both parallel and perpendicular directions (i.e., at $\cos \alpha = \pm 1,0$). These results are robust in both 2D and 3D turbulence. In both cases, the overall population of particles is dominated by the particles having pitch-angle cosine close to $\cos \alpha = \pm 1$, as the low-energy population controls the number census.

8. The different energization mechanisms are also responsible for producing a gyrotropic four-velocity distribution with distinct features in the direction pertaining to the mean magnetic field ${\bm{B}}_0  = B_0 {\bm{\hat z}}$. Specifically, the domain-averaged four-velocity distribution is elongated in the $\gamma \beta_z$ direction at low particle energies, due to the ${\bm{v}} \cdot {\bm{E}}_\parallel$ energization, while it becomes elongated in the direction perpendicular to the mean field at high particle energies, due to the ${\bm{v}} \cdot {\bm{E}}_\bot$ energization. At intermediate energies the distribution peaks at intermediate angles (i.e., at 45 degrees from the $\gamma \beta_z$ axis).

9. After the injection phase, particles exhibit a diffusive energy behavior in both 2D and 3D turbulence. We measured the diffusion coefficient in energy space directly from our PIC simulations, showing that $D_{\gamma} \propto \gamma^2$ for the  energy range of the power law. Furthermore, $D_{\gamma} \propto \sigma$, with $\sigma$ being the time-dependent magnetization. 
The estimated energy diffusion coefficient $D_{\gamma}  \sim  0.1 \sigma  ({c}/{l}) \, \gamma^2$ gives an acceleration timescale that can be very fast, $t_{acc} \sim (3/\sigma) (l/c)$, comparable to that of fast magnetic reconnection or even higher, depending on the plasma magnetization.

10. The mean energy gain of particles during the first acceleration phase (injection) is well described by linear acceleration by the typical reconnection electric field. Then, the subsequent mean energy gain due to  stochastic scattering off the turbulent fluctuations follows from the energy diffusion coefficient $D_{\gamma}$. In our simulations of decaying turbulence, as the plasma magnetization decreases due to the magnetic field annihilation, the stochastic acceleration timescale gets longer over time and the stochastic acceleration process eventually saturates.

\vspace{0.1in}

The aforementioned findings have implications for our understanding of the generation of nonthermal particles in high-energy astrophysical sources. The main astrophysical implications are: (i) the power-law slopes of the emitting particles, which are predicted to be harder for larger plasma magnetizations and stronger turbulent fluctuations, can potentially  explain the hard radio spectrum of the Crab Nebula \citep[e.g.][]{LyutikovMNRAS2019}; (ii) the anisotropy of the particle pitch-angle distribution, for which the synchrotron spectrum of the emitting particles is expected to be different than the commonly-assumed case of isotropic particles, has consequences for our understanding of emission from AGN jets \citep[e.g.][]{Tavecchio2019}; (iii)  magnetically-dominated  plasma turbulence leads to particle acceleration on rapid timescales, which can be even shorter than those associated with fast magnetic reconnection and are then capable to explain particle acceleration in the most extreme astrophysical accelerators \citep[e.g.][]{Takahashi09}.

\acknowledgments
We acknowledge fruitful discussions with Mikhail Medvedev, Jonathan Zrake, Vah\'{e} Petrosian, Martin Lemoine, Aaron Tran, Chuanfei Dong, Yi-Min Huang, and Maxim Lyutikov. This research acknowledges support from DoE DE-SC0016542, NSF ACI-1657507 and NASA ATP NNX17AG21G. The simulations were performed on Columbia University (Habanero and Terremoto), NASA-HEC (Pleiades), NERSC (Cori and Edison), TACC (Stampede2) and ORNL (Titan) resources.

\section*{Appendix} \label{Appendix}

\begin{figure}
\begin{center}
\hspace*{-0.035cm}\includegraphics[width=8.65cm]{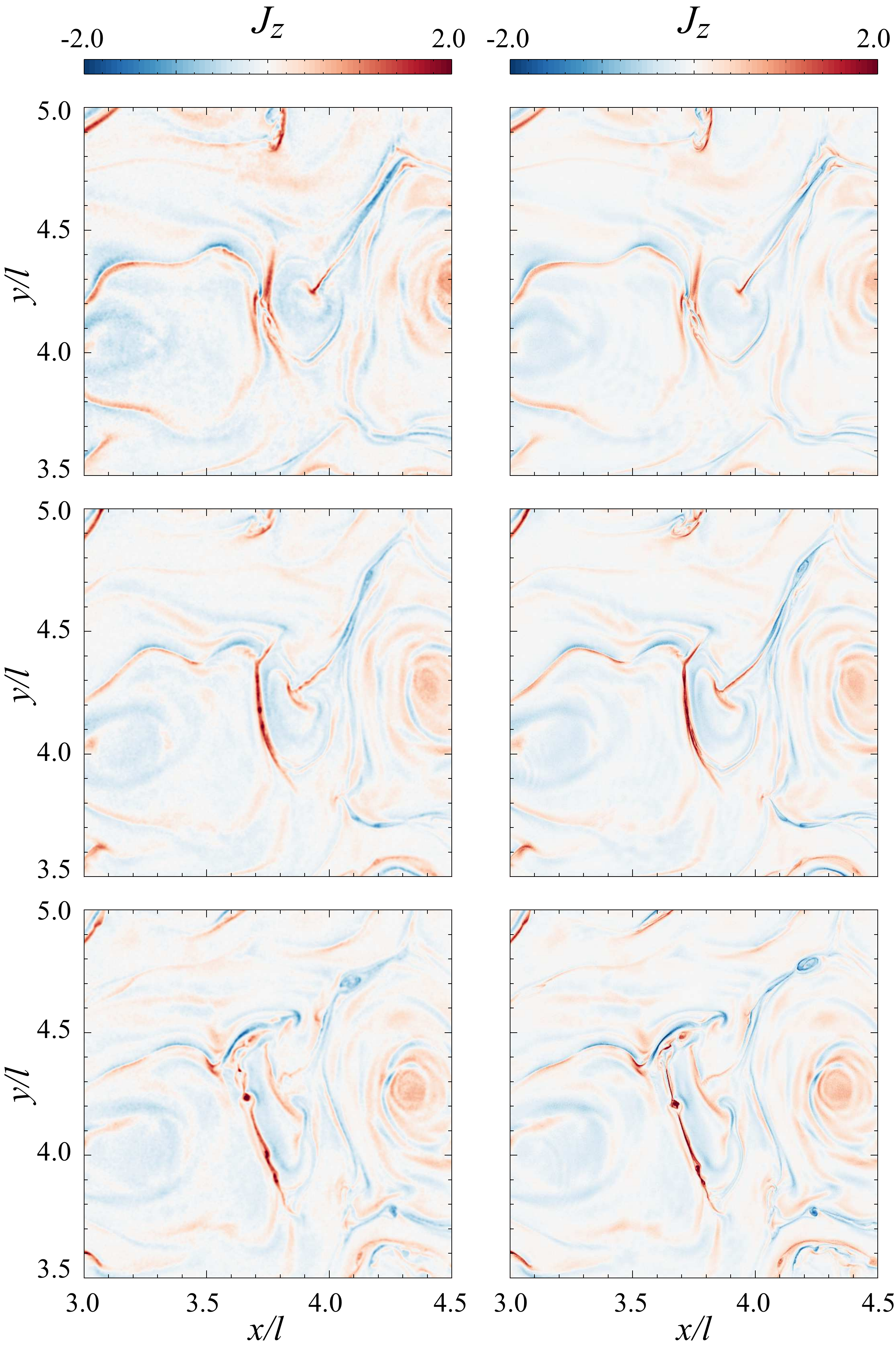}
\end{center}
\caption{Formation of current sheets and plasmoids (in the central part of the zoomed domain) from two 2D simulations where the initial plasma skin depth $d_{e0}$ is resolved with $3$ cells (left column) and $10$ cells (right column). Top, middle, and bottom panels refer to frames taken at $ct/l=1.8, \, 2.0, \, 2.2$. In both cases $\sigma_0=10$, $\delta B_{{\rm{rms}}0}/ B_0=1$, $L/d_{e0}=1640$, and $l=L/8$. In both cases we employ $16$ particles per cell.}
\label{fig40}
\end{figure}

\begin{figure}
\begin{center}
\hspace*{-0.185cm}\includegraphics[width=8.75cm]{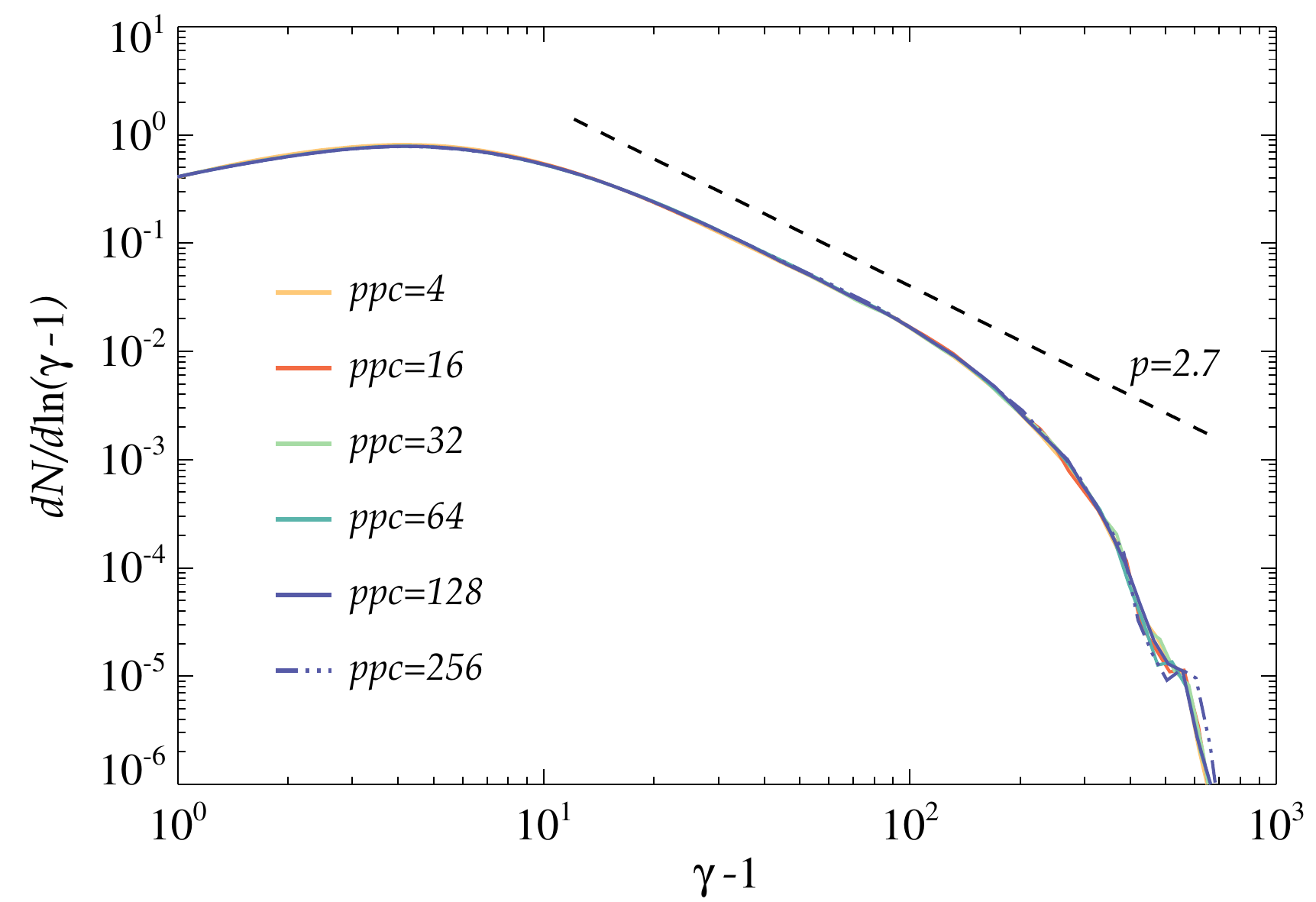}
\end{center}
\caption{Particle spectra $dN/d\ln(\gamma-1)$ at late times for 2D simulations with $\sigma_0=10$, $\delta B_{{\rm{rms}}0}/ B_0=1$, $L/d_{e0}=820$, and $l=L/8$, using different values of computational particles per cell, from ppc=4 to ppc=256.}
\label{ppc_convergence}
\end{figure}

In the magnetically-dominated regime ($\sigma_0 \gg 1$) studied here, we have verified that the presented results are converged with the adopted grid resolution and number of particles per cell.  

In this study, we presented results where the initial plasma skin depth $d_{e0}$ is resolved with $10$ cells in 2D and $3$ cells in 3D. However, $3$ cells per initial skin depth $d_{e0}$ (the skin depth increases during the simulation as the mean Lorentz factor increases as a result of magnetic field annihilation) are already sufficient to resolve current sheets and plasmoids. In Fig. \ref{fig4}, we show the electric current density $J_z$ taken at three different times ($ct/l=1.8, \, 2.0, \, 2.2$) from two 2D simulations where $d_{e0}$ is resolved with $3$ cells (left column) and $10$ cells (right column), which produce analogous structures.

We have also checked for convergence with respect to computational particles per cell. A comparison of the late time spectra from simulations employing different particles per cell (up to 256) is shown in Fig. \ref{ppc_convergence}, for simulations having domain size $L/d_{e0}=820$. We can see that the particle spectra are converged with the adopted particle resolution. Indeed, noise-level fluctuations are on small scales and do not affect the acceleration process in the regime investigated here.

\end{document}